\documentclass[default]{svmult}
\usepackage{hyperref}

\usepackage[final]{graphicx}
\usepackage{amssymb}
\usepackage{amsmath}

\usepackage{algorithm}
\usepackage{algorithmic}

\usepackage{caption}
\usepackage{colortbl}

\usepackage{enumerate}
\usepackage{enumitem}

\usepackage{tabto}
\usepackage{xcolor}
\newcommand{\comment}[1]{\tabto{6.5cm}\textcolor{gray}{// #1}}

\usepackage{wrapfig}

\usepackage{multirow}

\usepackage{multicol}        
\usepackage[bottom]{footmisc}

\usepackage{newtxtext}       %
\usepackage{newtxmath}       

\newif\ifTR\TRtrue
\newcommand{\tronly}[1]{\ifTR{#1}\else{}\fi}

\newcommand{\nat}{\mathbb{N}}
\newcommand{\zed}{\mathbb{Z}}
\newcommand{\bool}{\mathbb{B}}
\newcommand{\ptt}{\mathbb{S}}
\newcommand{\type}{\mathbb{T}}
\newcommand{\jstat}{\mathbb{J}}
\newcommand{\Kind}{\mathbb{K}}

\newcommand{\hasvalue}[1]{%
  \raisebox{-0.3ex}{$\xrightarrow{\smash{\raisebox{-0.1ex}{$\scriptstyle#1$}}}$}}

\newcommand{\pointsto}[1]{%
  \raisebox{-0.3ex}{$\xrightarrow{\smash{\raisebox{-0.1ex}{$\scriptstyle#1$}}}$}}

\newcommand{\nil}{\#}
\newcommand{\UNDEF}{\bot}

\newcommand{\first}{\mathtt{fst}}
\newcommand{\last}{\mathtt{lst}}
\newcommand{\all}{\mathtt{all}}
\newcommand{\reg}{\mathtt{reg}}
\newcommand{\dls}{\mathtt{dls}}
\newcommand{\ptr}{\mathtt{ptr}}

\newcommand{\GVar}{\textit{GVar}}
\newcommand{\MG}{\textit{MG}}
\newcommand{\MI}{\textit{MI}}
\newcommand{\SVar}{\textit{SVar}}
\newcommand{\Var}{\textit{Var}}
\newcommand{\addr}{\textit{addr}}
\newcommand{\bseq}{\textit{bseq}}
\newcommand{\cost}{\textit{cost}}
\newcommand{\false}{\textit{false}}
\newcommand{\hfo}{\textit{hfo}}
\newcommand{\kind}{\textit{kind}}
\newcommand{\len}{\textit{len}}
\newcommand{\lenThr}{\textit{lenThr}}
\newcommand{\level}{\textit{level}}
\newcommand{\nfo}{\textit{nfo}}
\newcommand{\of}{o\!f}
\newcommand{\ofprime}{\of'}                
\newcommand{\backspace}{\hspace{-0.30em}}  
\newcommand{\ty}{t}                        
\newcommand{\parent}{\textit{parent}}
\newcommand{\pfo}{\textit{pfo}}
\newcommand{\size}{\textit{size}}
\newcommand{\tg}{\textit{tg}}
\newcommand{\true}{\textit{true}}
\newcommand{\valid}{\textit{valid}}

\newcommand{\hfoBS}{\textit{hfo}\hspace{-0.05em}}
\newcommand{\nfoBS}{\textit{nfo}\hspace{-0.05em}}
\newcommand{\pfoBS}{\textit{pfo}\hspace{-0.05em}}

\newcommand{\iso}{\simeq}
\newcommand{\jsupset}{\sqsupset}
\newcommand{\jsubset}{\sqsubset}
\newcommand{\join}{\Join}

\newcommand{\miso}{$\iso$}
\newcommand{\mjsupset}{$\jsupset$}
\newcommand{\mjsubset}{$\jsubset$}
\newcommand{\mjoin}{$\join$}

\newcommand{\wrong}{{\large$\times$ }}

\newcommand{\diff}{dif\hspace*{-0.4mm}f}

\usepackage{listings}
\begin{document}

\title{Algorithmic Details behind the Predator Shape Analyser Based on Symbolic
Memory Graphs}

\titlerunning{Algorithmic Details behind the Predator Shape Analyser}


\author{Kamil Dudka\inst{1,2},
        Petr Muller\inst{1,2},
        Petr Peringer\inst{1},
        Veronika \v{S}okov\'{a}\inst{1},
        Tom\'{a}\v{s} Vojnar\inst{1}}

\institute{{Brno University of Technology, Faculty of Information Technology, 
  Czech~Republic}
\and
  {Red Hat Czech, Brno, Czech Republic}}


\maketitle


\vspace*{-2mm}\abstract{This chapter, which is an extended and revised version
of the conference paper \cite{predatorSAS}, concentrates on a detailed
description of the algorithms behind the Predator shape analyser based on
abstract interpretation and symbolic memory graphs. Predator is particularly
suited for formal analysis and verification of sequential non-recursive C code
that uses low-level pointer operations to manipulate various kinds of linked
lists of unbounded size as well as various other kinds of pointer structures of
bounded size. The tool supports practically relevant forms of pointer
arithmetic, block operations, address alignment, or memory reinterpretation. We
present the overall architecture of the tool, along with selected implementation
details of the tool as well as its extension into so-called Predator Hunting
Party, which utilises multiple concurrently-running Predator analysers with
various restrictions on their behaviour. Results of experiments with Predator
within the SV-COMP competition as well as on our own benchmarks are provided.}


\vspace*{-3mm}\section{Introduction}\vspace*{-3mm}

Dealing with \emph{pointers} and \emph{dynamic linked data structures} belongs
among the most challenging tasks of formal analysis and verification of software
due to a need to cope with infinite sets of reachable program configurations
having the form of complex graphs. This task becomes even more complicated when
considering low-level memory operations such as pointer arithmetic, safe usage
of pointers with invalid targets, block operations with memory, reinterpretation
of the memory contents, or address alignment.


\enlargethispage{5mm}

In this chapter, we present a fully-automated approach to formal verification of
list manipulating programs that is behind the \emph{Predator shape analyser} and
that is designed to cope with all of the above mentioned low-level memory
operations.  The approach is based on representing sets of heap graphs using the
so-called \emph{symbolic memory graphs} (SMGs). This representation is to some
degree inspired by works on separation logic with higher-order list predicates
\cite{InvaderCAV07}, but it is graph-based and uses a much more fine-grained
memory model. In particular, SMGs and the algorithms designed to make them
applicable in a fully-automated shape analysis based on abstract interpretation
allow one to deal with \emph{byte-precise} offsets of fields of objects, offsets
of pointer targets, as well as object sizes. As our experiments show, Predator
can successfully handle many programs on which other state-of-the-art
fully-automated approaches fail (by not terminating or by producing false
positives or even false negatives).

\vspace*{-3mm}\subsubsection*{Symbolic Memory Graphs}\vspace*{-2mm}

Going into slightly more detail, SMGs are directed graphs with two kinds of
nodes: \emph{objects} and \emph{values}. Objects represent allocated memory and
are further divided into \emph{regions} representing individual memory areas and
\emph{list segments} encoding linked sequences of $n$ or more regions
uninterrupted by external pointers (for some $n \geq 0$). Values represent
addresses and other data stored inside objects. Objects and values are linked by
two kinds of edges: \emph{has-value} edges from objects to values and
\emph{points-to} edges from value nodes representing addresses to objects. For
efficiency reasons, we represent equal values by a single value node.  We
explicitly track sizes of objects, byte-precise offsets at which values are
stored in them, and we allow pointers to point to objects with an arbitrary
offset, i.e., a pointer can point \emph{inside} as well as \emph{outside} an
object, not just at its beginning as in many current analyses.

SMGs allow us to handle possibly cyclic, nested (with an arbitrary depth),
and/or shared singly- as well as doubly-linked lists (for brevity, below, we
concentrate on doubly-linked lists only). Our analysis can fully automatically
recognise linking fields of the lists as well as the way they are possibly
hierarchically nested. Moreover, the analysis can easily handle lists in the
form common in system software (in particular, the Linux kernel), where list
nodes are linked through their middle, pointer arithmetic is used to get
to the beginning of the nodes, pointers iterating through such lists can
sometimes safely point to unallocated memory, the forward links are pointers to
structures while the backward ones are pointers to pointers to structures,~etc.

\enlargethispage{5mm}

To reduce the number of SMGs generated for each basic block of the analysed
program, we use a \emph{join operator} working over SMGs. Our join operator is
based on simultaneously traversing two SMGs while trying to merge the
encountered pairs of objects and values according to a set of rules carefully
tuned through many experiments to balance precision and efficiency (see
Section~\ref{sec:join} for details). Moreover, we use the join operator as the
core of our \emph{abstraction}, which is based on merging neighbouring objects
(together with their sub-heaps) into list segments. This approach leads to a
rather easy to understand and---according to our experiments---quite efficient
abstraction algorithm. In the abstraction algorithm, the join is not applied to
two distinct SMGs, but a single one, starting not from pairs of program
variables, but the nodes to be merged. Further, we use our join operator as a
basis for checking \emph{entailment} on SMGs too (by observing which kind of
pairs of objects and values are merged when joining two SMGs). In order to
handle lists whose nodes \emph{optionally} refer to some regions or sub-lists
(which can make some program analyses diverge and/or produce false alarms
\cite{InvaderTR07}), our join and abstraction support so-called \emph{0/1
abstract objects}, i.e., objects that may but need not be present.


Since on the low level, the same memory contents can be interpreted in different
ways (e.g., via unions or type-casting), we incorporate into our analysis the
so-called read, write, and join \emph{reinterpretation}. In particular, we
formulate general conditions on the reinterpretation operators that are needed
for soundness of our analysis, and then instantiate these operators for the
quite frequent case of dealing with blocks of nullified memory. Due to this, we
can, e.g., efficiently handle initialization of structures with tens or hundreds
of fields commonly allocated and nullified in practice through a single call of
\texttt{calloc}, at the same time avoiding false alarms stemming from that some
field was not explicitly nullified. Moreover, we provide a support for block
operations like \texttt{memmove} or \texttt{memcpy}. Further, we extend the
basic notion of SMGs to support pointers having the form of not just a single
address, but an interval of addresses. This is needed, e.g., to cope with
address alignment or with list nodes that are equal up to their incoming
pointers arriving with different offsets (as common, e.g., in memory
allocators).

\vspace*{-3mm}\subsubsection*{The Predator Analyser}\vspace*{-2mm}

The approach sketched above has been implemented in the Predator
analyser~\cite{predator}. Predator automatically proves absence of various
\emph{memory safety errors}, such as invalid dereferences, invalid free
operations, or memory leaks. Moreover, Predator can also provide the user with
the derived \emph{shape invariants}. Since SMGs provide a~rather detailed
memory model, Predator produces fewer false alarms than other tools,
and on the other hand, it can discover bugs that may be undetected by other
state-of-the-art tools (as illustrated by our experimental results). In
particular, Predator can discover \emph{out-of-bound dereferences} (including
buffer overflows on the stack, i.e., the so-called \emph{stack smashing}, which
can alter the execution flow and cause serious security vulnarabilities) as well
as nasty bugs resulting from dealing with \emph{overlapping blocks of memory} in
operations like \texttt{memcpy}.

\enlargethispage{5mm}

The Predator analyser is a basic building block of the so-called \emph{Predator
Hunting Party (PredatorHP)}. In PredatorHP, the Predator analyser is called a
\emph{verifier} due to its sound over-approximation of program semantics. It is
allowed to claim programs correct, but it is not allowed to warn about bugs
since it can produce false alarms due to the over-approximation it uses. For
detecting errors, PredatorHP contains several \emph{Predator hunters} which use
SMG-based algorithms but with no abstraction on pointer structures. The
different hunters differ in their search strategy (depth-first or breadth-first)
and the limits imposed on their search. They can warn about errors (not
producing any false alarms unless caused by abstraction of non-pointer data) but
cannot prove programs correct. The only exception is the case when the given
program has a finite state space and that is entirely explored. As our
experiments show, PredatorHP can indeed avoid many false alarms and it also
reduces the wall-clock time of the analysis (while usually increasing the CPU
time).

Predator has been successfully validated on a number of case studies, including
various operations on lists commonly used in the Linux kernel as well as code
taken directly from selected low-level critical applications (without any
changes up to adding a test environment). In particular, we present results of
our experiments with the memory allocator from the Netscape portable runtime
(NSPR), used, e.g., in Firefox, and the \texttt{lvm2} logical volume manager.
All of the case studies are available within the distribution of Predator. As we
show on the experimental results that we obtained, many of our case studies go
beyond what other currently-existing fully-automated program analysis and
verification tools can handle. We further present an experimental evaluation of
Predator and PredatorHP on the benchmarks of the International Software
Verification Competition 2019 (SV-COMP'19) where we concentrate mainly on the
effects of using PredatorHP and also on some recent extensions of Predator.

\vspace*{-3mm}\subsubsection*{Outline of the Chapter}\vspace*{-2mm}

The rest of the chapter is organised as follows. 
First, in Section~\ref{sec:intuition}, we extend the basic intuition on SMGs
provided above by a somewhat more detailed but still intuitive explanation.
Section~\ref{sec:exampleOfSMGs} provides a further illustration of SMGs through
two examples of SMGs representing data structures used in practice.
Then, we provide formal definitions of SMGs and related notions in
Section~\ref{sec:defOfSMGs}.

Subsequently, Section~\ref{sec:operations} describes principles of the
operations on SMGs that are needed for implementing a shape analyser on top of
SMGs.
The principles presented in Section~\ref{sec:operations} should suffice for
getting a decent understanding of all the operations.
However, for those more interested, Appendices
\ref{app:reinterpret}--\ref{app:condSymExec} provide a detailed description of
all the operations in the form of pseudo-code.

At this point, let us stress once more that the main computation loop of
Predator is that of \emph{abstract interpretation}.
It uses sets of symbolic program configurations based on SMGs, i.e., a
disjunctive extension of the basic abstract domain, to abstractly record sets of
program configurations reachable at program locations.
SMG abstraction is used to implement widening, and SMG join and entailment
(based itself on the SMG join) are used to reduce the number of SMGs tracked at
particular program locations.
Due to this very standard approach being employed, we do not provide a~further
detailed explanation of the basic computation loop itself.
However, in Section~\ref{sec:RunEx}, we provide a rather complete illustration
of the entire computation loop on an example program.

Section~\ref{sec:extensions} provides various extensions of the basic notion of
SMGs. 
Section~\ref{sec:implementation} discusses the architecture of Predator along
with its various implementation details.
Moreover, Section~\ref{sec:implementation} also introduces PredatorHP.
Section~\ref{sec:experiments} provides our experimental results.
In Section~\ref{sec:related}, we discuss related works.

Finally, Appendix~\ref{sec:tutorial} contains a brief tutorial on running and
configuring Predator and PredatorHP.

\enlargethispage{5mm}

\vspace*{-3mm}\section{Symbolic Memory Graphs}\label{sec:SMGs}\vspace*{-2mm}

\begin{figure}[t] \centering \resizebox{0.8\hsize}{!}{
\includegraphics{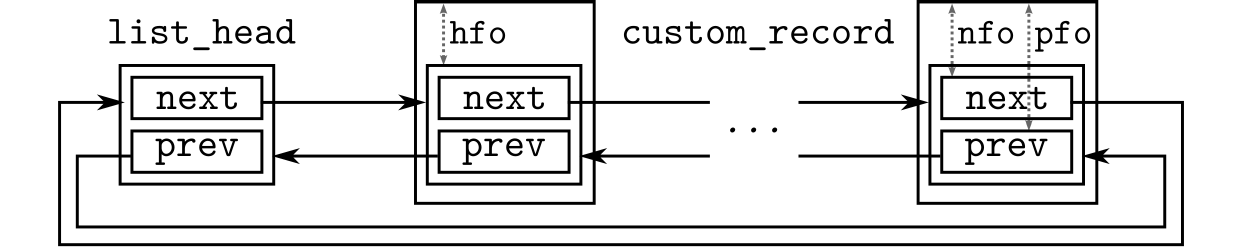} }\vspace{1em} \resizebox{0.8\hsize}{!}{
\includegraphics{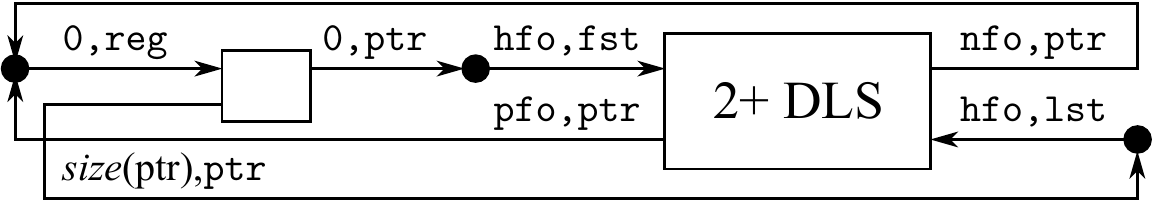} }
  
  \caption{A cyclic Linux-style DLL (top) and its SMG (bottom), with some SMG
  attributes left out for readability. For the meaning of the acronyms (e.g.,
  \texttt{hfo} stands for the head-field offset) see Section~\ref{sec:defOfSMGs}.}

  \vspace*{-2mm}
  \label{fig:linuxDLSandItsSMG}
\end{figure}

In this section, we introduce the notion of \emph{symbolic memory graphs} (SMGs)
that are intended---together with a mapping from global (static) and local
(stack) variables to their nodes---to encode (possibly infinite) sets of
configurations of programs with pointers and unbounded dynamic linked lists. We
start by an informal description of SMGs, followed by their formalisation. For
an illustration of the notions discussed below, we refer the reader to
Fig.~\ref{fig:linuxDLSandItsSMG}, which shows how SMGs represent cyclic
Linux-style DLLs. The head node of such lists has no data part (while all other
nodes include the head structure as well as custom data), and its next/prev
pointers point \emph{inside} list nodes, not at their beginning.

\vspace*{-2mm}\subsection{The Intuition behind SMGs}\label{sec:intuition}
\vspace*{-2mm}

An SMG consists of two kinds of nodes: \emph{objects} and \emph{values}
(in~Fig.~\ref{fig:linuxDLSandItsSMG}, they are represented by boxes and circles,
respectively). Objects are further divided to \emph{regions} and (doubly-linked)
\emph{list segments} (DLSs)\footnote{Our tool Predator supports
\emph{singly-linked list segments} too. Such segments can be viewed as
a~restriction of DLSs, and we omit them from the description in order to
simplify it.}. A region represents a contiguous area of memory allocated either
statically, on the stack, or on the heap. Each consistent SMG contains a special
region called the \emph{null object}, denoted $\nil$, which represents the
target of \texttt{NULL}. DLSs arise from abstracting sequences of doubly-linked
regions that are not interrupted by any external pointer. For example, in the
lower~part~of Fig.~\ref{fig:linuxDLSandItsSMG}, the left box is a region
corresponding to the \texttt{list\_head} from the upper part of the figure whereas the
right box is a DLS summarizing the sequence of \texttt{custom\_record} objects
from the upper part. Values are then used to represent \emph{addresses} and
other \emph{data} stored in objects. All values are abstract in that we only
distinguish whether they represent~equal or possibly different concrete values.
The only exception is the value $0$~that~is used to represent sequences of zero
bytes of any length, which includes the zeros of all numerical types, the
address of the null object, as well as nullified blocks of any size.  Zero
values are supported since they play a rather crucial role in C programs. In the
future, a~better distinction of values could be added.

\enlargethispage{5mm}

SMGs have two kinds of \emph{edges}: namely, \emph{has-value edges} leading from
objects to values and \emph{points-to edges} leading from addresses to objects
(cf. Fig.~\ref{fig:linuxDLSandItsSMG}). Intuitively, the edges express that
objects have values and addresses point to objects (non-address values have no
outgoing edge). Has-value edges are labelled by the \emph{offset} and
\emph{type} of the \emph{field} in which a~particular value is stored within an
object. Note that we allow the fields to overlap. This is used to represent
different \emph{interpretations} that a program can assign to a given memory
area and that we do not want the analyser to recompute again and again.
Points-to edges are labelled by~an~\emph{offset} and a~\emph{target specifier}.
The offset is used to express that the address from which the edge leads may, in
fact, point \emph{before, inside, or behind} an object. The target specifier is
only meaningful for list segments to distinguish whether a~given edge represents
the address (or addresses) of the first, last, or each concrete region
abstracted by the segment. The last option is used to encode links going to list
nodes from the structures nested~below~them (e.g., in a DLL of DLLs, each node
of the top-level list may be pointed from its nested list).

A key advantage of representing values (including addresses) as a separate
kind of nodes is that a single value node is then used to represent values which
are guaranteed to be equal in all concrete memory configurations encoded by a
given SMG. Hence, distinguishing between \emph{equal} values and \emph{possibly
different} values reduces to a simple identity check, not requiring a use of any
prover. Thanks to identifying fields of objects by offsets (instead of using
names of struct/union members), comparing their addresses for equality
simplifies to checking identity of the address nodes. For example, \texttt{(x
== \&x->next)} holds iff \texttt{next} is the first member of the structure
pointed by \texttt{x}, in which case both \texttt{x} and \texttt{\&x->next} are
guaranteed to be represented by a single address node in SMGs. Finally, the
distinction of has-value and points-to edges saves some space since the
information present on points-to edges would otherwise have to be copied
multiple times for a single target. 

Objects and values in SMGs are labelled by several \emph{attributes}. First,
each object is labelled by its \emph{kind}, allowing one to distinguish regions
and DLSs. Next, each object is labelled by its \emph{size}, i.e., the amount of
memory allocated for storing it. For DLSs, the size gives the size of their
nodes. All objects and values have the so-called \emph{nesting level} which is
an integer specifying at which level of hierarchically-nested structures the
object or value appears (level 0 being the top level). All objects are further
labelled by their \emph{validity} in order to allow for safe pointer arithmetic
over freed regions (which are marked invalid but kept as long as there is some
pointer~to~them).

\enlargethispage{5mm}

Next, each DLS is labelled by the \emph{minimum length} of the sequence of
regions represented by it.\footnote{Later, in Section~\ref{sec:extensions},
special list segments of length 0 or 1 are mentioned too.} In particular, the
notation ``$2+$'' used in Fig.~\ref{fig:linuxDLSandItsSMG} means that the
minimum length of the list segment is $2$. Further, each DLS is associated with
the offsets of the \emph{``next''} and \emph{``prev'' fields} through which the
concrete regions represented by the segment are linked forward and
backward\tronly{\footnote{The names ``next'' and ``prev'' (i.e., previous) are
used within our definition of list segments only. The concrete names of these
fields in the programs being analysed are irrelevant.}}. Each DLS is also
associated with the so-called \emph{head offset} at which a~sub-structure called
a \emph{list head} is stored in each list node (cf.
Fig.~\ref{fig:linuxDLSandItsSMG}). The usage of list heads is common in system
software. They are predefined structures, typically containing the next/prev
fields used to link list nodes. When a new list is defined, its node structure
contains the list head as a nested structure, its nodes are linked by pointers
pointing not at their beginning but inside of them (in particular, to the list
head), and pointer arithmetic is used to get to the beginning of the actual list
nodes.

Global and stack \emph{program variables} are represented by regions in a
similar way as heap objects and can thus be manipulated in a similar way
(including their manipulation via pointers, checking for out-of-bounds accesses
leading to stack smashing, etc.). Regions representing program variables are
tagged by their names and hence distinguishable whenever needed (e.g., when
checking for invalid frees of stack/global~memory,~etc.).


\vspace*{-3mm}\subsection{Further Illustration of the Notion of
SMGs}\label{sec:exampleOfSMGs} \vspace*{-2mm}

We now provide two more illustrative examples of how SMGs represent various data
structures common in practice.

\begin{figure}[p]
  \centering
  \resizebox{0.8\hsize}{!}{
    \includegraphics{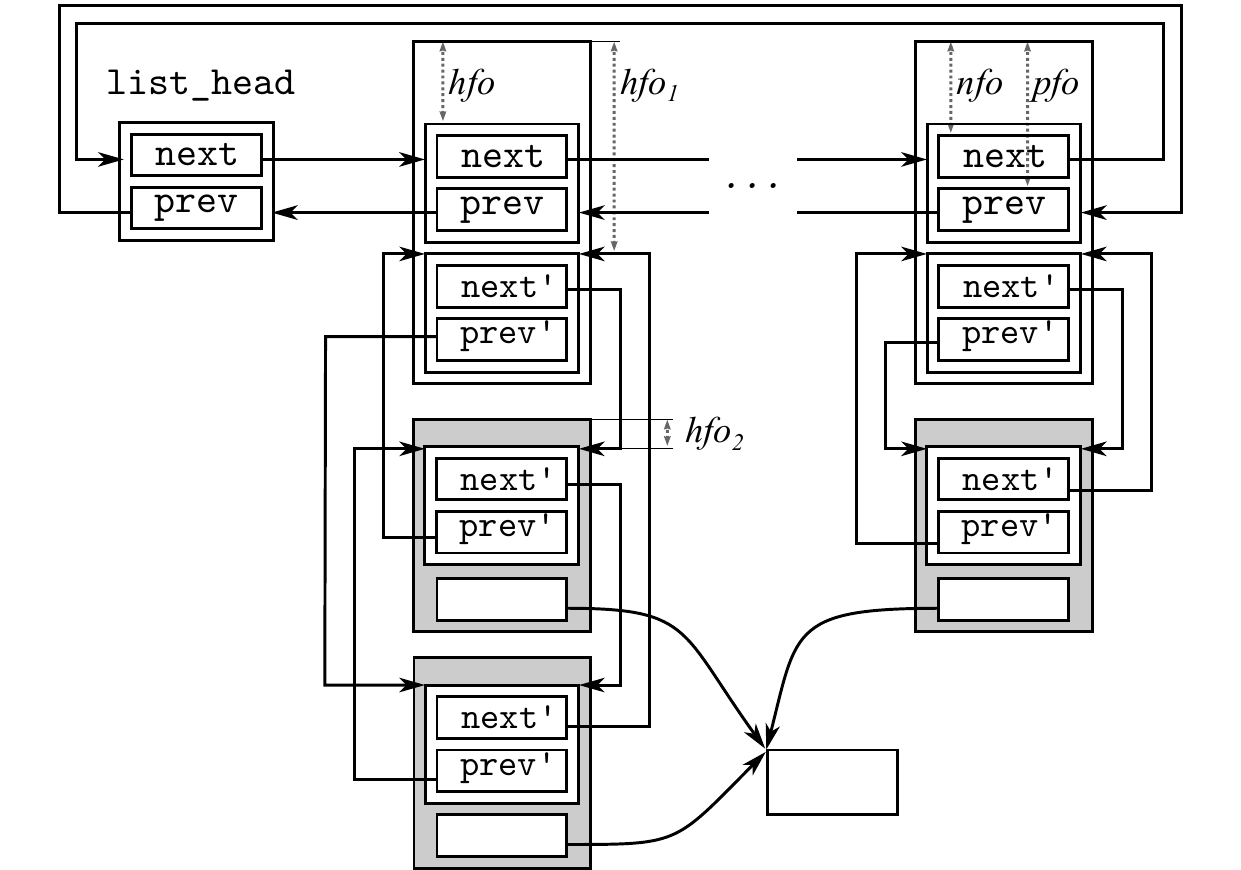}
  }
  \vspace{2em}
  \resizebox{0.8\hsize}{!}{
    \includegraphics{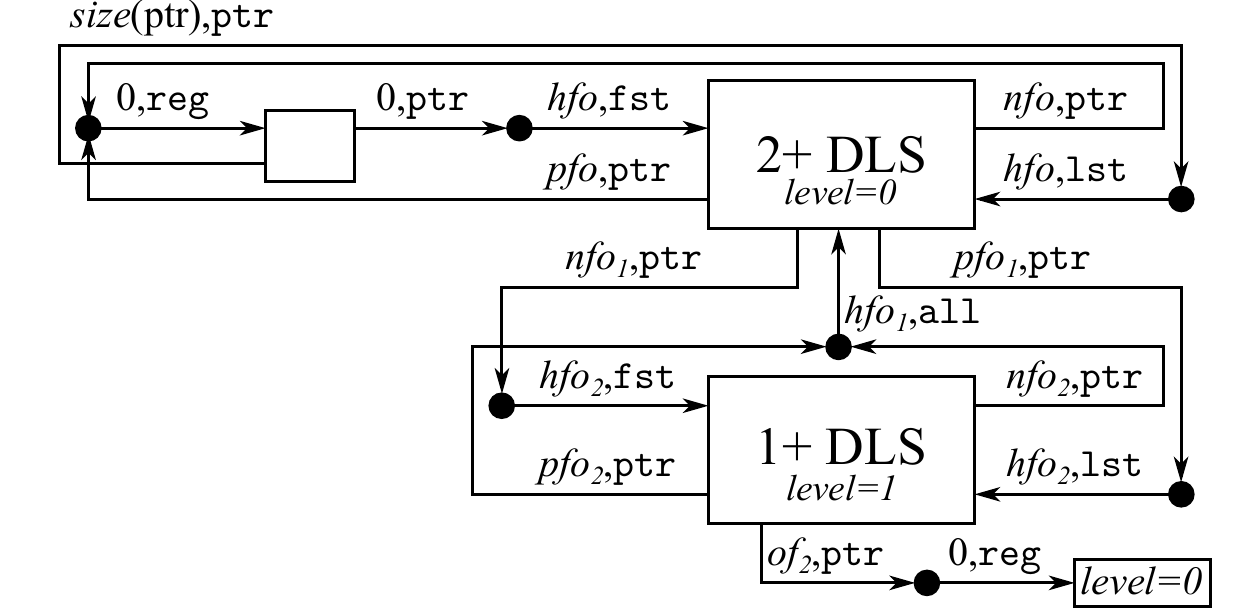}
  }
  \caption{A cyclic Linux-style DLL of DLLs with a shared data element (top) and
    its SMG (bottom).}
  \label{fig:linuxDLSofDLSandItsSMG}
\end{figure}

The upper part of Fig.~\ref{fig:linuxDLSofDLSandItsSMG} shows a Linux-style
cyclic DLL of cyclic DLLs. All nodes of all nested DLLs point to a shared memory
region. The lower part of the figure shows an SMG representing this structure.
Note that the top-level DLS as well as the shared region are on level 0 whereas
the nested DLSs are on level~1.

\begin{figure}
  \centering
  \resizebox{0.8\hsize}{!}{
    \includegraphics{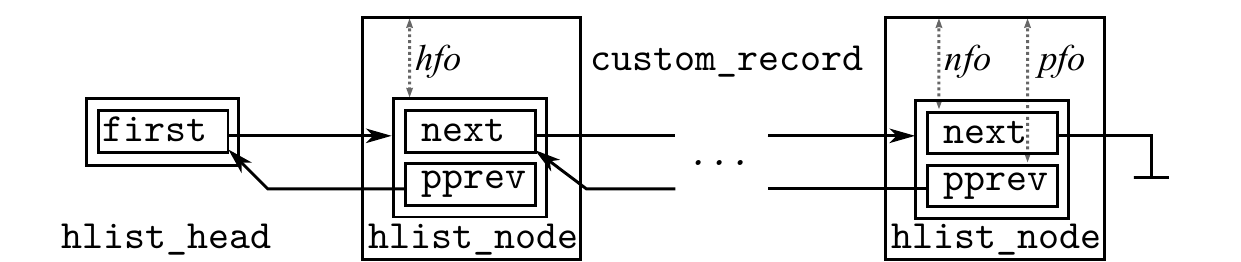}
  }
  \vspace{2em}
  \resizebox{0.8\hsize}{!}{
    \includegraphics{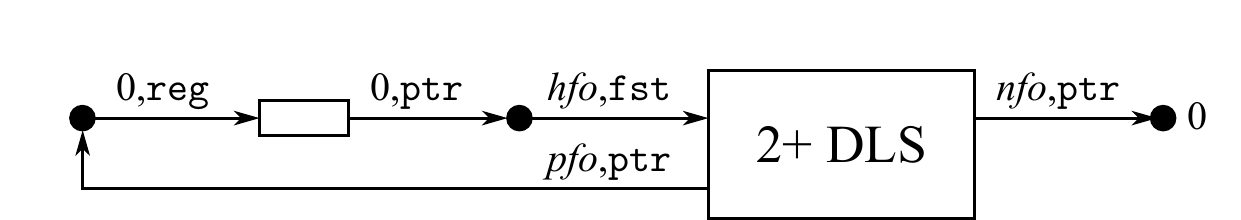}
  }
  \caption{A Linux-style list used in hash tables (top) and its SMG (bottom).}
  \label{fig:linuxHListAndItsSMG}
\end{figure}

The upper part of Fig.~\ref{fig:linuxHListAndItsSMG} shows another variant of
Linux-style DLLs which is optimised for use in hash tables. The lower part of
the figure shows an SMG representing this kind of lists. For lists used in hash
tables, the size of list headers determines the amount of memory allocated by an
empty hash table. That is why the lists presented in
Fig.~\ref{fig:linuxHListAndItsSMG} have headers reduced to the size of a single
field for the price of having forward and backward links of different types. In
particular, forward links are pointers to structures whereas backward links are
pointers to pointers to structures. This asymmetry may cause problems to
analysers that use a selector-based description of list segments, but it is not
a problem for us since our representation is purely offset-based.\footnote{A
need to use a special kind of list segments would arise in SMGs if the head and
next offsets were different, but that is unlikely to happen in this special case
since it would prevent the list head from having the size of a single pointer
only.}

\vspace*{-4mm}\subsection{Formal Definition of Symbolic Memory
Graphs}\label{sec:defOfSMGs} \vspace*{-2mm}


Let $\bool$ be the set of Booleans, $\type$ a set of types, $\size(\ty)$ the size
of instances of a type $\ty \in \type$, $\ptr \in \type$ a unique pointer
type\footnote{We assume $\size(\ptr)$ to be a constant, which implies that
separate verification runs are needed for verifying a program for target
architectures using different address sizes.}, $\Kind = \{\reg, \dls\}$ the set
of kinds of objects (distinguishing regions and DLSs), and $\ptt = \{ \first,
\last, \all, \reg \}$ the set of points-to target specifiers.

\paragraph{Symbolic Memory Graphs}

A~\emph{symbolic memory graph} is a~tuple $G = (O,V,\Lambda,H,P)$ where:
\begin{itemize}

  \item $O$ is a finite set of objects including the special null object $\nil$.

  \item $V$ is a finite set of \emph{values} such that $O \cap V = \emptyset$
  and $0 \in V$.

  \item $\Lambda$ is a tuple of the following labelling
  functions:\begin{itemize}

    \item The kind of objects $\kind: O \rightarrow \mathbb{K}$ where
    $\kind(\nil) = \reg$, i.e., $\nil$ is formally considered a region.  We let
    $R = \{r \in O \mid \kind(r) = \reg \}$ be the set of regions and $D = \{ d
    \in O \mid \kind(d) = \dls\}$ be the set of DLSs of $G$.

    \item The nesting level of objects and values $\level: O \cup V
    \rightarrow \nat$.

    \item The size of objects $\size: O \rightarrow \nat$.

    \item The minimum length of DLSs $len: D \rightarrow \nat$.

    \item The validity of objects $\valid: O \rightarrow \bool$.

    \item The head, next, and prev field offsets of DLSs $\hfo, \nfo, \pfo: D
    \rightarrow \nat$.

  \end{itemize}
 
  \item $H$ is a partial edge function $O \times \nat \times \type
  \rightharpoonup V$ which defines \emph{has-value edges} $o \hasvalue{\of,\ty}
  v$ where $o \in O$, $v \in V$, $\of \in \nat$, and $\ty \in \type$. We call
  $(\of,\ty)$ a \emph{field} of the object $o$ that stores the value $v$ of the
  type~$\ty$ at the offset $\of$.

  \item $P$ is a partial injective edge function $V \rightharpoonup \zed \times
  \ptt \times O$ which defines \emph{points-to edges} $v \pointsto{\of,\tg} o$
  where $v \in V$, $o \in O$, $\of \in \zed$, and $\tg \in \ptt$ such that $\tg =
  \reg$ iff $o \in R$. Here, $\of$ is an offset wrt the base address of
  $o$.\footnote{Note that the offset can even be negative, which happens, e.g.,
  when traversing a Linux list.} If $o$ is a DLS, $\tg$ says whether the edge
  encodes pointers to the \emph{first}, \emph{last}, or \emph{all} concrete
  regions represented by $o$.

\end{itemize} 

\begin{figure}[t]
  \centering
  \includegraphics[width=0.9\textwidth]{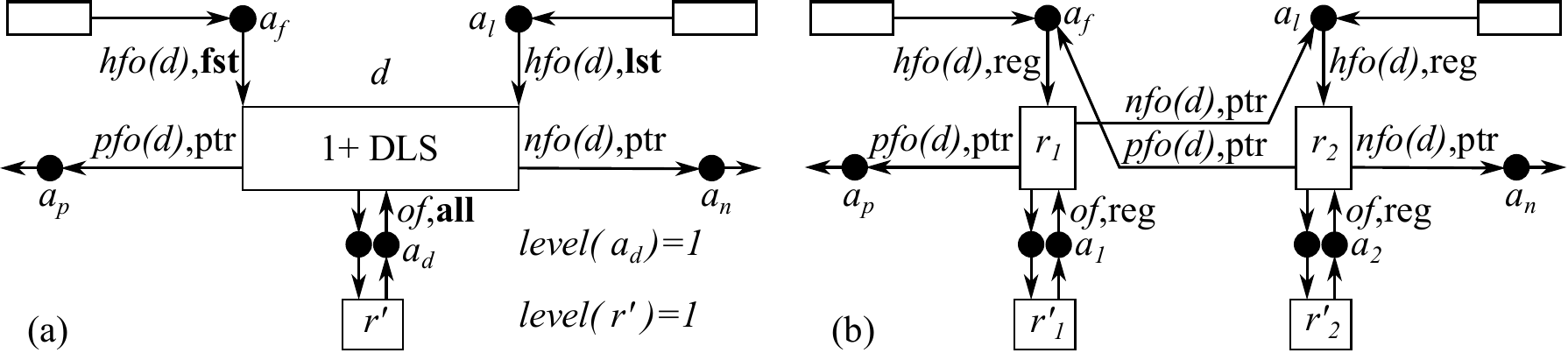}
  \caption{(a) An SMG and (b) its possible concretisation for the case when the
    DLS~$d$ represents exactly two regions (showing key attributes only).}
  \vspace*{-4mm}
  \label{fig:SMGdemo}
\end{figure}

We define the first node of a list segment such that the next
field of the node points inside the list segment (and the last node such that
the prev field of the node points inside the list segment). As already
mentioned, the~\emph{all} target specifier is used in hierarchically-nested list
structures where each nested data structure points back to the node of the
parent list below which it is nested. Fig.~\ref{fig:SMGdemo} illustrates how the
target specifier affects the semantics of points-to edges (and the
corresponding addresses): The DLS $d$ is concretized to the two regions $r_1$
and $r_2$, and the nested abstract region $r'$ to the two concrete regions
$r_1'$ and $r_2'$.  Note that if $r'$ was not nested, i.e., if it had $level(r')
= 0$, it would concretise into a single region pointed by both $r_1$ and $r_2$.

\enlargethispage{5mm}

\paragraph{Consistent Symbolic Memory Graphs}

In the following, we assume working with so-called consistent SMGs
only.\footnote{All the later presented algorithms will be such that they produce
a consistent SMG when they are applied on a consistent SMG (or SMGs). Therefore,
since our analysis will start from a consistent SMG, there is no need to check
the consistency on the fly.} In particular, we call an SMG $G =
(O,V,\Lambda,H,P)$ \emph{consistent} iff the~following~holds:\begin{itemize}

  \item \emph{Basic consistency of objects.} The null object is invalid,
  has size and level $0$, and its address is $0$, i.e., $\valid(\nil) = \false$,
  $\size(\nil) = \level(\nil) = 0$, and $0 \pointsto{0,\reg} \nil$. All DLSs are
  valid, i.e., $\forall d \in D: \valid(d)$. Invalid regions have no outgoing
  edges.

  \item \emph{Field consistency.} Fields do not exceed boundaries of objects,
    i.e., $\forall o \in O$ $\forall \of \in \nat$ ~$\forall \ty \in \type:~
  H(o,\of,\ty) \neq \bot \Rightarrow \of+\size(\ty) \leq \size(o)$.

  \item \emph{DLS consistency.} Each DLS $d \in D$ has a next pointer and a prev
  pointer, i.e., there are addresses $a_n, a_p \in A$ s.t. $H(d,\nfo(d),\ptr) =
  a_n$ and \mbox{$H(d,\pfo(d),\ptr) = a_p$} (cf.  Fig.~\ref{fig:SMGdemo}).  The
  next pointer is always stored in memory before the prev pointer, i.e., the
  next and prev offsets are s.t. $\forall d \in D: \nfo(d) < \pfo(d)$.
  Points-to edges encoding links to the first and last node of a~DLS $d$ are
  always pointing to these nodes with the appropriate head offset, i.e.,
  $\forall a \in A: \tg(P(a)) \in \{ \first, \last \} \Rightarrow \of(P(a)) =
  \hfo(d)$ where $d = o(P(a))$.\footnote{The last two requirements are not
  necessary, but they significantly simplify the below presented algorithms
  (e.g., the DLS materialisation given in Section~\ref{sec:semantics}).}
  Finally, in a consistent SMG there is no cyclic path containing $0+$ DLSs (and
  their addresses) only since its semantics would include an address not
  referring to any object.

  \item \emph{Nesting consistency.} Each nested object $o \in O$ of level $l =
  \level(o)>0$ has precisely one \emph{parent DLS}, denoted $\parent(o)$, that
  is of level $l-1$ and there is a~path from $\parent(o)$ to $o$ whose inner
  nodes are of level $l$ and higher (i.e., more nested) only---e.g., in
  Fig.~\ref{fig:SMGdemo}, $d$ is the parent of $r'$. Addresses with $\first$,
  $\last$, and $\reg$ targets are always of the same level as the object they
  refer to (as is the case for $a_f$, $a_l$, $a_1$, $a_2$ in
  Fig.~\ref{fig:SMGdemo}), i.e., $\forall a \in A: \tg(P(a)) \in \{ \first,
  \last, \reg \} \Rightarrow \level(a) = \level(o(P(a)))$. On the other hand,
  addresses with the $\all$ target go up one level in the nesting hierarchy,
  i.e., $\forall a \in A: \tg(P(a)) = \all \Rightarrow \level(a) =
  \level(o(P(a)))+1$ (cf. $a_d$ in Fig.~\ref{fig:SMGdemo}). Finally, edges
  representing back-pointers to all nodes of a list segment can only lead from
  objects (transitively) nested below that segment (e.g., in
  Fig.~\ref{fig:SMGdemo}, such an edge leads from the region $r'$ back to the
  DLS $d$, but it cannot lead from any other regions). Formally, for any $o, o'
  \in O$, $a \in H(o)$, $o(P(a))=o'$, and $\level(o)>\level(o')$,
  $\tg(P(a))=\all$ iff $o'=\parent^k(o)$ for some $k \geq 1$.

\end{itemize}

Let $G = (O,V,\Lambda,H,P)$ be an SMG with a set of regions $R$ and a set of
DLSs~$D$. We denote a DLS $d \in D$ of minimum length $n$, for which
$\len(d)=n$, as an $n+$~DLS. We use $\UNDEF$ to denote cases where $H$ or $P$ is
not defined. For any $v \in V$ for which \mbox{$P(v) \neq \UNDEF$}, we denote by
$\of(P(v))$, $\tg(P(v))$, and $o(P(v))$ the particular offset, target specifier,
and object of the triple $P(v)$, respectively. Further, for $o \in O$, we let
$H(o) = \{ H(o,\of,\ty) \mid \of \in \nat,\; \ty \in \type,\; H(o,\of,\ty) \neq
\UNDEF\}$. We let \mbox{$A = \{ v \in V \mid P(v) \neq \UNDEF \}$} be the set of
all \emph{addresses} used in $G$. Next, a~\emph{path} in $G$ is a sequence (of
length one or more) of values and objects such that there is an edge between
every two neighbouring nodes of the path. An object or value $x_2 \in O \cup V$
is \emph{reachable} from an object or value $x_1 \in O \cup V$ iff there is a
path from $x_1$ to $x_2$.

\enlargethispage{5mm}

\vspace*{-2mm}\paragraph{Symbolic Program Configurations}

Let $\GVar$ be a~finite set of global variables, $\SVar$ a countable set of
stack variables such that $\GVar \cap \SVar = \emptyset$, and let $\Var = \GVar
\cup \SVar$. A \emph{symbolic program configuration} (SPC) is a~pair $C =
(G,\nu)$ where $G$ is an SMG with a~set of regions $R$, and $\nu: \Var
\rightarrow R$~is a finite injective map such that $\forall x \in \Var:~
\level(\nu(x)) = 0 ~\wedge~ \valid(\nu(x))$. Note that $\nu$ evaluates to the
regions in which values of variables are stored, not directly the values
themselves. We call each object $o$ such that $\nu(x) = o$ for some $x \in
\GVar$ a \emph{static object}, and each object $o$ such that $\nu(x) = o$ for
some $x \in \SVar$ a~\emph{stack object}. All other objects are called
\emph{heap objects}.  An SPC is called \emph{garbage-free} iff all its heap
objects are reachable from static or stack objects.

\paragraph{Special Kinds of SMGs and SPCs}

We define the \emph{empty SMG} to consist solely of the null object, its address
$0$, and the points-to edge between them. The \emph{empty SPC} then consists of
the empty SMG and the empty variable mapping. An SMG $G' =
(O',V',\Lambda',H',P')$ is a~\emph{sub-SMG} of an SMG $G = (O,V,\Lambda,H,P)$
iff (1)~$O' \subseteq O$, (2)~$V' \subseteq V$, and (3)~$H'$, $P'$, and
$\Lambda'$ are restrictions of $H$, $P$, and $\Lambda$ to $O'$ and $V'$,
respectively. The sub-SMG of $G$ \emph{rooted at} an object or value $x \in O
\cup V$, denoted $G_x$, is the smallest sub-SMG of $G$ that includes $x$ and all
objects and values reachable from $x$. Given $F \subseteq \nat$, the
\emph{$F$-restricted} sub-SMG of $G$ rooted at an object $o \in O$ is the
smallest sub-SMG of $G$ that includes $o$ and all objects and values reachable
from $o$ apart from the addresses $A_F = \{ H(o,\of,\ptr) \mid \of \in F \}$ and
nodes that are reachable from $o$ through $A_F$ only.  Finally, the sub-SMG of
$G$ \emph{nested below} $d \in D$, denoted $\widehat{G}_d$, is the smallest
sub-SMG of $G$ including $d$ and all objects and values of level higher than
$\level(d)$ that are reachable from $d$ via paths that, apart from $d$, consist
exclusively of objects and values of a level higher than $\level(d)$.

\vspace*{-5mm}\subsection{The Semantics of SMGs}\label{sec:semantics}
\vspace*{-3mm}

We define the semantics of SMGs in two steps, namely, by first defining it in
terms of the so-called memory graphs whose semantics is subsequently defined in
terms of concrete memory images. In particular, a \emph{memory graph} (MG) is
defined exactly as an SMG up to that it is not allowed to contain any list
segments.  An SMG then represents the class of MGs that can be obtained (up to
isomorphism) by applying the following two transformations any number of times:
(1)~\emph{materialisation} of fresh regions from DLSs (i.e., intuitively,
``pulling out'' concrete regions from the beginning or end of segments) and
(2)~\emph{removal} of $0+$ DLSs (which may have become $0+$ due to the preceding
materialisation). Moreover, note that the operations of materialisation and
removal are used not only to define the semantics, but they will later be used
within symbolic execution of C statements over SMGs (and hence as a part of the
actual SMG-based analysis) too.

\enlargethispage{5mm}

\vspace*{-4mm}\subsubsection*{Materialisation and Removal of
DLSs}\vspace*{-2mm}

\begin{figure}[t]
  \centering
  \includegraphics[width=80mm]{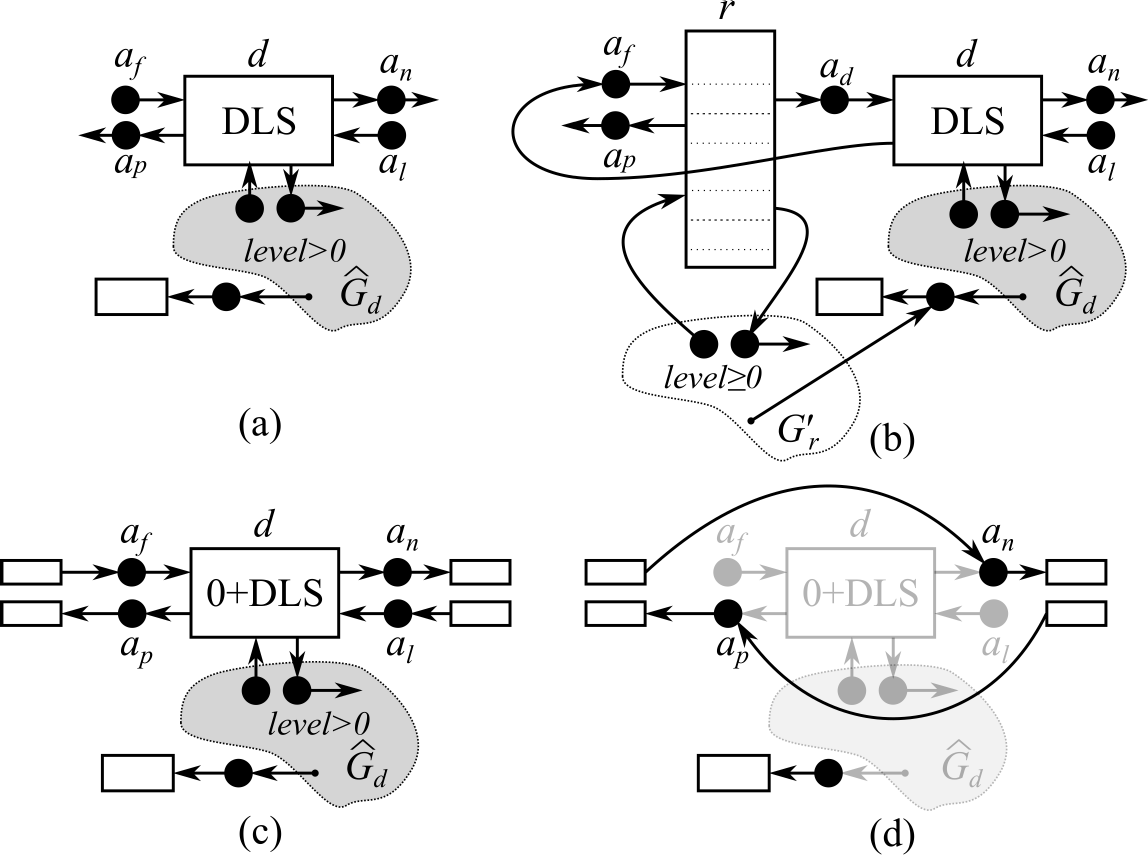}
  \vspace*{-2mm}
  \caption{Materialisation of a DLS: (a) input, (b) output (region $r$ got
    materialised from DLS $d$).  Removal of a DLS: (c) input, (d) output.
    Sub-SMGs $\widehat{G}_d$ and $G_r'$ are highlighted without their roots.}
  \vspace*{-2mm}
  \label{fig:mat-rem}
\end{figure}

Let $G = (O,V,\Lambda,H,P)$ be an SMG with the sets of regions $R$, DLSs $D$,
and addresses $A$. Let $d \in D$ be a DLS of level $0$. Further, let $a_n, a_p
\in A$ be the next and prev addresses of~$d$, i.e., $H(d,\pfo(d),\ptr) = a_p$ and
$H(d,\nfo(d),\ptr) = a_n$. The DLS $d$ can be \emph{materialised} as
follows---for an illustration of the operation, see the upper part of
Fig.~\ref{fig:mat-rem}:\begin{enumerate}

  \item \emph{Materialisation of the first region and its nested sub-SMG.} $G$
  is extended by a~fresh copy $G'_r$ of the sub-SMG $\widehat{G}_d$ nested below
  $d$. In $G'_r$, $d$~is replaced by a~fresh region $r$ such that
  $\size(r)=\size(d)$, $\level(r)=0$, and $\valid(r)=\true$. The nesting level
  of each object and value in $G'_r$ (apart from $r$) is decreased by~one.

  \item \emph{Interconnection of the materialised region and the rest of the
  segment.} Let $a_f \in A$ be the address pointing to the beginning of $d$,
  i.e., such that $P(a_f)=(\hfo(d),\first,d)$. If $a_f$ does not exist in $G$,
  it is added. Next, $A$ is extended by a~fresh address $a_d$ that will point to
  the beginning of the remaining part of $d$ after the concretisation (while
  $a_f$ will be the address of~$r$).  Finally, $H$ and $P$ are changed s.t.
  $P(a_f) = (\hfo(d),\reg,r)$, $H(r,\pfo(d),\ptr) = a_p$, $H(r,\nfo(d),\ptr) =
  a_d$, $P(a_d) = (\hfo(d),\first,d)$, and $H(d,\pfo(d),\ptr) = a_f$.

  \item \emph{Interconnection of the materialised sub-heap and non-nested
  objects.} For any object $o$ of $\widehat{G}_d$, let $o'$ be the corresponding
  copy of $o$ in $G'_r$ (for $o = d$, let $o' = r$). For each field $(\of,\ty)
  \in (\nat \times \type)$ of each object $o$ in $\widehat{G}_d$ whose value is
  of level $0$, i.e., $\level(H(o,\of,\ty))=0$, the corresponding field of $o'$
  in $G'_r$ is set to the same value, i.e., the set of edges is extended such
  that $H(o',\of,\ty) = H(o,\of,\ty)$.

  \item \emph{Adjusting the minimum length of the rest of the segment.} If
  $\len(d) > 0$, $\len(d)$ is decreased by one.

\end{enumerate}

\enlargethispage{5mm}

Next, let $d \in D$ be a DLS as above with the additional requirement
of $\len(d) = 0$ with the addresses $a_n$, $a_p$, $a_f$, and $a_l$ defined as in
the case of materialisation. The DLS~$d$ can be \emph{removed} as follows---for
an illustration, see the lower part of Fig.~\ref{fig:mat-rem}: (1)~Each
has-value edge $o \hasvalue{\of, \ty} a_f$ is replaced by the edge $o \hasvalue{\of,
\ty} a_n$. (2)~Each has-value edge $o \hasvalue{\of, \ty} a_l$ is replaced by the
edge $o \hasvalue{\of, \ty} a_p$. (3)~The subgraph $\widehat{G}_d$ is removed
together with the addresses $a_f$, $a_l$, and the edges adjacent with the
removed objects and values.

Given an SMG $G = (O,V,\Lambda,H,P)$ with a set of DLSs $D$, we denote by
$\MG(G)$ the class of all MGs that can be obtained (up to isomorphism) by
materializing each DLS $d \in D$ at least $\len(d)$ times and by subsequently
removing all DLSs.

!
\enlargethispage{5mm}
!

\vspace*{-4mm}\subsubsection*{Concrete Memory Images}\vspace*{-2mm}

The semantics of an MG $G = (R,V,\Lambda,H,P)$ is the set $\MI(G)$ of
\emph{memory images} $\mu: \nat \rightarrow \{ 0, \dots, 255 \}$ mapping
\emph{concrete addresses} to \emph{bytes} such that there exists a function
$\pi: R \rightarrow \nat$, called a \emph{region placement}, for which the
following holds: \begin{enumerate}

  \item Only the null object is placed at address zero, i.e., $\forall r \in R:~
  \pi(r) = 0 \Leftrightarrow r = \nil$.

  \item No two valid regions overlap, i.e., $\forall r_1, r_2 \in R:~ \valid(r_1)
  \wedge \valid(r_2) \Rightarrow \langle \pi(r_1), \\ \pi(r_1) + \size(r_1)) \cap
  \langle \pi(r_2), \pi(r_2)+\size(r_2)) = \emptyset$.

  \item Pointer fields are filled with the concrete addresses of the regions
  they refer to. Formally, for each pair of has-value and points-to edges $r_1
  \hasvalue{\of_1,\ptr} a \pointsto{\of_2,\reg} r_2$ in $H$ and $P$, resp.,
  $\addr(\bseq(\mu,\pi(r_1)+\of_1,\size(\ptr))) = \pi(r_2)+\of_2$ where
  $\bseq(\mu,p,\size)$ is the sequence of bytes $\mu(p) \mu(p+1)...
  \mu(p+\size-1)$ for any $p, \size > 0$, and $\addr(\sigma)$ is the
  concrete address encoded by the byte sequence~$\sigma$.

  \item Fields having the same values are filled with the same concrete values
  (up to nullified blocks that can differ in their length), i.e., for every two
  has-value edges $r_1 \hasvalue{\of_1,\ty_1} v$ and $r_2 \hasvalue{\of_2,\ty_2} v$ in
  $H$, where $v \neq 0$, $\bseq(\mu,\pi(r_1)+\of_1,\size(\ty_1)) =
  \bseq(\mu,\pi(r_2)+\of_2,\size(\ty_2))$.

  \item Finally, nullified fields are filled with zeros, i.e., for each
  has-value edge $r \hasvalue{\of,\ty} 0$ in $H$, $\mu(\pi(r)+\of+i)=0$ for all
  $0 \leq i < \size(\ty)$.

\end{enumerate} 

For an SMG $G$, we let $\MI(G)  = \bigcup_{G' \in \MG(G)} \MI(G')$.  Note that
it may happen that no concrete values satisfying the needed constraints exist.
In such a~case, the semantics of an (S)MG is empty. Note also that we restrict
ourselves to a flat address space, which is, however, sufficient for most
practical cases. \tronly{Finally, note that, for simplicity, we assume that each
sequence of bytes of length $\size(\ty)$ corresponds to some instance of the
type $\ty$, which can be an indeterminate value in the worst case.}

\vspace*{-3mm}\section{Operations on SMGs}\label{sec:operations}\vspace*{-2mm}

In this section, we propose algorithms for all operations on SMGs that are
needed for their application in program verification. In particular, we discuss
data reinterpretation (which is used for reading and writing from/to SMGs), join
of SMGs (which we use for entailment checking and as a part of the abstraction
too), abstraction, inequality checking, and symbolic execution of C programs.
More details can be found in Sections~\ref{app:reinterpret}--\ref{app:condSymExec}.

Below, we denote by $I(\of, \ty)$ the right-open integer interval $\left<\of,
\of + \size(\ty) \right)$, and, for a has-value edge $e: o \hasvalue{\of,\ty}
v$, we use $I(e)$ as the abbreviation of $I(\of, \ty)$.


\vspace*{-3mm}\subsection{Data Reinterpretation}\label{sec:reinterpretation}
\vspace*{-2mm}

SMGs allow fields of a single object to overlap and to even have the same offset
and size, in which case they are distinguishable by their types only. In line
with this feature of SMGs, we introduce the so-called \emph{read
reinterpretation} that can create multiple views (\emph{interpretations}) of a
single memory area without actually changing the semantics. On the other hand,
if we write to a field that overlaps with other fields, we need to reflect the
change of the memory image in the overlapping fields, for which the so-called
\emph{write reinterpretation} is used. These two operations form the basis of
all operations reading and writing memory represented by SMGs. Apart from them,
we also use \emph{join reinterpretation} which is applied when joining two SMGs
to preserve as much information shared by the SMGs as possible even when this
information is not explicitly represented in the same way in both the input
SMGs.

Defining reinterpretation for all possible data types (and all of their possible
values) is hard (cf. \cite{Tuch2009}) and beyond the scope of this work.
Instead of that, we define minimal requirements that must be met by the
reinterpretation operators so that our verification approach is sound. This
allows different concrete instantiations of these operators to be used in the
future. Currently, we instantiate the operators for the particular case of
dealing with nullified blocks of memory, which is essential for handling
low-level pointer manipulating programs that commonly use functions like
\texttt{calloc()} or \texttt{memset()} to obtain large blocks of nullified
memory.\footnote{Apart from the nullified blocks, our implementation also
supports tracking of uninitialized blocks of memory and certain manipulations of
null-terminated strings (cf. Section~\ref{sec:kernel}).}

\vspace*{-2mm}\subsubsection*{Read Reinterpretation}\vspace*{-1mm}

A read reinterpretation operator takes as input an SMG $G$ with a set of objects
$O$, an object $o \in O$, and a field $(\of, \ty)$ to be read from $o$ such that
$\of + \size(\ty) \le \size(o)$. The result is a pair $(G',v)$ where $G'$ is
an SMG with a~set of has-value edges $H'$ such that (1)~$H'(o,\of,\ty) = v \neq
\bot$ and (2)~$\MI(G) = \MI(G')$. The operator thus preserves the semantics of
the SMG $G$ but ensures that it contains a~has-value edge for the field being
read.  This edge can lead to a value already present in $G$ but also to a new
value derived by the operator from the edges and values existing in $G$. In the
extreme case, a fresh, completely unconstrained value node can be added,
representing an unknown value, which can, however, become constrained by further
program execution. In other words, read reinterpretation installs a~new view on
some part of the object $o$ without modifying the semantics of the SMG
in~any~way.


For the particular case of dealing with nullified memory, we use the following
concrete read reinterpretation (cf.  Section~\ref{app:rdReinterp}). If $G$
contains an edge $o \hasvalue{\of,\ty} v$, $(G,v)$ is returned. Otherwise, if
each byte of the field $(\of,\ty)$ is nullified by some edge \mbox{$o
\hasvalue{\ofprime,\ty'} 0$} present in~$G$, $(G',0)$ is returned where $G'$ is
obtained from $G$ by adding the edge \mbox{$o \hasvalue{\of,\ty} 0$}.
Otherwise, $(G',v)$ is returned with $G'$ obtained from $G$ by adding an edge
\mbox{$o \hasvalue{\of,\ty} v$} leading to a~fresh value $v$ (representing an
unknown value). It is easy to see that this is the most precise read
reinterpretation that is possible---from the point of view of reading nullified
memory---with the current support of types and values in SMGs.


\vspace*{-3mm}\subsubsection*{Write Reinterpretation}\vspace*{-1mm}

The write reinterpretation operator takes as input an SMG~$G$ with a set of
objects $O$, an object $o \in O$, a field $(\of, \ty)$ within $o$, i.e., such
that $\of + \size(\ty) \le \size(o)$, and a value $v$ that is to be written into
the field $(\of, \ty)$ of the object~$o$. The result is an SMG $G'$ with a~set
of has-value edges $H'$ such that (1)~$H'(o,\of,\ty) = v$ and (2)~$\MI(G)
\subseteq \MI(G'')$ where $G''$ is the SMG $G'$ without the edge $e: o
\hasvalue{\of,\ty} v$.  In other words, the operator makes sure that the
resulting SMG contains the edge $e$ that was to be written while the semantics
of $G'$ without $e$ over-approximates the~semantics of~$G$. Indeed, one cannot
require equality here since the new edge may collide with some other edges,
which may have to be dropped in the worst case.

For the case of dealing with nullified memory, we propose the following write
reinterpretation (cf.~Section~\ref{app:wrReinterp}, which includes an
illustration too). If $G$ contains the edge $e: o \hasvalue{\of,\ty} v$, $G$ is
returned. Otherwise, all has-value edges leading from $o$ to a~non-zero value
whose fields overlap with $(\of,\ty)$ are removed. Subsequently, if $v = 0$, the
edge $e$ is added, and the obtained SMG is returned. Otherwise, all remaining
has-value edges leading from $o$ to $0$ that define fields overlapping with
$(\of,\ty)$ are split and/or shortened such that they do not overlap with
$(\of,\ty)$, the edge $e$ is added, and the resulting SMG is returned. Again, it
is easy to see that this operator is the most precise write reinterpretation
from the point of view of preserving information about nullified memory that is
possible with the current support of types and values in SMGs.

\vspace*{-4mm}\subsection{Join of SMGs}\label{sec:join}\vspace*{-2mm}

Join of SMGs is a binary operation that takes two SMGs $G_1,G_2$ and returns an
SMG $G$ that is their common generalisation, i.e., $\MI(G_1) \subseteq \MI(G)
\supseteq \MI(G_2)$, and that satisfies the following further requirements
intended to minimize the involved information loss: If both input SMGs are
semantically equal, i.e., $\MI(G_1) = \MI(G_2)$, denoted $G_1 \iso G_2$, we
require the resulting SMG to be semantically equal to both the input ones, i.e.,
$\MI(G_1) = \MI(G) = \MI(G_2)$. If $\MI(G_1) \supset \MI(G_2)$, denoted $G_1
\jsupset G_2$, we require that $\MI(G) = \MI(G_1)$. Symmetrically, if $\MI(G_1)
\subset \MI(G_2)$, denoted $G_1 \jsubset G_2$, we require that $\MI(G) =
\MI(G_2)$. Finally, if the input SMGs are semantically incomparable, i.e.,
$\MI(G_1) \nsupseteq \MI(G_2) \; \wedge \; \MI(G_1) \nsubseteq \MI(G_2)$,
denoted $G_1 \join G_2$, no further requirements are put on the result of the
join (besides the inclusion stated above, which is required for the soundness of
our analysis). In order to distinguish which of these cases happens when joining
two SMGs, we tag the result of our join operator by the so-called \emph{join
status} with the domain $\jstat = \{\iso,\; \jsupset,\; \jsubset,\; \join\}$
referring to the corresponding relations above. Moreover, we allow the join
operation to fail if the incurred information loss becomes too big. Below, we
give an informal description of our join operator, for a full description
see~Section~\ref{app:join}.

\enlargethispage{5mm}

\begin{figure}[t]\sidecaption
  \resizebox{0.7\hsize}{!}{
    \includegraphics{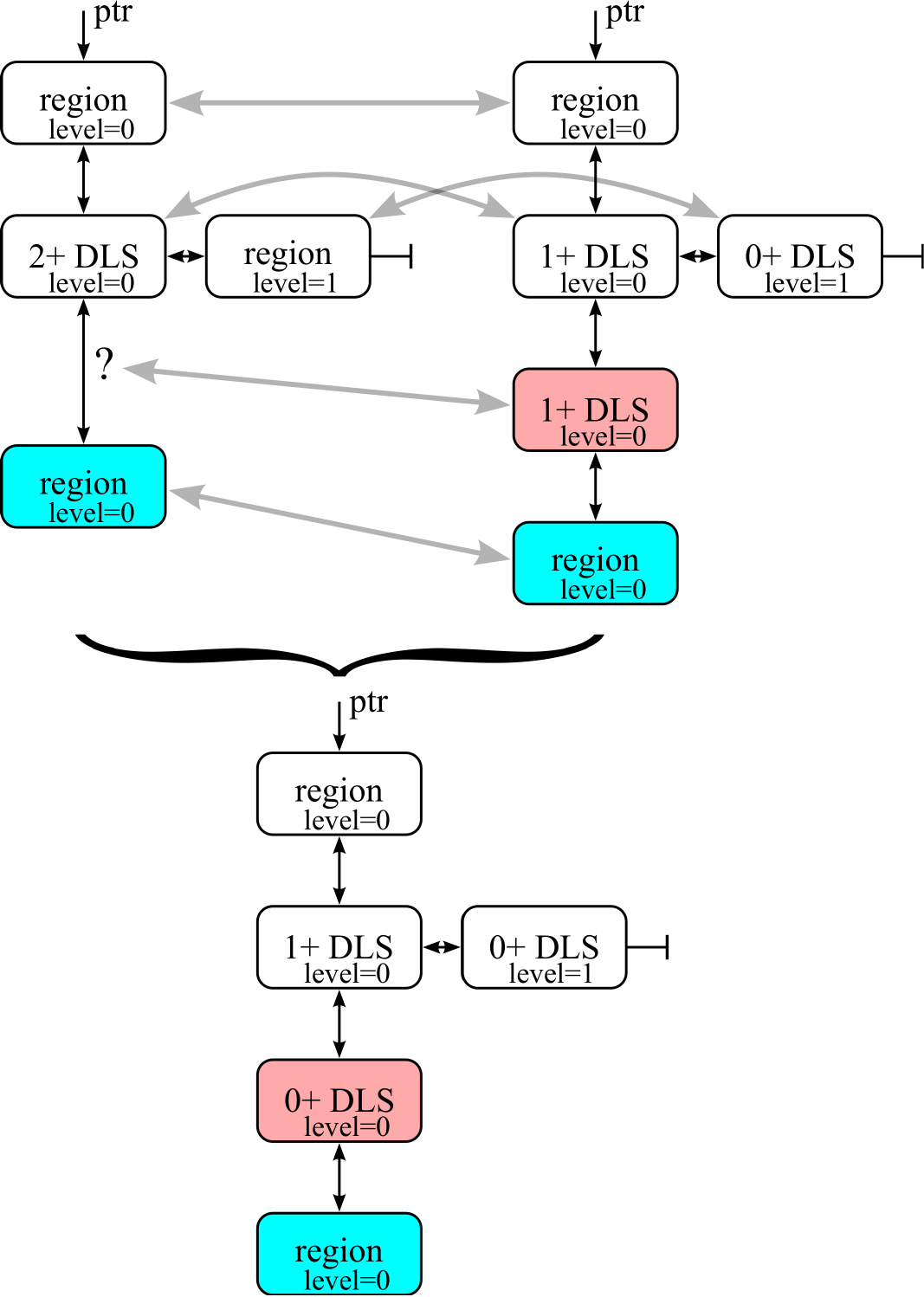}
  }\\\hspace{-10mm}
  \caption{An illustration of the basic principle of the join algorithm. In the
    figure, a simplified notation for describing SMGs is used, from which value
    nodes have been left out. The pair of input SMGs is at the top. The gray
    arrows show the pairs of objects joined by the algorithm during the
    simultaneous traversal of the input SMGs. The resulting SMG is at the bottom.}
  \vspace*{-4mm}
  \label{fig:join}
\end{figure}

The basic idea of our join algorithm, illustrated in Fig.~\ref{fig:join}, is the
following. The algorithm simultaneously traverses a given pair of source SMGs
and tries to join each pair of nodes (i.e., objects or values) encountered at
the same time into a~single node in the destination SMG. A~single node of one
SMG is not allowed to be joined with multiple nodes of the other SMG. This
preserves the distinction between different objects as well as between at least
possibly different values.\tronly{\footnote{Two separately allocated objects are
always different, values are only possibly different. Not to restrict the
semantics, different objects or (possibly) different values cannot be changed
into equal objects or values. Equal values could be changed into possibly
different ones, but we currently do not allow this either since this would
complicate the algorithm, and we did not see any need for that in our case
studies.}}

\enlargethispage{5mm}

The rules according to which it is decided whether a pair of objects
simultaneously encountered in the input SMGs can be joined are the following.
First, they must have the same size, validity, and in case of DLSs, the same
head, prev, and next offsets. It is possible to join DLSs of different lengths.
It is also possible to to join DLSs with regions that may be approximated as 1+
DLSs for that purpose. The result is a DLS whose length is the minimum of the
lengths of the joined DLSs (hence, e.g., joining a~region~with~a~2+~DLS gives a
1+ DLS). The levels of the joined objects must also be the~same up to the
following case. When joining a sub-SMG nested below a DLS with the corresponding
sub-SMG rooted at a~region (restricted by ignoring the next and prev links),
objects corresponding to each other appear on different levels: E.g., objects
nested right below a~DLS of level 0 are on level 1, whereas the corresponding
objects directly referenced by a region of level 0 are on level 0 (since for
regions, nested and shared sub-SMGs are not distinguished). This difference can,
of course, increase when descending deeper in a hierarchically-nested data
structure since the difference is essentially given by the different numbers of
DLSs passed on the different sides of the join. This difference is tracked by
the join algorithm, and only the objects whose levels differ in the appropriate
way are allowed to be joined.


When two objects are being joined, a \emph{join reinterpretation} operator is
used to ensure that they share the same set of fields and hence have the same
number and labels of outgoing edges (which is always possible albeit sometimes
for the price of introducing has-value edges leading to unknown values). A
formalization of join reinterpretation is available
in~Section~\ref{app:joinReinterp}, including a concrete join reinterpretation
operator designed to preserve maximum information on nullified blocks in both of
the objects being joined. The join reinterpretation allows the fields of the
joined objects to be processed in pairs of the same size and type. As for
joining values, we do not allow joining addresses with unknown
values.\footnote{Allowing a join of an address and an unknown value could lead
to a need to drop a part of the allocated heap in one of the SMGs (in case it
was not accessible through some other address too), which we consider to be a
too big loss of information.} Moreover, the zero value cannot be joined with
a~non-zero value. Further, addresses can be joined only if the points-to edges
leading from them are labelled by the same offset, and when they lead to DLSs,
they must have the same target specifier. On the other hand, apart from the
already above expressed requirement of not joining a~single value in one SMG
with several values in the other SMG, no further requirements are put on joining
non-address values, which is possible since we currently track their equalities
only.

To increase chances for successfully joining two SMGs, the basic algorithm from
above is extended as follows. When a pair of objects cannot be joined and at
least one of them is a DLS (call it $d$ and the other object $o$), the algorithm
proceeds as though $o$ was preceded by a 0+ DLS $d'$ that is up to its length
isomorphic with $d$ (including the not yet visited part of the appropriate
sub-SMG nested below $d$). Said differently, the algorithm virtually inserts
$d'$ before $o$, joins $d$ and $d'$ into a single 0+ DLS, and then continues by
trying to join $o$ and the successor of $d$. This extension is possible since
the semantics of a~0+~DLS includes the empty list, which can be safely assumed
to appear anywhere, compensating a missing object in one~of~the~SMGs.

Note, however, that the virtual insertion of a 0+ DLS implies a need to relax
some of the requirements from above. For instance, one needs to allow a join of
two different addresses from one SMG with one address in the other (the prev and
next addresses of~$d$ get both joined with the address preceding $o$). Moreover,
the possibility to insert 0+ DLSs introduces some non-determinism into the
algorithm since when attempting to join a pair of incompatible DLSs, a 0+ DLS
can be inserted into either of the two input DLSs, and we choose one of them.
The choice may be wrong, but for performance reasons, we never backtrack.
Moreover, we use the 0+ DLS insertion only when a join of two objects fails
locally (i.e., without looking at the successors of the objects being joined).
When a pair of objects can be locally joined, but then the join fails on their
successors, one could consider backtracking and trying to insert a 0+ DLS, which
we again do not do for performance reasons (and we did not see a need for that
in our cases studies so far).

The described join algorithm is used in two scenarios: (1)~When joining
garba\-ge-free SPCs to reduce the number of SPCs obtained from different paths
through the program, in which case the traversal starts from pairs of identical
program variables. (2) As a part of the abstraction algorithm for merging a pair
of neighbouring objects (together with the non-shared parts of the sub-SMGs
rooted at them) of a doubly-linked list into a single DLS, in which case the
algorithm is started from the neighbouring objects to be merged. In the join
algorithm, the join status is computed on-the-fly. Initially, the status is set
to~$\iso$. Next, whenever performing a step that implies a particular relation
between $G_1$ and $G_2$ (e.g., joining a $0+$ DLS from $G_1$ with a $1+$ DLS
from $G_2$ implies that $G_1 \jsupset G_2$, assuming that the remaining parts of
$G_1$ and $G_2$ are semantically equal), we appropriately update the join
status.

\vspace*{-4mm}\subsection{Abstraction}\label{sec:abstraction}\vspace*{-2mm}

Our abstraction is based on \emph{merging uninterrupted sequences} of
neighbouring objects, together with the $\{\nfo,\pfo\}$-restricted sub-SMGs
rooted at them, into a single DLS. This is done by repeatedly applying a slight
extension of the join algorithm on the $\{\nfo,\pfo\}$-restricted sub-SMGs
rooted at the neighbouring objects. The sequences to be merged are identified by
so-called \emph{candidate DLS entries} that consist of an object $o_c$ and next,
prev, and head offsets such that $o_c$ has a neighbouring object with which it
can be merged into a DLS that is linked through the given offsets. The
abstraction is driven by the \emph{cost} to be paid in terms of the loss of
precision caused by merging certain objects and the sub-SMGs rooted at them. In
particular, we distinguish joining of equal, entailed, or incomparable sub-SMGs.
The higher the loss of precision is, the longer sequence of mergeable objects is
required to enable a merge of the sequence.

In the extended join algorithm used in the abstraction (cf.
Section~\ref{app:mergeSubSMGs}), the two simultaneous searches are started from
two neighbouring objects $o_1$ and $o_2$ of the same SMG~$G$ that are the roots
of the $\{ \nfoBS_c, \pfoBS_c \}$-restricted sub-SMGs $G_1$, $G_2$ to be merged. The
extended join algorithm constructs the sub-SMG $G_{1,2}$ that is to be nested
below the DLS resulting from the join of $o_1$ and $o_2$.  The extended join
algorithm also returns the sets $O_1$, $V_1$ and $O_2$, $V_2$ of the objects and
values of $G_1$ and $G_2$, respectively, whose join gives rise to $G_{1,2}$.
Unlike when joining two distinct SMGs, the two simultaneous searches can get to
a single node at the same time. Clearly, such a node is shared by $G_1$ and
$G_2$, and it is therefore \emph{not} included into the sub-SMG $G_{1,2}$ to be
nested below the join of $o_1$ and $o_2$.

Below, we explain in more detail the particular steps of the abstraction. For
the explanation, we fix an SPC $C=(G,\nu)$ where $G = (O,V,\Lambda,H,P)$ is an
SMG with the sets of regions $R$, DLSs $D$, and addresses $A$.


\vspace*{-3mm}\subsubsection*{Candidate DLS Entries}\vspace*{-2mm}

A quadruple $(o_c, \hfoBS_c, \nfoBS_c, \pfoBS_c)$ where $o_c \in O$ and $\hfoBS_c,$ $ \nfoBS_c,
\pfoBS_c \in \nat$ such that $\nfoBS_c < \pfoBS_c$ is considered a \emph{candidate DLS
entry} iff the following holds: (1)~$o_c$ is a valid heap object. (2)~$o_c$ has
a neighbouring object \mbox{$o \in O$} with which it is doubly-linked through the
chosen offsets, i.e., there are $a_1, a_2 \in A$ such that $H(o_c,\nfoBS_c,\ptr) =
a_1$, $P(a_1)=(\hfoBS_c,\tg_1,o)$ for $\tg_1 \in \{ \first, \reg \}$,
$H(o,\pfoBS_c,\ptr) = a_2$, and $P(a_2)=(\hfoBS_c,\tg_2,o_c)$ for $\tg_2 \in \{ \last,
\reg \}$.

\vspace*{-3mm}\subsubsection*{Longest Mergeable Sequences\vspace*{-2mm}}

The \emph{longest mergeable sequence} of objects given by a candidate DLS entry
$(o_c, \hfoBS_c, \nfoBS_c, \pfoBS_c)$ is the longest sequence of distinct valid heap
objects whose first object is $o_c$, all objects in the sequence are of level
$0$, all DLSs that appear in the sequence have $\hfoBS_c$, $\nfoBS_c$, $\pfoBS_c$ as
their head, next, prev offsets, and the following holds for any two neighbouring
objects $o_1$ and $o_2$ in the sequence (for a~formal description, cf.
Section~\ref{app:longestMergSeq}): (1) The objects $o_1$ and
$o_2$ are doubly-linked through their $\nfoBS_c$ and $\pfoBS_c$ fields. (2) The
objects $o_1$ and $o_2$ are a part of a~sequence of objects that is not pointed
from outside of the detected list structure. (3) The $\{ \nfoBS_c, \pfoBS_c
\}$-restricted sub-SMGs $G_1$ and $G_2$ of $G$ rooted at $o_1$ and $o_2$ can be
joined using the extended join algorithm into the sub-SMG $G_{1,2}$ to be nested
below the join of $o_1$ and $o_2$. Let $O_1$, $V_1$ and $O_2$, $V_2$ be the sets
of non-shared objects and values of $G_1$ and $G_2$, respectively, whose join
gives rise to $G_{1,2}$. (4) The non-shared objects and values of $G_1$ and
$G_2$ (other than $o_1$ and $o_2$ themselves) are reachable via $o_1$ or $o_2$,
respectively, only. Moreover, the sets $O_1$ and $O_2$ contain heap objects
only.

\enlargethispage{5mm}

\vspace*{-3mm}\subsubsection*{Merging Sequences of Objects into
DLSs}\vspace*{-2mm}

\begin{figure}[t]
  \centering
  \resizebox{0.85\hsize}{!}{
    \includegraphics{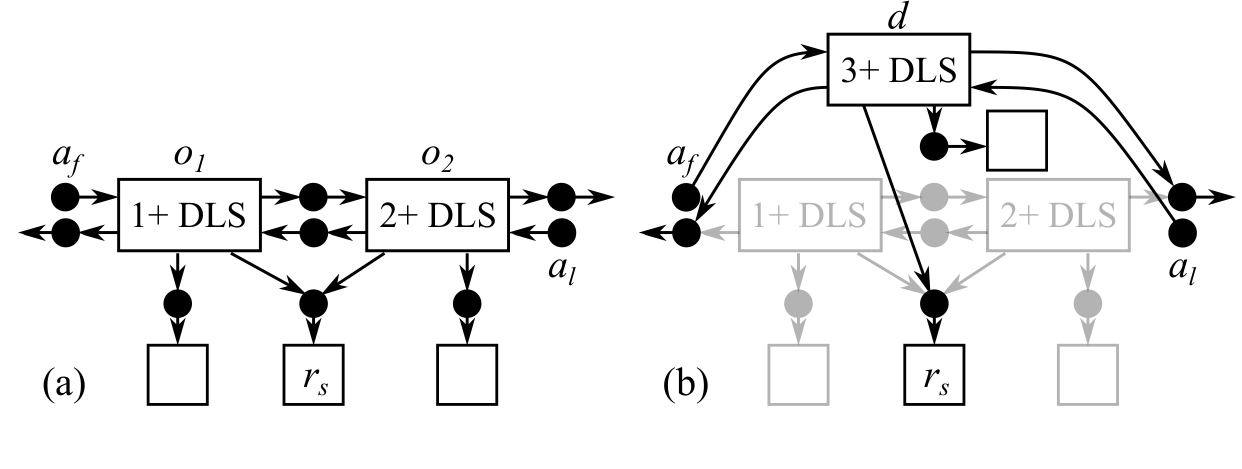}
  }
  \vspace*{-3mm}
  \caption{The elementary merge operation: (a) input (b) output.}
  \vspace*{-4mm}
  \label{fig:merge}
\end{figure}

Sequences of objects are merged into a~single DLS \emph{incrementally}, i.e.,
starting with the first two objects of the sequence, then merging the resulting
new DLS with the third object in the sequence, and so on. Each of the
\emph{elementary merge operations} is performed as follows (see
Fig.~\ref{fig:merge} for an illustration).

Assume that $G$ is the SMG of the current SPC (i.e., the initial SPC or the SPC
obtained from the last merge) with the set of points-to edges $P$ and the set of
addresses $A$. Further, assume that $o_1$ is either the first object in the
sequence to be merged or the DLS obtained from the previous elementary merge,
$o_2$ is the next object of the sequence to be processed, and $\hfoBS_c$,
$\nfoBS_c$, $\pfoBS_c$ are the offsets from the candidate DLS entry defining the
sequence to be merged. First, we merge $o_1$ and $o_2$ into a DLS $d$ using
$\hfoBS_c$, $\nfoBS_c$, and $\pfoBS_c$ as its defining offsets, which is a part
of our extended join operator (cf.~Section~\ref{app:mergeSubSMGs}). The sub-SMG
nested below $d$ is created using the extended join algorithm too. Next, the
DLS-linking pointers arriving to $o_1$ and $o_2$ are redirected to $d$. In
particular, if there is $a_f \in A$ such that $P(a_f) = (o_1,\hfoBS_c,\tg)$ for
some $\tg \in \{ \first, \reg \}$, then $P$ is changed such that $P(a_f) =
(d,\hfoBS_c,\first)$. Similarly, if there is $a_l \in A$ such that $P(a_l) =
(o_2,\hfoBS_c,\tg)$ for some $\tg \in \{ \last, \reg \}$, then $P$ is changed
such that $P(a_l) = (d,\hfoBS_c,\last)$.  Finally, each heap object and each
value (apart from the null address and null object) that are not reachable from
any static or stack object of the obtained SPC are removed from its SMG together
with all the edges adjacent to them.

\vspace*{-3mm}\subsubsection*{The Top-Level Abstraction Algorithm}\vspace*{-2mm}

Assume we are given an SMG $G$, and a candidate DLS entry $(o_c, \hfoBS_c,
\nfoBS_c, \pfoBS_c)$ defining the longest mergeable sequence of objects $\sigma
= o_1 o_2 \ldots o_n$ in $G$ of length $|\sigma| = n \geq 2$.  We define the
\emph{cost} of merging a~pair of objects $o_1, o_2$, denoted $\cost(o_1, o_2)$,
as follows.  First, $\cost(o_1, o_2) = 0$ iff the
$\{\nfoBS_c,\pfoBS_c\}$-restricted sub-SMGs $G_1$ and $G_2$ rooted at $o_1, o_2$
are equal (when ignoring the kinds of $o_1$ and $o_2$, i.e., when not
distinguishing whether $o_1$, $o_2$ are regions or DLSs as well as ignoring the
minimum length constraints on $o_1$, $o_2$). This is indicated by the $\iso$
status returned by the modified join algorithm applied on $G_1, G_2$.  Further,
$\cost(o_1, o_2) = 1$ iff $G_1$ entails $G_2$, or vice versa, which is indicated
by the status \mjsupset~or~\mjsubset. Finally, $\cost(o_1, o_2) = 2$ iff $G_1$
and $G_2$ are incomparable, which is indicated by status \mjoin.  The cost of
merging a~sequence of objects $\sigma = o_1 o_2 \ldots o_n$, denoted
$\cost(\sigma)$, is defined as the maximum of $\cost(o_1, o_2),\; \cost(o_2,
o_3), ..., \cost(o_{n-1}, o_n)$.

Our abstraction is parameterized by associating each cost $c \in \{ 0, 1, 2 \}$
with the \emph{length threshold}, denoted $\lenThr(c)$, defining the minimum
length of a sequence of mergeable objects allowed to be merged for the given
cost. Intuitively, the higher the cost is, the bigger loss of precision is
incurred by the merge, and hence a bigger number of objects to be merged is
required to compensate the cost. In our experiments discussed in
Section~\ref{sec:implementation}, we, in particular, found as optimal the
setting $\lenThr(0) = \lenThr(1) = 2$ and $\lenThr(2) = 3$. Our tool, however,
allows the user to tweak these values.

\enlargethispage{5mm}

Based on the above introduced notions, the process of \emph{abstracting an SPC}
can now be described as follows. First, all candidate DLS entries are
identified, and for each of them, the corresponding longest mergeable sequence
is computed. Then, each longest mergeable sequence $\sigma$ for which $|\sigma|
< \lenThr(\cost(\sigma))$ is discarded. Out of the remaining ones, we select
those that have the lowest cost. From them, we then select those that have the
longest length. Finally, out of them, one is selected arbitrarily. The selected
sequence is merged, and then the entire abstraction process repeats until there
is no sequence that can be merged taking its length and cost~into~account.


\vspace*{-3mm}\subsection{Checking Equality and Inequality of
Values}\label{sec:inconsistency-check}\vspace*{-2mm}

Checking equality of values in SMGs amounts to simply checking their identity.
For checking inequality, we use an algorithm which is sound and efficient but
incomplete. It is designed to succeed in most common cases, but we allow it to
fail in some exceptional cases (e.g., when comparing addresses out of bounds of
two distinct objects) in order not to harm its efficiency. The basic idea of the
algorithm is as follows (cf. Section~\ref{app:condSymExec}): Let $v_1$ and
$v_2$ be two distinct values of level~0 to be checked for inequality (other
levels cannot be directly accessed by program statements). First, if the same
value or object can be reached from $v_1$ and $v_2$ through 0+ DLSs only (using
the next/prev fields when coming through the $\first$/$\last$ target specifiers,
respectively), then the inequality between $v_1$ and $v_2$ is not established.
This is due to $v_1$ and $v_2$ may become the same value when the possibly empty
0+ DLSs are removed (or they may become addresses of the first and last node of
the same 0+ DLS, and hence be equal in case the list contains a single node).
Otherwise, $v_1$ and $v_2$ are claimed different if the final pair of values
reached from them through a sequence of 0+ DLSs represents different addresses
due to pointing (1)~to different valid objects (each with its own unique
address) with offsets inside their bounds, (2)~to the null object and a~non-null
object (with an in-bound offset), (3)~to the same object with different offsets,
or (4)~to the same DLS with length at least~2 using different target specifiers.
Otherwise, the inequality is not established.

\vspace*{-4mm}\subsection{A Note on Symbolic Execution over
SMGs/SPCs}\vspace*{-2mm}

The symbolic execution algorithm based on SPCs is similar to
\cite{InvaderCAV07}. It uses the read reinterpretation operator for memory
lookup (as well as type-casting) and the write reinterpretation operator for
memory mutation. Whenever a DLS is about to be accessed (or its address with a
non-head offset is about to be taken), a~materialisation (as described in
Section \ref{sec:semantics}) is performed so that the actual program statements
are always executed over concrete objects.\footnote{A DLS can be materialised
from its last element too, which is analogous to the materialisation from the
first element as described in Section \ref{sec:semantics}.} If the minimum
length of the DLS being materialised is zero, the computation is split into two
branches---one for the empty segment and one for the non-empty segment. In the
former case, the DLS is removed (as described in Section~\ref{sec:semantics})
while in the latter case, the minimum length of the DLS is incremented. When
executing a conditional statement, the algorithm for checking (in)equality of
values from Section~\ref{sec:inconsistency-check} is used. If neither equality
nor inequality are established, the execution is split into two branches, one of
them assuming the compared values to be equal, the other assuming them not to be
equal. This may again involve removing 0+ DLSs in one of the branches and
incrementing their minimum length in the other
(cf.~Section~\ref{app:condSymExec}).

\enlargethispage{5mm}

To reduce the number of SPCs generated by the symbolic execution, the
\emph{join} operator introduced in Section~\ref{sec:join} can be used to join an
SPC that was newly generated for some particular program location with some SPC
generated for that location sooner (e.g., joining a region with a 2+ DLS into a
single 1+ DLS), hence reducing the number of SPCs remembered for that location
(and explored from that location). Trading speed for precision, the operator can
be applied at every control location, at the beginning of every basic block, at
every loop head location, or not at all.  Apart from the join operator, the
\emph{abstraction} mechanism from Section~\ref{sec:abstraction} is to be applied
to ensure termination on unbounded list structures. Again, the use of the
abstraction may be restricted as in the case of the join operator (it needs not
even be used if no unbounded data structures are used). As the terminating
criterion, one can use isomorphism or entailment checking between a newly
generated SPC and those already known for a given program location. Checking for
\emph{isomorphism} can be done using the join operator of
Section~\ref{sec:join}, making sure that it succeeds and returns the $\iso$
status. Checking for \emph{entailment} can also be done using the proposed join
operator, this time checking for the $\jsubset$ or $\jsupset$ status (allowing
one to discard either a sooner generated SPC or the new SPC). 

\vspace*{-3mm}\subsubsection*{Soundness of the Analysis}\vspace*{-2mm}

In the described analysis, program statements are always executed on
concrete objects only, closely following the C semantics. The read
reinterpretation is defined such that it cannot change the semantics of the
input SMG, and the write reinterpretation can only over-approximate the
semantics in the worst case. Likewise, our abstraction and join algorithms are
allowed to only over-approximate the semantics---indeed, when joining a
pair of nodes, the semantics of the resulting node is always generic enough to
cover the semantics of both of the joined nodes (e.g., the join of a
$2+$~DLS with a compatible region results in a $1+$~DLS, etc.).
Moreover, the entailment check used to terminate the analysis is based on
the join operator and consequently conservative. Hence, it is not difficult to
see that the proposed analysis is sound (although a full proof of
this fact would be rather technical).

\enlargethispage{5mm}

\vspace*{-4mm}\subsection{Running Example}\label{sec:RunEx}\vspace*{-2mm}

We now illustrate some of the main presented concepts on a running example. In
particular, we consider the C code shown in Fig.~\ref{fig:RunExCode}. For its
five designated locations \texttt{L1}--\texttt{L5}, Fig.~\ref{fig:RunEx} shows
the SPCs generated by our analysis. We use a~simplified notation for the SPCs
similar to that already used in Fig.~\ref{fig:join}. The ``bar''-terminated
edges denote pointers to \texttt{NULL}. We draw edges corresponding to
\texttt{prev} pointers to the left of objects and those corresponding to
\texttt{next} pointers to the right of objects. Bidirectional edges denote a
pair of \texttt{prev}/\texttt{next} pointers. 

\begin{figure}[t]
\begin{lstlisting}
#include <stdlib.h>                     // calloc(), free()

struct list_item {
    struct list_item *next;
    struct list_item *prev;
};

int main(void)
{
    struct list_item *l = NULL;         // pointer to a DLL
    struct list_item *x = NULL;         // auxiliary pointer  

    while (__VERIFIER_nondet_int() != 0) {
      L1:
        x = calloc(sizeof(struct list_item),1);
        x->next = l;
        if (l != NULL)
            l->prev = x;
        l = x;
      L2:
    }
  L3:
    while (l != NULL) {
        x = l;
        l = l->next;
        free(x);
      L4:
    }
  L5: return 0;
}
\end{lstlisting}
\vspace*{-4mm}
\caption{A running example creating and destroying a DLL.}
\vspace*{-4mm}
\label{fig:RunExCode}
\end{figure}

\begin{figure}[t]
  \centering
  \resizebox{0.8\hsize}{!}{
    \includegraphics{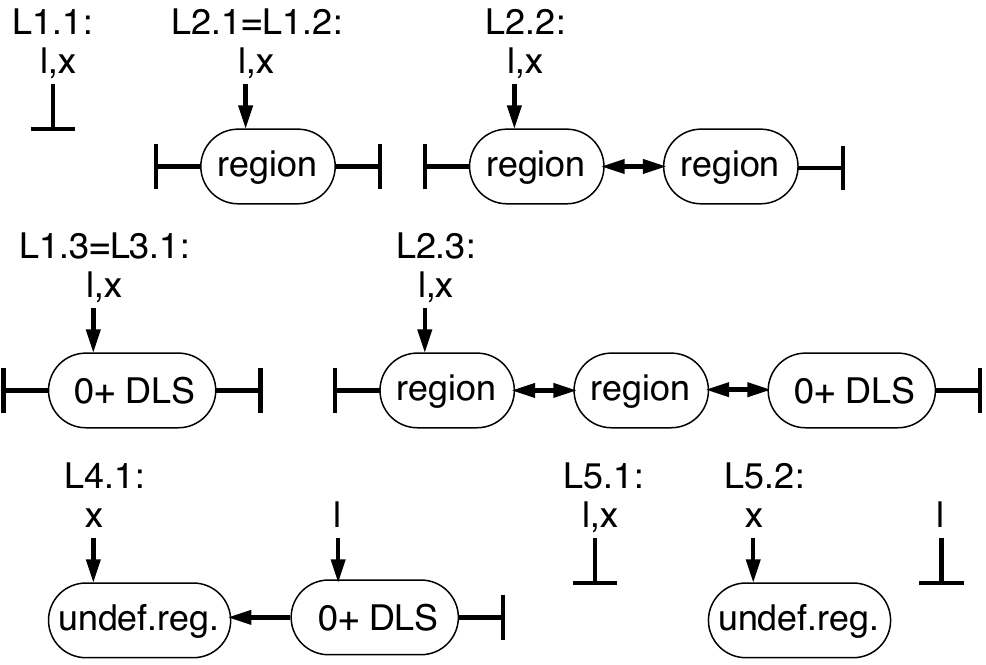}
  }
  \vspace*{0mm}
  \caption{Some of the SPCs generated during analysis of the code from
    Fig.~\ref{fig:RunExCode}.}
  \vspace*{-4mm}
  \label{fig:RunEx}
\end{figure}

The way the different SPCs are obtained is discussed below. However, first, note
that all objects of all SPCs are at level 0 (i.e., there are no nested sub-SMGs
here). Assuming that the code is compiled for a 64-bit architecture with 8B-long
pointers, all objects are of size 16B. The \texttt{next} pointers start at
offset 0 while the \texttt{prev} pointers at offset 8, and they all have the
target offset 0.

The first SPC that gets generated at the location \texttt{L1} is denoted as
\texttt{L1.1}. It sets both the \texttt{l} and \texttt{x} program variables to
\texttt{NULL}. From \texttt{L1.1}, the SPC \texttt{L2.1} is generated at the
location \texttt{L2}. As we can see, it consists of a single region with the
\texttt{prev} field implicitly nullified (as \texttt{calloc} was used).
\texttt{L2.1} is propagated as the second SPC for the location \texttt{L1},
i.e., it becomes \texttt{L1.2}. The join operator presented above cannot join
\texttt{L1.1} and \texttt{L1.2}, and so both of them are kept (for the time
being). Through another iteration of the first \texttt{while} loop,
\texttt{L2.2} is obtained from \texttt{L1.2} in a natural way.

Now, several interesting changes happen. First, when closing the second
iteration of the first \texttt{while} loop, \texttt{L2.2} gets abstracted to an
SPC containing a single 2+ DLS that is \texttt{NULL}-terminated at both of its
ends: indeed, \texttt{L2.2} contains an uninterrupted sequence of two
equally-sized regions linked in the fashion of a DLL. Second, the resulting SPC
can be joined both with \texttt{L1.2}, yielding an SPC consisting of a 1+ DLS,
and then with \texttt{L1.1}, yielding the new SPC \texttt{L1.3} containing a 0+
DLS, which is kept as the sole SPC at the location \texttt{L1}.

\enlargethispage{5mm}

The SPC \texttt{L1.3} is once more passed through the first \texttt{while} loop.
Notice that the \texttt{if (l != NULL)} statement will split the 0+ DLS present
in \texttt{L1.3} to two cases: the SPC \texttt{l}=\texttt{x}=\texttt{NULL} and
an SPC containing a 1+ DLS. Only the latter SPC is then subject to the
\texttt{l->prev = x} statement. For this statement to be symbolically executed,
the 1+ DLS is materialised to a sequence of a region, pointed to by \texttt{l},
and a 0+ DLS segment linked to it through a pair of \texttt{next}/\texttt{prev}
pointers. By linking the region newly allocated by \texttt{calloc} and pointed
to by \texttt{x} with the materialised region pointed to by \texttt{l}, and by
moving the \texttt{l} pointer, the \texttt{L2.3} SPC will eventually arise at
the end of the body of the first \texttt{while} loop. \texttt{L2.3} entails
\texttt{L2.2}, and so the latter does not need to be kept at the location
\texttt{L2}. More importantly, by abstraction applied when going back to the
loop header, \texttt{L2.3} will be transformed into an SPC consisting of a 2+
DLS, which will then be found to be entailed by \texttt{L1.3}.

Assuming that the analysis explores the program such that the first
\texttt{while} loop is first completely explored and only then the analysis
continues further on, the only SPC generated for location \texttt{L3} is
\texttt{L3.1}, which is the same as \texttt{L1.3.} 

Due to the condition of the second \texttt{while} loop, \texttt{L3.1} is split
to two SPCs: once again the SPC \texttt{l}=\texttt{x}=\texttt{NULL}, which then
appears as the SPC \texttt{L5.1} at the end of the program, and an SPC
containing a 1+ DLS. Only the latter enters the first iteration of the second
\texttt{while} loop. For the \texttt{l = l->next} statement to be symbolically
executed, the 1+ DLS is materialised into a sequence of a region, pointed to by
both \texttt{l} and \texttt{x}, and a~0+~DLS. The \texttt{l} pointer is then
moved to point to the beginning of the 0+ DLS. The \texttt{x}-pointed region is
subsequently freed, after which all its outgoing edges are removed, and the
region itself is marked as undefined, taking us to the \texttt{L4.1} SPC.

The condition of the second \texttt{while} loop splits \texttt{L4.1} to two
cases again: the case where \texttt{l}=\texttt{NULL}, yielding the SPC
\texttt{L5.2} at the end of the program, and an SPC where \texttt{l} points to a
1+ DLS whose \texttt{prev} pointer points to the freed region pointed to by
\texttt{x}. For the \texttt{l = l->next} statement to be symbolically executed,
the 1+ DLS is materialised to a sequence of an \texttt{l}-pointed region and a
0+ DLS. The \texttt{prev}-pointer of the \texttt{l}-pointed region points to the
previously freed region (with \texttt{x} already pointing to the same region as
\texttt{l}), and its \texttt{next} pointer points to the 0+ DLS. Subsequently,
\texttt{l} is moved to point to the beginning of the 0+ DLS, and the
\texttt{x}-pointed region is freed. Due to that, its outgoing edges are removed,
and the previously freed region, which is now completely inaccessible, is
removed too. Hence, \texttt{L4.1} is again obtained.

\vspace*{-3mm}\section{Extensions of SMGs}\label{sec:extensions}\vspace*{-2mm}

Next, we point out that the above introduced notion of SMGs can be easily
extended in various directions, and we briefly discuss several such extensions
(including further kinds of abstract objects), most of which are implemented in
the Predator tool.

\vspace*{-4mm}\subsubsection*{Explicit Non-equivalence Relations}\vspace*{-2mm}

When several objects have the same concrete value stored in some of their
fields, this is expressed by making the appropriate has-value edges lead from
these objects to the same value node in the SMG. On the other hand, two
different value nodes in an SMG do not necessarily represent different concrete
values. To express that two abstract values represent distinct concrete values,
SMGs can be extended with a symmetric, irreflexive relation over values, which
we call an \emph{explicit non-equivalence relation.}\footnote{This is similar to
the equality and non-equality constraints in separation logic, but only
non-equality constraints are kept explicit.} Such a relation can be introduced
when the analysis proceeds to some branch of a conditional statement such that
the condition of the statement (negated for the false branch) implies the
non-equivalence relation. The introduced non-equivalence relations are then to
be taken into account in further operations, including symbolic execution of
conditional statements and the join operator. In the latter case, when joining
two SMGs where some non-equality edge exists in one of them only, it may be
dropped for the price of appropriately worsening the resulting join status.

Clearly, SMGs can be quite naturally extended by allowing more predicates on
data, which is, however, beyond the scope of this work (up to a~small extension
by tracking not only the $0$ value but also intervals with constant bounds that
is mentioned below) and has so far been not implemented in the Predator tool
either.

\enlargethispage{5mm}

\vspace*{-4mm}\subsubsection*{Checking Equivalence of Valid and Invalid
Objects}\vspace*{-2mm}

Testing of inequality described in Section~\ref{sec:inconsistency-check}
concerns inequality of pointers pointing to different valid objects, null and
non-null objects, the same object with different offsets, or to different ends
of a doubly-linked list segment with at least two elements. However, there is
one more way how inequality can be established, namely, when comparing pointers
to a valid region and to an invalid region where the invalid one was allocated
later than the valid one (which we can check due to the way the objects are
numbered in Predator).  Indeed, in such a case, both of the regions must have
existed at the same time, the valid object is continuously valid since then, and
the two objects must lie on different addresses since the address of a
continuously valid object could not have been recycled.

\vspace*{-4mm}\subsubsection*{Singly-Linked List Segments (SLSs)}\vspace*{-2mm}

Above, we have presented all algorithms on SMGs describing doubly-linked lists
only. Nevertheless, the algorithms work equally well with singly-linked lists
represented by an additional kind of abstract objects, SLSs, that have no $\pfo$
offset, and their addresses are allowed to use the $\first$ and $\all$ target
specifiers only. The algorithm looking for DLS entry candidates then simply
starts looking for SLS entry candidates whenever it does not discover the
back-link.

\vspace*{-4mm}\subsubsection*{0/1 Abstract Objects}\vspace*{-2mm}

In order to enable summarization of lists whose nodes can \emph{optionally}
point to some region or that point to nested lists whose
length never reaches 2~or more, we introduce the so-called \emph{0/1
abstract objects}. We distinguish three kinds of them with different numbers of
neighbour pointers. The first of them represents 0/1 SLSs with
one neighbour pointer, another represents 0/1 DLSs with two neighbour
pointers\tronly{\footnote{If a DLL consists of exactly one node, the value of
its next pointer is equal to the value of its prev pointer. There is no point in
distinguishing them, so we call them both neighbour pointers.}}. These objects
can be later joined with compatible SLSs or DLSs. The third kind has no
neighbour pointer, and its address is assumed to be NULL when the region is not
allocated. This kind is needed for optionally allocated regions referred from
list nodes but never handled as lists themselves. The 0/1 abstract objects are
created by the join algorithm when a region in one SMG cannot be matched with an
object from the other SMG and none of the above described join
mechanisms~applies.


\vspace*{-4mm}\subsubsection*{Offset Intervals and Address
Alignment}\vspace*{-2mm}

The basic SMG notion labels points-to edges with scalar offsets within the
target object. This labelling can be generalized to \emph{intervals of offsets}.
The intervals can be allowed to arise by joining objects with incoming pointers
compatible up to their offset. This feature is useful, e.g., to handle lists
arising in higher-level memory allocators discussed in the next section where
each node points to itself with an offset depending on how much of the node has
been used by sub-allocation. Offset intervals also naturally arise when the
analysis is allowed to support \emph{address alignment}, which is typically
implemented by masking several lowest bits of pointers to zero, resulting in a
pointer whose offset is in a certain interval wrt the base address. Similarly,
one can allow the \emph{object size} to be given by an interval, which in turn
allows one to abstract lists whose nodes are of a variable size.

\vspace*{-4mm}\subsubsection*{Integer Constants and Intervals}\vspace*{-2mm}

The basic SMG notion allows one to express that two fields have the same value,
which is represented by the corresponding has-value edges leading to the same
value node, or that their values differ, which is represented using the above
mentioned explicit non-equivalence relation. In order to improve the support of
dealing with integers, SMGs can be extended by associating value nodes with
concrete integer numbers. These can be respected by the join algorithm up to
some given bound and replaced by the unknown value when the bound is exceeded
(as done in Predator). Alternatively, they can be abstracted to intervals with
bounds being concrete integer constants up to some bound or plus/minus infinity
(also supported by Predator), or some other abstract numerical domains may be
used (unsupported~by~Predator).

\vspace*{-4mm}\section{Implementation}\label{sec:implementation}\vspace*{-2mm}

We have implemented the above described algorithms (including most of the
mentioned extensions) in the Predator
tool.\footnote{\url{https://www.fit.vutbr.cz/research/groups/verifit/tools/predator/}}
Below, we describe the architecture of the tool---in fact, a \emph{tool suite}
centred around the Predator analysis kernel---and various further extensions,
optimisations, and implementation details related to it. The description is
partly based on the tool paper \cite{predatorHVC}.

\vspace*{-4mm}\subsection{Architecture}\label{sec:architecture}\vspace*{-2mm}

\begin{figure}[t]
\begin{center}
  \includegraphics[width=1.0\textwidth]{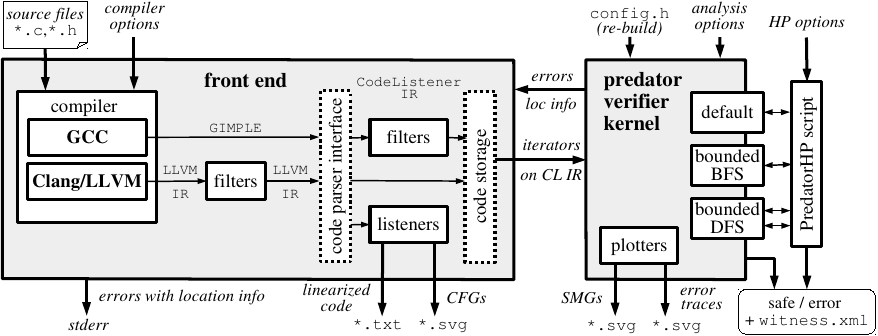}
  \vspace*{-3mm}
  \caption{Architecture of the Predator tool suite.}
  \vspace*{-6mm}
  \label{fig:arch}
\end{center}
\end{figure}

The architecture of the Predator tool suite is shown in Fig.~\ref{fig:arch}.
Its \emph{front end} is based on the \emph{Code Listener} (CL) infrastructure
\cite{codelistener11} that can accept input from both the GCC and Clang/LLVM
compilers. CL is connected to both GCC and LLVM as their plug-in (or
``pass'').

When used with \emph{GCC}, CL reads in the GIMPLE intermediate representation
(IR) from GCC and transforms it into its own \emph{Code Listener IR} (CL IR),
based on simplified GIMPLE. The resulting CL IR can be
\emph{filtered}---currently there is a filter that replaces \texttt{switch}
instructions by simple conditions---and stored into the code storage. When used
with \emph{Clang/LLVM}, CL reads in the LLVM IR and uses LLVM's AddressSanitizer
for a use-after-scope instrumentation of the lifetime of variables, removes LLVM
\texttt{switch} instructions, and (optionally)~simplifies~the IR through
a~number of other \emph{filters} in the form of LLVM optimisation passes, both
LLVM native or newly added. These filters can in-line functions, split composed
initialisation of global variables, and/or change memory references to register
references (removing unnecessary \texttt{alloca} instructions). These
transformations can be used independently of Predator to simplify the LLVM IR to
have a~simpler starting point for developing new analysers. Moreover, CL offers
a \emph{listeners architecture} that can be used to further process CL IR.
Currently, there are listeners that can print out the CL IR or produce
a~graphical form of the control flow graphs (CFGs) present in~it.


The \emph{code storage} stores the obtained CL IR and makes it available to the
Predator verifier kernel through a special API. This API allows one to easily
iterate over the types, global variables, and functions defined in the code. For
each function, one can then iterate over its parameters, local variables, and
its CFG. Of course, other verifier kernels than the one of Predator can be
linked to the code storage.


\vspace*{-2mm}\subsection{The Kernel of Predator}\label{sec:kernel}
\vspace*{-2mm}

The kernel of Predator (written in C++ like its front end) implements a form of
an abstract interpretation loop over the SMG domain where the widening takes
into account the newest computed SMG for a given location only and is based on
the abstraction mechanism from Section~\ref{sec:abstraction}.

\vspace*{-4mm}\subsubsection*{Programs To Be Verified}\vspace*{-2mm}

A program to be verified by Predator must be \emph{closed} in that it must
allocate and initialize all the data structures used. 
%
%
By default, Predator disallows calls to \emph{external functions} in order to
exclude any side effect that could potentially break memory safety. The only
allowed external functions are those that Predator recognizes as built-in
functions. From the C standard, the following functions are currently included
and properly modelled wrt proving memory safety (there are a few more,
GCC-/LLVM-specific ones):\begin{itemize}

  \item \texttt{malloc}, \texttt{alloca}, \texttt{calloc}, \texttt{free}, and
  \texttt{realloc};

  \item \texttt{exit} and \texttt{abort};

  \item \texttt{memset}, \texttt{memcpy}, and \texttt{memmove};

  \item \texttt{printf} and \texttt{puts}; and

  \item \texttt{strlen}, \texttt{strncpy}, and \texttt{strcmp}.

\end{itemize} Models of further functions can be added by the user for the price
of recompiling the analyser. Predator also provides several built-in functions
that are specific to its verification approach, e.g., functions to dump SMGs or
program traces to files.

\enlargethispage{5mm}

\vspace*{-3mm}\subsubsection*{Interprocedural Features}\vspace*{-2mm}

Predator supports indirect calls via \emph{function pointers}, which is
necessary for verification of programs with callbacks (e.g., Linux drivers).
Predator does not support \emph{recursive programs}, but it implements an
\emph{inter-procedural analysis} based on \emph{function summaries}
\cite{reps95}. The summaries consist of pairs of (sub-)SPCs that appeared at the
input/output of a given function during the so-far performed analysis. Depending
on the configuration of Predator, input parts of the summaries are created
either (1)~by taking the entire SPC encountered at a function call or (2)~by
carving out the part of the SPC that is reachable from function parameters and
global variables. The summaries are stored in a call cache. When testing whether
a summary for a~call with a certain input SPC has already been created (or
covered by another SPC), Predator can compare the current input SPC (or its
relevant part) with those stored in the call cache either by isomorphism or
entailment, depending on its configuration. Predator monitors how many
consecutive cache misses are encountered for each function after the last cache
hit, and if that number gets above some configurable threshold, the cache will
not be used for the given function.

Regions for \emph{stack variables} are created automatically as needed and
destroyed as soon as they become dead according to a \emph{static live variables
analysis}\footnote{If a program variable is referenced by a pointer, its
destruction needs to be postponed.}, performed before running the symbolic
execution. When working with \emph{initialised
variables}\tronly{\footnote{According to the C99 standard, all static variables
are initialised, either explicitly or implicitly.}}, we take advantage of our
efficient representation of nullified blocks---we first create a has-value edge
\mbox{$o \hasvalue{0, \mathtt{char[}\size(o)\mathtt{]}} 0$} for each initialised
variable represented by a region $o$, then we execute all explicit initialisers,
which themselves automatically trigger the write reinterpretation.  The same
approach is used for \texttt{calloc}-based heap allocation. Thanks to this, we
do not need to initialise each structure member explicitly, which would incur a
significant overhead. 

\vspace*{-4mm}\subsubsection*{Various Optimisations}\vspace*{-2mm}

As an optimisation, a \emph{copy-on write} approach is used when creating new
SMGs by modifying the already existing ones. Also, the algorithms for
abstraction and join implemented in Predator use some further optimisations of
the basic algorithms described in Section~\ref{sec:operations}.  While objects
in SMGs are type-free, Predator tracks their \emph{estimated type} given by the
type of the pointers through which objects are manipulated. The estimated type
is used during abstraction to postpone merging a~pair of objects with
incompatible types. Note, however, that this is really a heuristic only---we
have a case study that constructs list nodes using solely \texttt{void}
pointers, and it can still be successfully verified by Predator. Another
heuristic is that certain features of the join algorithm (e.g., insertion of a
non-empty DLS or introduction of a 0/1 abstract object) are disabled when
joining SMGs while enabled when merging nodes during abstraction.

\enlargethispage{6mm}

Predator iteratively computes sets of SMGs for each basic block entry of the
control-flow graph of the given program, covering all program configurations
reachable at these program locations. Termination of the analysis is aided by
the abstraction and join algorithms described above. Since the join algorithm is
expensive, it is used at loop boundaries only. When updating states of other
basic block entries, we compare the SMGs for equality\tronly{\footnote{The join
algorithm can be easily restricted to check for equality only---if any action
that would imply inequality of the input SMGs is about to be taken, the join
operation fails immediately.}} only, which makes the comparison way faster,
especially in case a pair of SMGs cannot be joined. Similarly, the abstraction
is by default used at loop boundaries only in order not to introduce abstract
objects where not necessary (reducing the space for false positives that can
arise due to breaking assumptions sometimes used by programmers for code inside
loops as witnessed by some of our case studies).

\vspace*{-4mm}\subsubsection*{Non-Pointer Data}\vspace*{-2mm}

Predator's support of \emph{non-pointer data} is currently limited. As mentioned
already above, Predator can track \emph{integer data} precisely up to a
predefined bound ($\pm 10$ by default), and once the bound is reached, the
values are abstracted out. Optionally, Predator can use intervals with constant
bounds (which may be widened to infinity) while also tracking some simple
dependences between intervals, such as a shift by a constant and
a~multiplication by $-1$. Reinterpretation is used to handle \emph{unions}.
\emph{String} and \emph{float constants} can be assigned, but any operations on
these data types conservatively yield an undefined value. \emph{Arrays} are
handled as allocated memory blocks with their entries accessible via field
offsets much like in the case of structures.

\vspace*{-3mm}\subsubsection*{Dealing with Integer Intervals}
\label{sec:int-size-reg} \vspace*{-2mm}

As can be seen from the previous text, integer intervals may arise in Predator
in multiple contexts: (1) One can get pointers with \emph{interval-based target
offsets}, e.g., by address alignment or when joining pointers pointing to the
same object but with different offsets. (2) Integer variables can get
interval-based values, e.g., by joining results of the analysis on different
branches or when restricting unknown values by some conditions on integer
variables. (3) Finally, there can also arise memory regions of
\emph{interval-based size}, e.g., when allocating structures or arrays whose
size is given by a variable whose value is given by an interval or when joining
SMGs where a pair of corresponding memory regions differs in the size. However,
the introduction of intervals can not only lead to some loss of precision (by
loosing relations of the individual values in the interval with values of other
variables), but Predator currently also imposes many restrictions on how the
intervals can subsequently be used. For example, it does not allow one to
dereference interval-sized regions, due to which the basic version of Predator
behaves poorly when analysing programs with structures or arrays of unknown size
(dependent on program input or abstracted away).

To at least partially improve on this situation, Predator has been extended by
the following pragmatic heuristic. Namely, whenever it hits a conditional
statement that would normally yield an interval value with fixed bounds (e.g.,
executing the statement \texttt{if (n
>=0 \&\& n<10)} where \texttt{n} is unconstrained before the statement), it will
split the further run of the analysis into as many branches as the number of
values in the interval is, each of them evaluating for a concrete value from the
interval. After the split, no interval-based allocations and dereferences (nor
any other problematic interval-based operations), which Predator would fail on,
happen. Though this solution is rather simple, it works nicely in some cases. Of
course, it can lead to a memory explosion when the intervals are large, but then
the analysis fails with no answer as it would fail without the heuristic too.

The above modification of Predator concerns dealing with intervals with finite
bounds. For the case when one of the bounds is infinite, Predator has been
extended to \emph{sample} the interval and perform the further analysis with the
sampled values. The sampling is done by simply taking some number of concrete
values from the given interval starting/ending with the bound that is fixed
(intervals with both bounds infinite correspond to the unknown value). The
number of considered samples is currently set to 3. Of course, this strategy
cannot be used to soundly verify correctness of programs, and so it is used for
detecting bugs only.

\vspace*{-3mm}\subsubsection*{Errors Sought and Error Reporting}\vspace*{-2mm}

Predator is able to discover or prove absence of various kinds of \emph{memory
safety errors}, including various forms of illegal dereferences (null
dereferences, dereferences of freed or unallocated memory, out-of-bound
dereferences), illegal free operations (double free operations, freeing non-heap
objects), as well as memory leakage. \tronly{Memory leakage checks are optimized
by collecting sets of lost addresses for each operation that can introduce a
memory leak (write reinterpretation, \texttt{free}, etc.), followed by checking
whether reachability of allocated objects from program variables depend on the
collected addresses.} Moreover, Predator also uses the fact that SMGs allow for
easy checking whether a~given pair of memory areas overlap. Indeed, if both of
them are inside of two distinct valid regions, they have no overlaps, and if
both of them are inside the same region, one can simply check their offset
ranges for intersection. Such checks are used for reporting invalid uses of
\texttt{memcpy} or the C-language assignment, which expose undefined behavior if
the destination and source memory areas (partially) overlap with each other.
Predator also looks for violations of \emph{assertions} written in the code.

Predator can also detect \emph{invalid dereferences} of \emph{objects local to a
block} from outside of the block. For that, it tracks usage of the
\emph{clobber} instruction of CL IR, which is used to terminate the life time of
local variables of code blocks. Whenever the instruction is encountered, the
concerned memory region is marked as deallocated, and further dereferences of
that region are detected as erroneous.

Predator reports discovered errors together with their location in the code in
the standard GCC format, and so they can be displayed in standard editors or
IDEs. Predator can also produce error traces in a textual or graphical format or
in the XML format of SV-COMP (cf. Section~\ref{sec:svcomp}). Predator also
supports \emph{error recovery} to report multiple program errors during one run.
For example, if a memory leak is detected, Predator only reports a warning, the
unreachable part of the SMG is removed, and the symbolic execution then
continues.


\vspace*{-3mm}\subsubsection*{Options}\vspace*{-2mm}

The kernel supports many \emph{options}. Some of them can be set in the
\texttt{config.h} file (requiring the kernel to be re-compiled) and some when
starting the analysis (cf. the tutorial in Section~\ref{sec:tutorial}). Apart
from various debugging options and some options mentioned already above, one
can, e.g., decide whether the abstraction and join should be performed after
every basic block or at loop points only (abstraction can also be performed when
returning from function calls). One can specify the maximum call depth, choose
between various search orders, switch on/off the use of function summaries and
destruction of dead local variables, control error recovery, and control
re-ordering of lists of SMGs kept for program locations (based on their hit
ratio) and/or their pruning wrt entailment and their location in CFGs.

\vspace*{-2mm}\subsection{Predator Hunting Party}\label{sec:hp} \vspace*{-2mm}

\begin{figure}[t]
  \centering
    \includegraphics{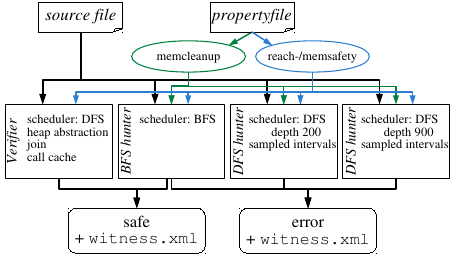}
  \vspace*{-1mm}
  \caption{The flow of control of Predator Hunting Party as used in SV-COMP'19.}
  \vspace*{-2mm}
  \label{fig:PredatorHP}
\end{figure}

The \emph{Predator Hunting Party} (or \emph{PredatorHP} for short) uses the
original Predator analyser to prove programs safe while at the same time using
several bounded versions of Predator for \emph{bug hunting}. PredatorHP, whose
flow of control is shown in Fig.~\ref{fig:PredatorHP}, is implemented as a
Python script that runs several instances of Predator in parallel and composes
the results they produce into the final verification verdict.

In particular, PredatorHP starts four Predators: One of them is the original
Predator that soundly overapproximates the behaviour of the input program---we
denote it as the \emph{Predator verifier} below. Apart from that, three further
Predators, modified as follows, are started: Their join operator is reduced to
joining SMGs equal up to isomorphism, they use no list abstraction, and they are
also prohibited from using the call cache, which is otherwise used for
implementing fuction summaries.\footnote{The use of function summaries is
prohibited since they may yield over-approximation, in particular, when testing
applicability of summaries by entailment.} Two of them use a bounded depth-first
search to traverse the state space, and so we call them \emph{Predator DFS
hunters}. They use bounds of 200 and 900 GIMPLE instructions, respectively. The
third of them---the \emph{Predator BFS hunter}---uses a~breadth-first search to
traverse the state space. The DFS hunters are allowed to use the above described
heuristic based on sampling intervals when looking for bugs in interval-sized
memory regions.

\enlargethispage{5mm}

If the Predator verifier claims a program correct, so does PredatorHP, and it
kills all other Predators. If the Predator verifier claims a program incorrect,
its verdict is ignored since it can be a false alarm (and, moreover, it is
highly non-trivial to check whether it is false or not due to the involved use
of list abstractions and joins). If one of the Predator DFS hunters finds an
error, PredatorHP kills all other Predators and claims the program incorrect,
using the trace provided by the DFS hunter who found the error as a violation
witness.\footnote{The obtained trace can still be spurious due to abstraction
applied on non-pointer data.} One of the DFS hunters searches quickly for bugs
with very short witnesses, and one then searches for longer but still not very
long witnesses. If a DFS hunter claims a program correct, its verdict is ignored
as it may be unsound. If a BFS hunter manages to find an error within the time
budget, PredatorHP claims the program incorrect (without a time limit, the BFS
hunter is guaranteed to find every error). If the BFS hunter finishes and does
not find an error, the program is claimed correct. Otherwise, the verdict
``unknown'' is obtained.

The main strength of PredatorHP is that---unlike various bounded model
checkers---it treats unbounded heap manipulation in a \emph{sound} way. At the
same time, it is also quite \emph{efficient}, and the use of various
concurrently running Predator hunters greatly decreases chances of producing
\emph{false alarms} (there do not arise any due to heap manipulation, the
remaining ones are due to abstraction on~other~data~types).

\vspace*{-3mm}\section{Experiments}\label{sec:experiments}\vspace*{-3mm}

In this section, we present results of experiments with Predator both outside of
SV-COMP as well as within SV-COMP. In the latter case, we concentrate in
particular on the 2019 edition of SV-COMP and on the influence of using
PredatorHP as well as one of the later introduced optimisations of Predator,
namely, that of dealing with intervals of values
(Section~\ref{sec:int-size-reg}).

\vspace*{-3mm}\subsection{Experiments with Predator Outside of
SV-COMP}\label{sec:exp1}\vspace*{-2mm}

Already when SMGs and their implementation in Predator were first published in
\cite{predatorSAS}, Predator had been successfully tested on a number of case
studies. Among them there were more than 256 case studies (freely available with
Predator) illustrating various programming constructs typically used when
dealing with linked lists. These case studies include various advanced kinds of
lists used in the Linux kernel and their typical manipulation, typical error
patterns that appear in code operating with Linux lists, various sorting
algorithms (insert sort, bubble sort, merge sort), etc. These case studies have
up to 300 lines of code, but they consist almost entirely of complex memory
manipulation (unlike larger programs whose big portions are often ignored by
tools verifying memory safety). Next, Predator was also successfully tested on
the driver code snippets distributed with SLAyer~\cite{slayer} as well as on the
\texttt{cdrom} driver originally checked by Space Invader \cite{InvaderCAV08}.
In some of these programs, Predator identified errors not found by the other
tools due to their more abstract (not byte-precise) treatment of memory
\cite{predatorSAS}.\footnote{Invader did not check memory manipulation via array
subscripts, and SLAyer did not check size of the blocks allocated on the heap.
The case studies of SLAyer were later updated, and so they do not contain the
identified problems any more.}

\enlargethispage{5mm}

Further, we also considered two real-life low-level programs: a memory allocator
from the Netscape portable runtime (NSPR) and a module taken from the
\texttt{lvm2} logical volume manager. The NSPR allocator allocates memory from
the operating system in blocks called \emph{arenas}, grouped into singly-linked
lists called \emph{arena pools}, which can in turn be grouped into lists of
arena pools (giving lists of lists of arenas). User requests are then satisfied
by sub-allocation within a suitable arena of a given arena pool. We consider a
fixed size of the arenas and check safety of repeated allocation and
deallocation of blocks from arena pools as well as lists of arena pools. The
blocks are of aligned size chosen randomly and ranging up to the arena size.
For this purpose, a support for offset intervals as described above is needed.
The intervals arise from abstracting lists whose nodes (arenas) point with
different offsets to themselves (one byte behind the last sub-allocated block
within the arena) and from address alignment, which the NSPR-based allocator is
also responsible for. Our approach allows us to verify that pointers leading
from each arena to its so-far free part never point beyond the arena and that
arena headers never overlap with their data areas, which are the original
assertions checked by NSPR arena pools at run-time (if compiled with the debug
support). Our \texttt{lvm2}-based case studies then exercise various functions
of the module implementing the volume metadata cache. As in the case of NSPR
arenas, we use the original (unsimplified) code of the module, but we use a
simplified test harness where the \texttt{lvm2} implementation of hash tables is
replaced by the \texttt{lvm2} implementation of doubly-linked lists. 

The original results of Predator, Invader, and SLAyer on the above described
case studies are available in \cite{predatorSAS}. In Table \ref{tbl:invpredres},
we present results on some of these case studies obtained from a wider selection
of tools, complementing the originally considered tools with more tools selected
out of those scoring well in heap-related categories of SV-COMP.\footnote{We
have also considered MemCAD 1.0.0 \cite{rival15}, but we were unable to make it
work on the chosen programs, and so we do not include it into the results.} All
experiments were run on a computer with an Intel Core~i7-3770K processor at 3.5
GHz with 32 GiB RAM.  However, due to problems with installing some of the tools
and due to some of them not being maintained any more, we had to consider
different environments for running the experiments and consider tools made
available in different years. Namely, Invader and SLAyer (marked by ``*'' in the
table) were taken in their versions from years 2008 and 2011, respectively.
Forester and CPA-kInd (marked by ``**'' in the table) were taken in their
versions from SV-COMP'17 and run in a virtual machine with Ubuntu 16.04 (which
restricted the available memory to 17.5 GB) and BenchExec 1.14 \cite{benchexec}.
The remaining tools were taken in their versions from SV-COMP'19 and run in a
virtual machine with Ubuntu 18.04 (with the available memory again restricted to
17.5 GB) and BenchExec 1.17.

All the tools were run in their default configurations. Better results can
sometimes be obtained for particular case studies by tweaking certain
configuration options (abstraction threshold, call cache size, etc.).  However,
while such changes may improve the performance in some case studies, they may
harm it in others, trigger false positives, or even prevent the analysis from
termination.


The results show that Predator provides the best results out of the considered
tools. Indeed, the other considered tools often even crash, timeout, provide
false positives, or even false negatives. Note also that PredatorHP provides
worse results than the original Predator analyser on the chosen test cases. This
is due to the stress on avoiding false positives (and not allowing the Predator
verifier to announce errors) and due to running Predator hunters in parallel
with the verifier (causing overall higher time consumption).

\begin{table*}[p]

 \begin{center}

  \caption{Experimental results on the Invader's \texttt{cdrom} test case and
  selected Predator's test cases showing either the verification time or one of
  the following outcomes: \textbf{FP}~=~false positive, \textbf{FN}~=~false
  negative, \textbf{F}~=~the expected error not found, another potential error
  reported (may be spurious: not checked), \textbf{T}~=~time out (900~s),
  \textbf{oom}~=~out of memory (15~GB), \textbf{seg}~=~segmentation fault,
  \textbf{x}~=~parsing problems, \textbf{xx}~=~internal error,
  \textbf{U}~=~inconclusive verification result (some form of ``don't know''
  explicitly produced by the tool), \textbf{?}~=~unknown failure of the tool (we
  were unable to closer diagnose the failure).}

  {\fontsize{9}{10}\selectfont
  \begin{tabular}{l|c|c|c|c|c|c|c|c|c|c|c|c|}

    & 
	\rotatebox[origin=l]{90}{\tt cdrom\_false-valid-deref.c} & 
	\rotatebox[origin=lc]{90}{\tt five-level-sll-destroyed-bottom-up.c} & 
	\rotatebox[origin=l]{90}{\tt five-level-sll-destroyed-top-down.c} & 
	\rotatebox[origin=l]{90}{\tt linux-dll-of-linux-dll.c} & 
	\multicolumn{2}{c|}{\rotatebox[origin=l]{90}{\tt list-of-arena-pools-with-alignment.c}} & 
	\rotatebox[origin=l]{90}{\tt lvmcache\_add\_orphan\_vginfo\_false-valid-memtrack.c} & 
	\multicolumn{2}{c|}{\rotatebox[origin=l]{90}{\tt merge\_sort.c}} & 
	\rotatebox[origin=l]{90}{\tt merge\_sort\_false-unreach-call.c} & 
	\rotatebox[origin=l]{90}{\tt scope-goto\_false-valid-deref.c} & 
	\rotatebox[origin=l]{90}{\tt cmp-freed-ptr\_false-unreach-call.c} \\ \hline

    Predator &
    \ 0.63 &\ 1.17 &\ 0.12 &\ 0.11 &\ 0.80 &\ 0.80 &\ 1.32 &\ 0.26 &\ 0.26 &\ 0.06 &\ 0.04 &\ 0.04 \\ \hline
    PredatorHP &
T & 4.53 &\ 0.78 &\ 0.73 & T & T & x &\ 1.02 &\ 1.01 &\ 0.32 &\ 0.23 &\ 0.23 \\ \hline
    Invader* &
FN & FP & FP & T & FP & - & x & FP & - & - & - & - \\ \hline
    SLAyer* &
x & x & x & x & x & - & x & x & - & - & - & - \\ \hline
    Forester** &
xx & xx & T & T & xx & xx & x & xx & xx & xx & xx & U \\ \hline
    CPA-Seq &
T & T & T & T & T & T & U & T & T & 75.78 & T & x \\ \hline
    CPA-kInd** &
- & - & - & - & - & T & - & - & seg & T & - & x \\ \hline
    DepthK &
oom & T & T & ? & T & T & ? & T & T &\ 1.11 & ? & ? \\ \hline
    2LS &
xx & T & T & xx & xx & xx & xx & T & T & T & x & FN \\ \hline
    ESBMC-kind &
U & oom & oom & FP & oom & oom & x & T & T &\ 0.68 & FN & FN \\ \hline
    Map2Check &
? & ? & ? & T & ? & ? & ? & T & ? & ? & T & ? \\ \hline
    UAutomizer &
xx & T & T & xx & xx & xx & x & T & T & 12.99 &\ 6.28 &\ 5.60 \\ \hline
    UKojak &
xx & T & T & xx & xx & xx & x & T & T & 14.18 &\ 6.52 &\ 5.64 \\ \hline
    UTaipan &
xx & T & T & xx & xx & xx & x & T & T & 13.55 &\ 6.13 &\ 5.72 \\ \hline
    DIVINE &
10.88 & T & T & T & U & U & x & FP & T & 11.06 & seg & FN \\ \hline
    DIVINE-explicit &
10.42 & T & T & T & FP & xx & x & U & U &\ 9.97 & U & FN \\ \hline
    Symbiotic &
F & xx & xx & T & xx & FP & x & T & T &\ 0.40 &\ 0.34 & FN \\ \hline

   \end{tabular}}


  \label{tbl:invpredres}

 \end{center}

\end{table*}

\vspace*{-2mm}\subsection{Predator and SV-COMP}\label{sec:svcomp}\vspace*{-2mm}

Predator participated in SV-COMP since its beginning. In the first to third editions
(i.e., from SV-COMP'12 to SV-COMP'14), the basic Predator analyser was involved.
Since the 4th edition (SV-COMP'15), PredatorHP (see Section~\ref{sec:hp}) was
used. Its usage very significantly reduced the number of false alarms. On the
other hand, although PredatorHP decreased the overall wall time of the
verification (and some verification tasks were speeded up even in terms of the
CPU time), the overall CPU time consumption increased. 


In Table~\ref{tbl:compare}, we present an analysis of the performance of
PredatorHP and its components, i.e., Predator hunters and the Predator verifier
on benchmarks of several heap- and memory-related categories of SV-COMP'19. To
get the data, we ran the experiments on a machine with an Intel Core~i7-3770K
processor at 3.5 GHz with 32 GiB RAM. As in the previous subsection, they were
run in a virtual machine with Ubuntu 18.04 (with the available memory restricted
to 17.5 GB) and~BenchExec~1.17.

\begin{table*}[t]

 \begin{center}

  \caption{An analysis of results of PredatorHP and its component Predators on
  SV-COMP'19 benchmarks. The meaning of the columns is as follows: T~=~correct
  true, F~=~correct false, \wrong = FN (false negative) / FP (false positive).}

  \vspace*{-2mm}

  {\fontsize{9}{10}\selectfont
  \begin{tabular}{l|rrr|rrr|rrr|rrr|rrr}

    \multirow{3}{*}{Benchmarks}     &
    \multicolumn{3}{c|}{\multirow{2}{*}{PredatorHP}} &
    \multicolumn{3}{c|}{Predator}   &
    \multicolumn{9}{c}{Predator hunters}    \\

     &
     & & &
    \multicolumn{3}{c|}{verifier}   &
    \multicolumn{3}{c|}{BFS}        &
    \multicolumn{3}{c|}{DFS 200}    &
	\multicolumn{3}{c}{DFS 900}    \\

     &
    \multicolumn{1}{c}{F} & \multicolumn{1}{c}{T} & \multicolumn{1}{c|}{\wrong} &
    \multicolumn{1}{c}{F} & \multicolumn{1}{c}{T} & \multicolumn{1}{c|}{\wrong} &
    \multicolumn{1}{c}{F} & \multicolumn{1}{c}{T} & \multicolumn{1}{c|}{\wrong} &
    \multicolumn{1}{c}{F} & \multicolumn{1}{c}{T} & \multicolumn{1}{c|}{\wrong} &
    \multicolumn{1}{c}{F} & \multicolumn{1}{c}{T} & \multicolumn{1}{c}{\wrong} \\\hline

    {\it MemSafety} & 
    ~172 & ~153 &       &
    ~134 & ~110 & 52/0  &
    ~154 & ~121 &       &
    ~124 &  ~64 &  0/1  &
    ~166 & ~148 &  0/3  \\

    {~~~~Arrays} & 
     15 &   0 &       &
      7 &   0 &       &
      7 &   0 &       &
     15 &  29 &  0/1  &
     15 &  46 &  0/1  \\

    {~~~~Heap} & 
     81 &  67 &       &
     83 &  59 & 16/0  &
     76 &  51 &       &
     55 &  29 &       &
     76 &  40 &  0/2  \\

    {~~~~LinkedLists} & 
     28 &  66 &       &
     27 &  32 & 36/0  &
     27 &  49 &       &
     16 &   3 &       &
     28 &  48 &       \\

    {~~~~Other} & 
     16 &  20 &       &
     17 &  19 &       &
     16 &  21 &       &
     12 &   3 &       &
     15 &  14 &       \\

    {~~~~MemCleanup} & 
     32 &   - &       &
      0 &   - &       &
     28 &   - &       &
     26 &   - &       &
     32 &   - &       \\
	 
    {\it ReachSafety} & 
      & & &
      & & &
      & & &
      & & &
      & & \\
	 
    {~~~~Heap} & 
     71 & 129 &  ~4/0  &
     70 &  93 & ~43/0  &
     67 & 108 &  ~4/0  &
     54 &  56 &  ~1/0  &
     71 & 108 &  ~1/0  \\


   \end{tabular}}

   \vspace*{-4mm}

  \label{tbl:compare}

 \end{center}

\end{table*}

The use of PredatorHP allowed us to avoid 52 false alarms generated by Predator
Verifier in the MemSafety category: Under PredatorHP, 40 of these benchmarks are
successfully verified, 10 benchmarks end by a timeout, and the expected error is
not reported in 2 benchmarks (instead a false alarm about another issue is
raised). Likewise, in the ReachSafety category, 39 false alarms are avoided: 34
of the concerned benchmarks are successfully verified, 5 benchmarks end with a
timeout (and there still remain 4 false alarms due to imprecise treatment of
integers).

\enlargethispage{5mm}

Moreover, the BFS hunter managed to verify various benchmarks with small finite
state spaces, some of them with arithmetic operations on data fields of nodes of
bounded-length lists (e.g., \verb|list-simple/dll2*|, \verb|list-simple/sll2*|,
\verb|ldv-memsafety/memleaks_test23_{1,3}_true-valid-memsafety.i|, or also
\verb|heap-data/shared_mem*|). In their case, the list abstraction is not
needed, and, by not using it, we avoid interval abstraction on data fields,
which causes imprecision and makes the Predator verifier to announce a false
alarm (which PredatorHP ignores).
%
%
In those cases where one needs to track data fields of unbounded lists, a
timeout is hit since the verifier produces false alarms and hunters run forever
(this happens, e.g., in \verb|heap-data/process_queue_true-unreach-call.c|,
\verb|heap-data/min_max_true-unreach-call.c|,
\texttt{list-ext-properties/list-\linebreak
ext\_flag\_1\_true-valid-memsafety.c}). Another cause of timeouts is then
missing abstraction for non-list data structures (such as, e.g., trees in the
benchmarks of \verb|memsafety-ext/tree*|), which prevents both the hunters and
the verifier from terminating.


As for the time consumption, the original Predator used 25,200 seconds of
CPU/wall time (that are equal in this case) to handle all benchmarks considered
in Table~\ref{tbl:compare} in the MemSafety category.  Further, it needed 14,700
seconds in the ReachSafety category. On the other hand, PredatorHP needed 17,100
seconds of wall time and 36,800 seconds of CPU time for the MemSafety category
and 9,850 seconds of wall time and 20,000 seconds of CPU time for the
ReachSafety category. This shows what we mentioned already at the beginning of
the section, i.e., the fact that PredatorHP decreased the wall time but
increased the CPU time.


However, even in terms of the CPU time, PredatorHP was faster in 6 correct and 6
erroneous benchmarks of the MemSafety category (with the correct cases and 3 of
the erroneous cases handled by the BFS hunter, and with the 3 remaining
erroneous cases handled by DFS hunters). In the ReachSafety category, PredatorHP
was faster in 1 correct and 1 erroneous benchmark (both handled by the BFS
hunter).
The reason for that is that the list abstraction introduces some overhead that
is not necessary in some cases: in particular, benchmarks on locks
(\texttt{locks/test\_locks\_*}), and benchmarks with arrays
(\texttt{ldv-regression/test23\_\{true,false\}-unreach-call.c}), and
%
%
benchmarks on lists with bounded-length and data fields
(\texttt{list-ext-properties/} \texttt{simple-ext\_1\_true-valid-memsafety.c},
\texttt{list-ext2-properties/simple\_} \texttt{and\_skiplist\_}
\texttt{2lvl\_false-unreach-call.c}).

\bigskip

Next, we briefly discuss influence of the pragmatic heuristics for dealing with
intervals of values discussed in Section~\ref{sec:int-size-reg}. The first of
them replaces dealing with an interval of a bounded size by performing the
verification independently for each element of the interval. This approach
resolved unknown results of the basic Predator in the following two cases:
\texttt{list-ext3-properties/sll\_nondet\_insert\_}
\texttt{true-unreach-call\_true-valid-memsafety.c}, which inserts a node at a
specific index given by an interval into a list of an unknown but finite length
(namely, two to five elements), and
\texttt{ldv-regression/test24\_true-unreach-call\_true-} \texttt{termination.c},
which indexes an array by an interval.
Apart from that, due to the heuristic sampling of unbounded intervals, DFS
hunters found errors in 9 test cases (\verb|array-memsafety/*|) when looking for
bugs in interval-sized memory regions allocated by \verb|alloca| .


\bigskip

\enlargethispage{6mm}

The efficiency of SMGs together with all the optimisations allowed Predator to
win 7 gold medals, 5 silver medals, and 1 bronze medal at SV-COMP'12--19. In
2018 and 2019, it did not win any gold medal, which was caused to a large degree
by that SV-COMP merged benchmarks targeting at programs with arrays with those
focusing on pointers, dynamic memory, and dynamic linked data structures.
However, even in SV-COMP'19, Predator was the first in the MemSafety-Heap and
MemSafety-LinkedLists subcategory. 

For SV-COMP'20, Predator was further improved in several relatively minor ways
(e.g., its SMG-based analysis has been extended to support \emph{memory
reallocation} on the heap, Predator's handling of intervals has been fine-tuned,
etc.)---for more details, see \cite{svcomp20-predator}. This allowed Predator to
once again win a gold medal in the MemSafety category. Moreover, in SV-COMP'20,
Predator has been integrated as an auxiliary tool of Symbiotic where it either
helps it to prove some programs correct (if it manages to do so quickly enough),
or information contained in its bug reports is combined with results of static
pointer analysis implemented in Symbiotic to get a more precise (i.e. smaller)
set of potentially misbehaving instructions on which Symbiotic subsequently
concentrates its further analysis based on symbolic execution
\cite{svcomp20-symbiotic}.

Predator did not officially participate in SV-COMP'21 due to insufficient
manpower for keeping it up-to-date with various changes in the competition's
rules, formats, and with various specific features of new verification tasks.
However, it participated ``hors concours'' with absolutely no change wrt the
previous year, and the results indicate that it would still be capable of
scoring quite favourably in some categories or sub-categories. Moreover, the
integration of Predator with Symbiotic was improved, and Predator appeared in
SV-COMP'21 as a part of Symbiotic too.

\vspace*{-4mm}\section{Related Work}\label{sec:related}\vspace*{-2mm}

Many approaches to formal analysis and verification of programs with dynamic
linked data structures have been proposed. They differ in their generality,
level of automation, as well as the formalism on which they are based. As we
said already in the introduction of the chapter, SMGs and the shape analysis
based on them are inspired by the fully-automated approaches based on
\emph{separation logic} with higher-order list predicates implemented in the
Space Invader \cite{InvaderCAV07,InvaderCAV08} and SLAyer~\cite{slayer} tools.
Compared with them, however, we use a purely graph-based memory
representation.\footnote{In fact, a graph-based representation was used already
in the first version of Predator \cite{predator}. However, that representation
was a rather straightforward graph-based encoding of separation logic formulae,
which is not the case anymore for the representation described here.} Our heap
representation is finer, which---on one hand---complicates its formalization but
allows for treating the different peculiarities of low-level memory manipulation
on the other hand. Moreover, somewhat surprisingly, although our heap
representation is rather detailed, it still allowed us to propose algorithms for
all the needed operations such that they are quite efficient.\footnote{Indeed,
the version of Predator based on SMGs as presented here turned out to be much
faster than the first one of \cite{predator} while at the same time producing
fewer false positives.}

\enlargethispage{5mm}

Compared with Space Invader and SLAyer, Predator is not only faster, but also
terminates more often, avoids false positives and, in particular, is able to
detect more classes of program errors (as illustrated in the section on
experiments). Both Space Invader and SLAyer provide some support for pointer
arithmetic, but their systematic description is---to the best of our
knowledge---not available, and, moreover, the support seems to be rather basic
as illustrated by our experimental results. A~support for pointer arithmetic in
combination with separation logic appears in \cite{calcagno06} too, which is,
however, highly specialised for a~particular kind of linked lists with variable
length entries used in some memory allocators.


As for the memory model, probably the closest to the notion of SMGs is
\cite{Laviron2010}, which uses the so-called \emph{separating shape graphs}.
They support tracking of the size of allocated memory areas, pointers with
byte-precise offsets wrt addresses of memory regions, dealing with offset
ranges, as well as multiple views on the same memory contents. A major
difference is that \cite{Laviron2010} and the older work \cite{rival07}, on
which \cite{Laviron2010} is based, use so-called summary edges annotated by
\emph{user-supplied} data structure invariants to summarize parts of heaps of an
unbounded size. This approach is more general in terms of the supported shapes
of data structures but less automated because the burden of describing the shape
lies on the user. We use abstract objects (list segments) instead, which are
capable of encoding various forms of hierarchically nested-lists (very often
used in practice) and are carefully designed to allow for \emph{fully-automated}
and \emph{efficient} learning of the concrete forms of such lists (the concrete
fields used, the way the lists are hierarchically-nested, their possible
cyclicity, possibly shared nodes, optional nodes, etc.). Also, the level of
nesting is not fixed in advance---our list segments are labelled by an integer
nesting level, which allows us to represent hierarchically-nested data
structures as flattened graphs. Finally, although \cite{Laviron2010} points out
a need to reinterpret the memory contents upon reading/writing, the
corresponding operations are not formalized there. 
%

A graph-based abstraction of sets of heap configurations is used in
\cite{marron08} too. On one hand, the representation allows one to deal even
with tree-like data structures, but, on the other hand, the case of
doubly-linked lists is not considered. Further, the representation does not
consider the low-level memory features covered by SMGs. Finally, the abstraction
and join operations used in \cite{marron08} are more aggressive and hence less
precise than in our case.

%


The work \cite{Kreiker2010}, which is based on an instantiation of the
\emph{TVLA framework} \cite{tvla02}, focuses on analysis of Linux-style lists,
but their approach relies on an implementation-dependent way of accessing list
nodes, instead of supporting pointer arithmetics, unions, and type-casts in a
generic way. Finally, the work \cite{Tuch2009} provides a detailed treatment of
low-level C features such as alignment, byte-order, padding, or type-unsafe
casts in~the context of theorem proving based on separation logic. Our
reinterpretation operators provide a lightweight treatment of these features
designed to be used in the context of a~fully-automated analysis based on
abstraction.


Another tool that can handle some features of pointer arithmetic is Forester
\cite{forester}. It is based on hierarchically-nested \emph{forest automata},
i.e., tuples of interconnected tree automata, and the approach of abstract
regular tree model checking \cite{artmc}. Forester can handle
fully-automatically more general classes of dynamic linked data structures than
Predator (trees, trees with additional pointers, skip lists\footnote{Adding a
support of such non-list dynamic data structures to Predator is non-trivial. For
that, new kinds of heap segments would have to be added, together with
algorithms for all the needed operations.  In fact, an attempt to add \emph{tree
segments} was once done. It was realised that one would need tree segments with
and without ``holes'' on the leaf level, possibly even a variable number of
them, through which the tree segments would link to the rest of the heap.
However, algorithms that would be capable of handling such segments, combine
them reasonably with list segments (one needs to handle appropriately questions
such as whether a tree degenerated to the left-most branch is a list or a tree),
and prevent state space explosion stemming from introducing a number of
different kinds of tree segments have never been finished.}), has a more
flexible abstraction (which can adjust to various non-standard shapes of data
structures), and a support for dealing with ordered data in one of its versions
\cite{bengt16} as well as a support for finite data together with a
counterexample-guided abstraction refinement loop in another version
\cite{panda17}. However, as our experimental results in Section~\ref{sec:exp1}
show, Forester's support of low-level features is much more limited and it is
often less efficient too (though not always as shown in \cite{forester}).

Interestingly, there have appeared two tools that attempt to reimplement SMGs in
the context of \emph{configurable program analysis} (CPA) \cite{cpa07}: namely,
CPAlien \cite{cpalien14} and CPAchecker \cite{cpaSMG14}. CPAlien was an
experimental tool with a partial implementation of SMGs only: in particular,
abstraction was missing. The support of SMGs in CPAchecker is---as far as we
know---more complete, but so far it is also used without abstraction for
efficiency reasons (at least that was the case up to SV-COMP'20). Without
abstraction, however, CPAchecker cannot successfully verify correct programs in
case dealing with unbounded lists is needed for the verification.

\enlargethispage{5mm}

The above problem of not being able to soundly verify programs whose
verification requires dealing with unbounded dynamic data structures is much
more common among the tools that participated in heap- and memory-related
(sub-)categories of the different editions of SV-COMP up to SV-COMP'20. This
problem manifests, e.g., in Symbiotic \cite{symbiotic4} or Ultimate Automizer
\cite{uautomizer}. In particular, in order not to sacrifice soundness, the
analysis implemented in these tools cannot successfully terminate on such
programs (while still other tools perform bounded analysis only and produce in
principle unsound answers). A tool that participates in SV-COMP and that can
handle verification on unbounded list structures in a sound way is 2LS
\cite{2ls18}. This tool is based on a combination of \emph{template-based
invariants}, $k$-induction, abstract domains (for representing suitable
parameters of the template-based invariants), and SAT solving. 2LS can handle
unbounded list structures and can even reason (to some degree) about data stored
in them \cite{fmcad18}. However, 2LS currently has no support for pointer
arithmetic.

Finally, it is worth mentioning that the above mentioned works on separation
logic and Space Invader later led to the so-called \emph{bi-abductive analysis}
of programs with dynamic linked data structures \cite{infer11}. This approach
was implemented in a tool called Infer that concentrated on (nested) dynamic
linked lists---despite that the approach of \cite{infer11} itself is more
general. An advantage of the approach is that it can handle \emph{open code
fragments} (i.e., there is no need to model the environment of the code fragment
under verification) and it can perform the analysis \emph{modularly}, analysing
functions along the call tree, starting from the leaves (which is quite scalable
though it may involve some loss of precision). A generalisation of the approach
appeared in relation with the S2 tool \cite{s2:15}. This tool can handle
programs with very complex data structures (e.g., trees with linked leaves). On
the other hand, the approach is rather fragile in that it relies on the program
to handle data structures in a way that is well aligned with their inductive
definitions and sometimes it fails even on rather simple programs. Moreover,
despite Infer contained some support of pointer arithmetic, the support of
low-level pointer features in bi-abductive analyses is quite limited and remains
an open problem for the future.

Indeed, for the future, it would be very useful to have an abductive analyser
supporting truly low-level features of memory manipulation and at the same time
capable of analysing code fragments (since analysis of such fragments is
probably the most welcome in real life according to our experience). As for the
low-level features, one can, of course, go even further than Predator does:
e.g., sometimes even bit-precision is needed---for instance, when some bits of
pointers are used to store non-pointer information by bit-masking (which is used
sometimes, e.g., to store the colour of nodes in red-black trees).




\paragraph*{Acknowledgement.} The work was supported by the Czech Science
Foundation project 20-07487S and the FIT BUT project FIT-S-23-8151.




\vspace*{-3mm}\section*{Appendix}

\enlargethispage{5mm}

\vspace*{-3mm}\section{Data Reinterpretation of Nullified Blocks}
\label{app:reinterpret} \vspace*{-2mm}

In this appendix, we present a detailed description of the algorithms for read
and write reinterpretation of nullified blocks, which were briefly introduced in
Section~\ref{sec:reinterpretation}. Given an SMG $G = (O, V, \Lambda, H, P)$, we
define $H_{ov}(o, \of, \ty)$ as the set of all has-value edges leading from
$o$~whose fields overlap with the field $(\of, \ty)$, i.e.: $$H_{ov}(o, \of,
\ty) = \{ (o \hasvalue{\ofprime\backspace,\ty'} v) \in H \mid I(\of, \ty) \;
\cap \; I(\ofprime\backspace,\ty') \; \ne \; \emptyset \}.$$ Further, we define
$H_{zr}(o, \of, \ty)$ as the subset of $H_{ov}(o, \of, \ty)$ containing all its
edges leading to $0$, i.e.:\vspace*{-2mm} $$H_{zr}(o, \of, \ty) = \{ (o
\hasvalue{\ofprime\backspace,\ty'} 0) \in H_{ov}(o, \of, \ty) \}.$$

\vspace*{-8mm}\subsection{Read Reinterpretation of Nullified Blocks}
\label{app:rdReinterp} \vspace*{-2mm}

Algorithm~\ref{alg:readValue} gives the algorithm of read reinterpretation
instantiated for dealing with nullified blocks of memory as precisely as
possible.\vspace*{-2mm}

\begin{algorithm}[H]

  \caption{\small$\mathit{readValue}(G, o, \of, \ty)$}

  \label{alg:readValue}

  \smallskip

  \textbf{Input:}\begin{itemize}[nosep]

    \item An SMG $G = (O,V,\Lambda,H,P)$.

    \item An object $o \in O$.

    \item A field $(\of, \ty)$ within $o$, i.e., $\of + \size(\ty) \le \size(o)$.

  \end{itemize}

  \textbf{Output:}\begin{itemize}[nosep]

    \item A tuple $(G', v)$ that is the result of read
    reinterpretation of $G$ wrt the object $o$ and the field $(\of,\ty)$ such that
    fields representing nullified memory are read as precisely as the notion of
    SMGs allows (i.e., the operator recognises that the field to be read is
    nullified iff the input SMG guarantees that each byte of the field is indeed
    zero).

  \end{itemize}

  \textbf{Method:}\begin{enumerate}[nosep]

    \item Let $v := H(o, \of, \ty)$.

    \item If $v \ne \UNDEF$, return $(G,v)$.

    \item If the field to be read is covered by nullified blocks, i.e., if
    $\forall \of \le i < \of + \size(\ty) ~\exists e \in H_{zr}(o, \of, \ty):~ i \in
    I(e)$, let $v := 0$. Otherwise, extend $V$ by a fresh value node $v$.

    \item Extend $H$ by the has-value edge $o \hasvalue{\of, \ty} v$ and return
    $(G,v)$ based on the obtained SMG $G$.

  \end{enumerate}

\end{algorithm}

\enlargethispage{5mm}

\vspace*{-8mm}\subsection{Write Reinterpretation of Nullified Blocks}
\label{app:wrReinterp} \vspace*{-2mm}

\begin{figure}[h]
  \resizebox{\hsize}{!}{
    \includegraphics{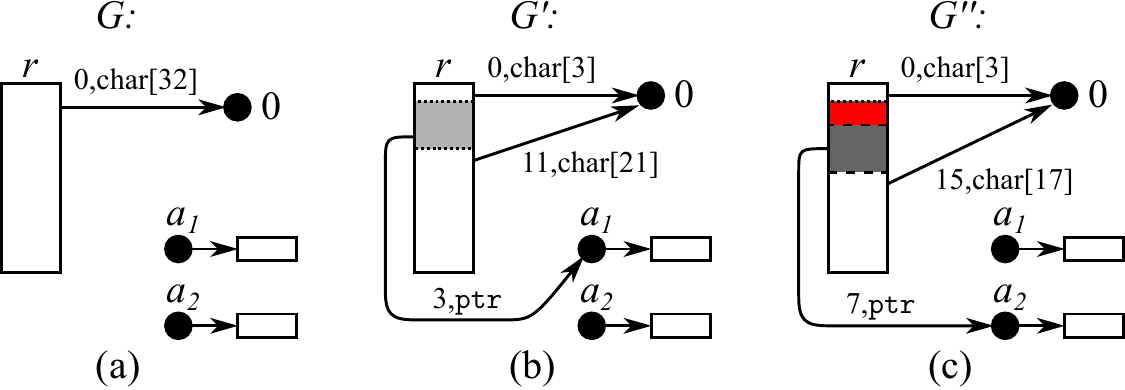}
  }

  \caption{An illustration of write reinterpretation: (a) an initial SMG $G$,
  (b) the SMG $G'$ obtained by $\mathit{writeValue}(G,r,3,ptr,a_1)$, and (c) the
  SMG $G''$ obtained by $\mathit{writeValue}(G',r,7,ptr,a_2)$. Note that $G''$
  contains an undefined value in a field of size of 4 bytes at offset 3.}
  \label{fig:reinterpretation}

\end{figure}

Algorithm~\ref{alg:writeValue} gives the algorithm of write reinterpretation
instantiated for dealing with nullified blocks of memory as precisely as
possible. An illustration of how the algorithm works can be found in
Fig.~\ref{fig:reinterpretation}.

\begin{algorithm}[t]

  \caption{\small$\mathit{writeValue}(G, o, \of, \ty, v)$}

  \label{alg:writeValue}

  \smallskip

  \textbf{Input:}\begin{itemize}[nosep] 

    \item An SMG $G = (O,V,\Lambda,H,P)$.

    \item An object $o \in O$.

    \item A field $(\of, \ty)$ within $o$, i.e., $\of + \size(\ty) \le \size(o)$.

    \item A value $v$ such that $v \not\in O$ (needed so that $v$ can be safely
    added into $V$).

  \end{itemize}

  \textbf{Output:}
  \begin{itemize}[nosep]

    \item An SMG $G'$ that is the result of write
    reinterpretation of $G$ wrt the object $o$, the field $(\of,\ty)$, and the
    value $v$ such that as much information on nullified memory as is allowed by
    the notion of SMGs is preserved (i.e., each byte that is nullified in the
    input SMG will stay nullified in the output SMG unless it is overwritten by
    a possibly non-zero value).

  \end{itemize}

  \textbf{Method:}\begin{enumerate}[nosep]

    \item If $H(o,\of,\ty) = v$, return $G$.

    \item Let $V := V \cup \{ v \}$.

    \item Remove from $H$ all edges leading from $o$ to non-zero values whose
    fields overlap with the given field, i.e., the edges in $H_{ov}(o, \of, \ty)
    \setminus H_{zr}(o, \of, \ty)$.

    \item If $v \neq 0$, then for each edge $(e_z: o \hasvalue{\of_z,\ty_z} 0)
    \in H_{zr}(o, \of, \ty)$ do:\begin{enumerate}[nosep]

      \item Remove the edge $e_z$ from $H$.

      \item Let $\ofprime := \of + \size(\ty)$ and $\ofprime_z := \of_z + 
      \size(\ty_z)$.

      \item If $\of_z < \of$, extend $H$ by the edge $o \hasvalue{\of_z,
      \mathtt{char[}\of - \of_z\mathtt{]}} 0$.

      \item If $\ofprime < \ofprime_z$, extend $H$ by the edge
        $o \hasvalue{\ofprime\backspace, \mathtt{char[}\ofprime_z - 
        \ofprime\mathtt{]}} 0$.

    \end{enumerate}

    \item Extend $H$ by the has-value edge $o \hasvalue{\of, \ty} v$ and return
    the obtained SMG.

  \end{enumerate}

\end{algorithm}

\vspace*{-4mm} \section{The Join Algorithms} \label{app:join} \vspace*{-2mm}

This appendix provides a detailed description of the join algorithms introduced
in~Section~\ref{sec:join}. We first describe the $\mathit{joinSubSMGs}$
function, which implements the core functionality on top of which both joining
garbage-free SPCs (to reduce the number of SPCs obtained from different paths
through the program) as well as merging a~pair of neighbouring objects of
a~doubly-linked list into a single DLS within abstraction are built.
Subsequently, we describe the functions $\mathit{joinValues}$,
$\mathit{joinTargetObjects}$, and
$\mathit{insertLeftDlsAndJoin}\,/\,\mathit{insertRightDlsAndJoin}$ on which
$\mathit{joinSubSMGs}$ is based. In fact, $\mathit{joinSubSMGs}$ calls
$\mathit{joinValues}$ on pairs of corresponding values that appear below the
roots of the sub-SMGs to be joined, $\mathit{joinValues}$ then calls
$\mathit{joinTargetObjects}$ on pairs of objects that are the target of value
nodes representing addresses, and the $\mathit{joinTargetObjects}$ algorithm
recursively calls $\mathit{joinSubSMGs}$ to join the sub-SMGs of the objects to
be joined. The $\mathit{insertLeft}$($\mathit{Right}$)$\mathit{DlsAndJoin}$
functions are called from $\mathit{joinValues}$ when the given pair of addresses
cannot be joined since their target objects are incompatible, and an attempt to
save the join from failing is done by trying to compensate a~DLS missing in one
of the SMGs by inserting it with the minimum length being 0 (which is possible
since a 0+ DLS is a possibly empty list segment). Finally, we describe the
$\mathit{joinSPCs}$ and $\mathit{mergeSubSMGs}$ functions implemented on top of
the generic $\mathit{joinSubSMGs}$ function. The $\mathit{joinSPCs}$ function
joins garbage-free SPCs into a single SPC that semantically covers both.  The
$\mathit{mergeSubSMGs}$ function merges a pair of objects during abstraction
into a single DLS while the non-shared part of the sub-SMGs rooted at them is
joined into the nested sub-SMG of the resulting DLS.  


\begin{wraptable}[6]{r}{25mm}
  \begin{center}
  \vspace*{-9mm}
  \begin{tabular}{|c|cccc|}
     \hline
	$s_1\backslash s_2$  & \miso & \mjsupset & \mjsubset & \mjoin\\\hline
     \miso     & \miso     & \mjsupset & \mjsubset & \mjoin\\
     \mjsupset & \mjsupset & \mjsupset & \mjoin    & \mjoin\\
     \mjsubset & \mjsubset & \mjoin    & \mjsubset & \mjoin\\
     \mjoin    & \mjoin    & \mjoin    & \mjoin    & \mjoin\\\hline
  \end{tabular}
  \label{tbl:updateJoinStatus}

  \end{center}
\end{wraptable}

As mentioned in~Section~\ref{sec:join}, the join algorithm computes on the fly
the so-called \emph{join status} which compares the semantics of the SMGs being
joined (with the semantics being either equal, in an entailment relation, or
incomparable). For the purpose of maintaining the join status, the table shown
on the right defines the function $\mathit{updateJoinStatus}: \jstat \times
\jstat \longrightarrow \jstat$ that combines the current join status $s_1$,
obtained from joining the so-far explored parts of the SMGs being joined, with a
status $s_2 \in \jstat$ comparing the semantics of the objects/values being
currently joined. Note that the function is monotone in that once the status,
which is initially \miso, becomes \mjsupset~ or \mjsubset, it can never get back
to~\miso, and once the status becomes \mjoin, it cannot change any more.

In case of the $\mathit{joinSPCs}$ function, $\mathit{joinSubSMGs}$ needs to be
called multiple times for a~single pair of SPCs (starting from different program
variables), and it is necessary to keep certain state information between the
calls. Besides the join status mentioned above, the algorithm maintains a
mapping of values and objects between the source SMGs and the destination SMG.
This is needed in order to identify potentially conflicting mappings arising
when starting the join from different program variables as well as to identify
parts of SMGs that have already been processed. The mapping is encoded as a pair
of partial injective functions $m_1: (O_1 \rightharpoonup O) \cup (V_1
\rightharpoonup V)$ and $m_2: (O_2 \rightharpoonup O) \cup (V_2 \rightharpoonup
V)$. Additionally, we assume that the $\nil$ object and the $0$ address, which
have their pre-defined unique roles in all SMGs, never appear in the ranges of
$m_1, m_2$. In case of the $\mathit{mergeSubSMGs}$ function, the mapping of
objects and values is used to obtain the sets of nodes recognized as nested data
structures.

In the following, we write $\kind_1$, $\size_1$, $\level_1$, $\len_1$,
$\valid_1$, $\nfoBS_1$, $\pfoBS_1$, and $\hfoBS_1$ to denote $\kind$, $\size$,
$\level$, $\len$, $\valid$, $\nfo$, $\pfo$, and $\hfo$ from $\Lambda_1$, i.e.,
the labelling function of the first SMG being joined. Likewise for $\Lambda_2$.
We further define $\len'$ as a wrapper function of $\len$ such that $\len'(o) =
\len(o)$ if $\kind(o) = \dls$, and $\len'(o) = 1$ if $\kind(o) = \reg$.


\enlargethispage{5mm}

\vspace*{-3mm}\subsection{Join Reinterpretation}\label{app:joinReinterp}
\vspace*{-2mm}

The read and write reinterpretations described in Section
\ref{sec:reinterpretation} operate on a single object of a~single SMG. However,
when joining a pair of SMGs, we need to compare pairs of their objects, figure
out what they have semantically in common, and modify their sets of fields such
that they become the same (even if for the price of loosing some information),
allowing one to subsequently attempt to join their corresponding sub-SMGs. For
that purpose, we introduce \emph{join reinterpretation}.

A~join reinterpretation operator inputs a pair of SMGs $G_1$ and $G_2$, whose
sets of objects are $O_1$ and $O_2$, respectively, and a pair of objects $o_1
\in O_1$ and $o_2 \in O_2$ such that $\size_1(o_1)=\size_2(o_2)$. The operator
returns a~triple $(s,G'_1,G'_2)$ where $s \in \jstat$ is a join status, and
$G_1'$, $G_2'$ are two SMGs with sets of \mbox{has-value edges $H'_1$, $H'_2$,
respectively,}

\begin{wraptable}[6]{r}{57mm}
  \begin{center}
    \vspace*{-9mm}
\begin{tabular}{|c|c|c|}\hline

  $s$        & semantics of $G'_1$        & semantics of $G'_2$\\

  \hline

  $\iso$     & $\MI(G_1) =       \MI(G'_1)$ & $\MI(G_2) =       \MI(G'_2)$ \\

  $\jsupset$ & $\MI(G_1) =       \MI(G'_1)$ & $\MI(G_2) \subset \MI(G'_2)$ \\

  $\jsubset$ & $\MI(G_1) \subset \MI(G'_1)$ & $\MI(G_2) =       \MI(G'_2)$ \\

  $\join$    & $\MI(G_1) \subset \MI(G'_1)$ & $\MI(G_2) \subset \MI(G'_2)$ \\

  \hline

\end{tabular}
  \end{center}
\end{wraptable}


\noindent such that: (1)~The sets of fields of $o_1$ and $o_2$ are the same,
i.e., $\forall \of \in \nat ~\forall \ty \in \type:~ H'_1(o_1,\of,\ty) \neq \bot
\Leftrightarrow H'_2(o_2,\of,\ty) \neq \bot$. (2)~The status $s$ and the
semantics of $G'_1$ and $G'_2$ are defined according to the table shown on the
right. Intuitively, the status $\jsupset$ means that some aspects of $G_1$ are
less restrictive than those of $G_2$, implying a~need to lift these restrictions
in $G_2$ to keep chance for its successful join with $G_1$, which enlarges the
semantics of $G_2$ while that of $G_1$ stays the same. Likewise for the other
symbols of $\jstat$.

\begin{algorithm}[b]

  \caption{\small$joinFields(G_1, G_2, o_1, o_2)$}

  \label{alg:joinFields}

  \smallskip

  \textbf{Input:}\begin{itemize}[nosep]

    \item SMGs $G_1 = (O_1,V_1,\Lambda_1,H_1,P_1)$ and
    $G_2=(O_2,V_2,\Lambda_2,$ $H_2,P_2)$ with sets of addresses $A_1$ and $A_2$,
    respectively.

    \item Objects $o_1 \in O_1$ and $o_2 \in O_2$, s.t. $\size_1(o_1) =
    \size_2(o_2)$.

  \end{itemize}

  \textbf{Output:}\begin{itemize}[nosep]

    \item A tuple $(s', G_1', G_2')$ consisting of a join status and two SMGs
    that is the result of join reinterpretation of $G_1$ and $G_2$ wrt $o_1$
    and $o_2$ which reorganizes the nullified fields of $o_1$ and $o_2$ with the
    aim of obtaining the smallest possible number of nullified fields such that
    either (a)~each byte of such fields is nullified in both $G_1$ and $G_2$, or
    (b)~the field is nullified in one of them, and in the other, it contains a
    non-null address.

  \end{itemize}

  \textbf{Method:}\begin{enumerate}[nosep]

    \item Let $H_1' := H_1$, ~$H_2' := H_2$.

    \item Process the set $H_{1,0} = \{ o_1 \hasvalue{\of, \ty} 0 \in H_1' \}$
      of edges leading from $o_1$ to $0$ in $G_1$ as follows:

    \begin{enumerate}[nosep]
      \item Remove the edges that are in $H_{1,0}$ from $H_1'$.

      \item Extend $H_1'$ by the smallest set of edges $H'_{1,0}$ in which for
        each $0 \leq i < \size_1(o_1)$ there is an edge $o_1
        \hasvalue{\ofprime\backspace,\ty'} 0 \in H'_{1,0}$ such that 
        $i \in I(\ofprime\backspace, \ty')$ where
        $\ty' = \mathtt{char[}n\mathtt{]}$ for some $n > 0$ iff $\exists (o_1
        \hasvalue{\of_1, \ty_1} 0) \in H_{1,0}~ \exists (o_2
        \hasvalue{\of_2,\ty_2} 0)\in H_2':~ i \in I(\of_1,\ty_1) \cap I(\of_2,
        \ty_2)$.

      \item For each address $a_2 \in A_2 \setminus \{ 0 \}$ and each edge $(o_2
        \hasvalue{\of,\ptr} a_2) \in H_2'$ for which there is no $a_1 \in A_1
        \setminus \{ 0 \}$ such that $(o_1 \hasvalue{\of,\ptr} a_1) \in H_1'$,
        but $I(\of,\ptr) \subseteq \bigcup_{e \in H_{1,0}} I(e)$, extend $H_1'$
        by the edge $o_1 \hasvalue{\of,\ptr} 0$.

    \end{enumerate}

    Then do the same for $o_2$ with swapped sets of edges $H_1'$ and $H_2'$,
    using $A_1$ instead of $A_2$, and $H_{2,0}$ and $H'_{2,0}$ instead of
    $H_{1,0}$ and $H'_{1,0}$, respectively.

    \item Let $s := \;\iso$.

    \item For each $0 \le i < \size_1(o_1)$:
      \begin{itemize}[nosep]
        \item If $\exists (e:\; o_1 \hasvalue{\of,\ty} 0) \in H_1$ such that $i
          \in I(e)$ and $\forall (e': o_1 \hasvalue{\ofprime\backspace,\ty'} 0) 
          \in H_1': i \notin I(e')$, let $s := 
          \mathit{updateJoinStatus}(s, \jsubset)$.

        \item If $\exists (e:\; o_2 \hasvalue{\of,\ty} 0) \in H_2$ such that $i
          \in I(e)$ and $\forall (e': o_2 \hasvalue{\ofprime\backspace,\ty'} 0) 
          \in H_2': i \notin I(e')$, let $s := \mathit{updateJoinStatus}(s, 
          \jsupset)$.

      \end{itemize}

    \item For all fields $(\of,\ty)$ such that $H_1'(o_1,\of,\ty) \neq \bot 
      ~\wedge~ H_2'(o_2,\of,\ty) = \bot$, extend $H_2'$ such that 
      $H_2'(o_2,\of,\ty) = v$ for some fresh $v$ added into $V_2$. Proceed 
      likewise for non-nullified fields of $o_2$ not defined in $o_1$.

    \item Return $(s, G_1, G_2)$ where $G_1 = (O_1, V_1, \Lambda_1, H_1', P_1)$
      and $G_2 = (O_2, V_2, \Lambda_2, H_2', P_2)$.

\end{enumerate}

\end{algorithm}


For the particular case of dealing with nullified memory, we implement the join
reinterpretation as follows (cf. Algorithm~\ref{alg:joinFields}). First,
nullified fields are shortened, split, and/or composed in each of the objects
with the aim of obtaining the smallest possible number of nullified fields such
that either (1)~each byte of such fields is nullified in both original SMGs, or
(2)~the field is nullified in one SMG, and, in the other, it contains a non-null
address. The former is motivated by preserving as much information about
nullified memory as possible when joining two objects. The latter is motivated
by the fact that a null pointer may be interpreted as a special case of
a~null-terminated 0+ DLS and hence possibly joined with an address in the other
SMG if its target is a DLS. Finally, whenever a field $(\of,\ty)$ remains
defined in $o_1$ but not in $o_2$ after the described transformations, i.e., if
$H_1(o_1,\of,\ty) \neq \bot$ and $H_2(o_2,\of,\ty) = \bot$, $H_2$ is extended
such that $H_2(o_2,\of,\ty) = v'$ for some fresh $v'$ (and likewise the other
way around). Note that the join status is not updated since it will be updated
later on when joining the appropriate values.

\vspace*{-3mm}\subsection{Join of Sub-SMGs} \label{app:joinSubSMGs}
\vspace*{-2mm}

The $\mathit{joinSubSMGs}$ function (cf.~Alg.~\ref{alg:joinSubSMGs}) is
responsible for joining a pair of sub-SMGs rooted at a~given pair of objects and
for constructing the resulting sub-SMG within the given destination SMG. The
function inputs a triple of SMGs $G_1$, $G_2$, $G$ (two source SMGs and one
destination SMG) and a triple of equally-sized objects $o_1$, $o_2$, $o$ from
the SMGs $G_1$, $G_2$, $G$, respectively. If the $\mathit{joinSubSMGs}$ function
fails in joining the given sub-SMGs, it returns $\bot$. Otherwise, it returns a
triple of SMGs $G_1'$, $G_2'$, $G'$ such that:\begin{itemize} 

  \item $\MI(G_1) \subseteq \MI(G_1')$ and $\MI(G_2) \subseteq \MI(G_2')$ where
  $G_1$ and $G_2$ can differ from $G'_1$ and $G'_2$, respectively, due to an
  application of join reinterpretation on some of the pairs of objects being
  joined only.

  \item The sub-SMGs $G_1''$ and $G_2''$ of $G_1'$ and $G_2'$ rooted at $o_1$
  and $o_2$, respectively, are joined into the sub-SMG $G''$ of $G'$ rooted at
  $o$, i.e., it is required that $\MI(G_1'') \subseteq \MI(G'') \supseteq
  \MI(G_2'')$.

  \item The sub-SMG $G' \setminus G''$ is exactly the sub-SMG of $G$ that
  consists of objects and values that are not removed in
  Step~\ref{removalForDelayedJoin} of the $\mathit{joinTargetObjects}$ function
  due to using the principle of delayed join of sub-SMGs described in
  Section~\ref{app:DelayedJoin} (which will take care of the fact that some
  sub-SMG may be reachable along some access path in one of the SMGs $G_1$/$G_2$
  only, due to some optional nested sub-heap missing in the other SMG, in which,
  however, it may be reachable via some other access path and may thus be
  discovered and joined later on).
  
\end{itemize}

The $\mathit{joinSubSMGs}$ function first applies the join reinterpretation
operator (denoted $joinFields$) on $G_1$, $G_2$ and $o_1$, $o_2$, which ensures
that the sets of fields of $o_1$ and $o_2$ are identical. The function then
iterates over the set of fields of these objects, and for each field $(\of,\ty)$
does the following:\begin{itemize} 

  \item finds the pair of values $v_1$ and $v_2$ which the has-value edges of
  $o_1$ and $o_2$ labelled by $(\of,\ty)$ lead to,
    
  \item calls the $\mathit{joinValues}$ function for $v_1$ and $v_2$, which is
  responsible for recursively joining the remaining parts of the sub-SMGs rooted
  at them (before the call, the nesting level difference between $G_1$ and $G_2$
  is possibly updated as described below), and
    
  \item extends the set of edges of $G$ by $o \hasvalue{\of,\ty} v$ where $v$ is
  the value returned by the $\mathit{joinValues}$ function.

\end{itemize}

\vspace*{-3mm}\subsubsection*{Adjusting the Nesting Level Difference}
\vspace*{-2mm}

In Section~\ref{sec:join}, it is said that the levels of the objects being
joined can differ since the objects may sometimes appear below a DLS and
sometimes below a region (and while an object that appears below a DLS may be
considered nested---provided that each node of the segment has a separate copy
of such an object---there is no notion of nesting below regions since for
regions which represent concrete objects there is no need to distinguish private
and shared sub-SMGs). The functionality of $\mathit{joinSubSMGs}$ therefore
includes tracking of the difference in levels (denoted $l_{\diff}$) at which
objects and values to be joined within some sub-SMG can appear. When objects
$o_1$ and $o_2$ are being joined, the difference is computed as follows: If
$o_1$ is a~DLS and $o_2$ is a region, the current value of~$l_{\diff}$ is
increased by one. Symmetrically, if $o_1$ is a region and $o_2$ is a DLS, the
value of~$l_{\diff}$ is decreased by one. Note that when going to sub-heaps in
both $G_1$ and $G_2$, the change of the difference is 0. The new difference is
then used when joining the values of the fields of $o_1$ and $o_2$ (apart from
the next and prev fields of course).

\enlargethispage{5mm}

\begin{algorithm}[t]
  \caption{\small$\mathit{joinSubSMGs}$($s$, $G_1$, $G_2$, $G$, $m_1$, $m_2$, 
    $o_1$, $o_2$, $o$, $l_{\diff}$)}
  \label{alg:joinSubSMGs}
  \smallskip

  \textbf{Input:}\begin{itemize}[nosep]

    \item Initial join status $s \in \jstat$.

    \item SMGs $G_1 = (O_1,V_1,\Lambda_1,H_1,P_1)$, $G_2 =
      (O_2,V_2,\Lambda_2,$ $H_2,P_2)$, and $G = (O, V, \Lambda, H, P)$.

    \item Injective partial mappings of nodes $m_1, m_2$ as defined
      in~Section~\ref{app:join}.

    \item Objects ~$o_1 \in O_1$, ~$o_2 \in O_2$, ~$o \in O$.

    \item Nesting level difference $l_{dif\hspace*{-0.4mm}f} \in \zed$.

  \end{itemize}

  \textbf{Output:}\begin{itemize}[nosep]

    \item $\bot$ in case the sub-SMGs of $G_1$ and $G_2$ rooted at $o_1$ and
      $o_2$ cannot be joined.

    \item Otherwise, a tuple $(s', G_1', G_2', G', m_1', m_2')$ where:
      \begin{itemize}[nosep]

        \item $s' \in \jstat$ is the resulting join status.

        \item $G_1', G_2', G'$ are SMGs as defined in Section
          \ref{app:joinSubSMGs}.

        \item $m_1'$, $m_2'$ are the resulting injective partial mappings of
          nodes.

      \end{itemize}

  \end{itemize}

  \textbf{Method:}\begin{enumerate}[nosep]

    \item Let $res := joinFields(G_1, G_2, o_1, o_2)$.  If $res = \bot$, return
      $\bot$.  Otherwise let\\$(s', G_1, G_2) := res$ ~and~ $s :=
      \mathit{updateJoinStatus}(s, s')$.

    \item Collect the set $F$ of all pairs $(\of,\ty)$ occurring in has-value edges
      leading from $o_1$ or $o_2$.

    \item For each field $(\of,\ty) \in F$ do:
      \begin{itemize}[nosep]

        \item Let $v_1 = H_1(o_1, \of, \ty)$, $v_2 = H_2(o_2, \of, \ty)$, and
          $l_{\diff}' := l_{\diff}$.

        \item If $\kind_1(o_1) = \dls$ and $(\of,\ty)$ is not the next/prev
          field of $o_1$, let $l_{\diff}' := l_{\diff}' + 1$.

        \item If $\kind_2(o_2) = \dls$ and $(\of,\ty)$ is not the next/prev
          field of $o_2$, let $l_{\diff}' := l_{\diff}' - 1$.

        \item Let $res := \mathit{joinValues}(s, G_1, G_2, G, m_1, m_2, v_1, v_2,
          l_{\diff}')$. If $res = \bot$, return $\bot$. Otherwise, let $(s, G_1,
          G_2, G, m_1, m_2, v) := res$.

        \item Introduce a new has-value edge $o \hasvalue{\of, \ty} v$ in $H$.

      \end{itemize}

    \item Return $(s, G_1, G_2, G, m_1, m_2)$.

  \end{enumerate}
\end{algorithm}

\enlargethispage{5mm}

\vspace*{-4mm} \subsection{Join of Values} \label{app:joinValues} \vspace*{-2mm}

The $\mathit{joinValues}$ function (cf.~Alg.~\ref{alg:joinValues}) joins a pair of
sub-SMGs rooted at a given pair of values and returns a single value node that
represents both the input values in the destination SMG. The function inputs a
triple of SMGs $G_1$, $G_2$, $G$ (two source SMGs and one destination SMG) and a
pair of values $v_1$ and $v_2$ from $G_1$ and $G_2$, respectively. If the
function fails in joining the given values, it returns $\bot$. Otherwise, it
returns a triple of SMGs $G_1'$, $G_2'$, $G'$, and a value $v$ from $G'$ such
that:\begin{itemize}

  \item $\MI(G_1) \subseteq \MI(G_1')$ and $\MI(G_2) \subseteq \MI(G_2')$ where
  $G_1$ and $G_2$ can differ from $G'_1$ and $G'_2$, respectively, due to an
  application of join reinterpretation on some of the pairs of objects being
  joined only.

  \item The sub-SMGs $G_1''$ and $G_2''$ of $G_1'$ and $G_2'$ rooted at $v_1$
  and $v_2$, respectively, are joined into the sub-SMG $G''$ of $G'$ rooted at
  $v$, i.e., it is required that \mbox{$\MI(G_1'') \subseteq \MI(G'') \supseteq
  \MI(G_2'')$}.

 \item The sub-SMG $G' \setminus G''$ is exactly the sub-SMG of $G$ that
 consists of objects and values that are not removed in
 Step~\ref{removalForDelayedJoin} of the $\mathit{joinTargetObjects}$ function
 due to using the principle of delayed join of sub-SMGs described in
 Section~\ref{app:DelayedJoin}.

\end{itemize}

\enlargethispage{5mm}

\begin{algorithm}[t]
  \caption{\small$\mathit{joinValues}$($s$, $G_1$, $G_2$, $G$, $m_1$, $m_2$,
  $v_1$, $v_2$, $l_{\diff}$)}
  \label{alg:joinValues}
  \textbf{Input:}\begin{itemize}[nosep]

    \item Initial join status $s \in \jstat$.

    \item SMGs $G_1 = (O_1,V_1,\Lambda_1,H_1,P_1)$, $G_2 =
      (O_2,V_2,\Lambda_2,H_2,P_2)$, and $G = (O, V, \Lambda, H, P)$.

    \item Injective partial mappings of nodes $m_1, m_2$ as defined
      in~Section~\ref{app:join}.

    \item Values $v_1 \in V_1$ and $v_2 \in V_2$.

    \item Nesting level difference $l_{\diff} \in \zed$.

  \end{itemize}

  \textbf{Output:}\begin{itemize}[nosep]

    \item $\bot$ in case the sub-SMGs of $G_1$ and $G_2$ rooted at $v_1$ and
      $v_2$ cannot be joined.

    \item Otherwise, a tuple $(s', G_1', G_2', G', m_1', m_2', v')$ where:
      \begin{itemize}[nosep]

        \item $s' \in \jstat$ is the resulting join status.

        \item $G_1', G_2', G'$ are SMGs as defined in Section
          \ref{app:joinValues}.

        \item $m_1'$, $m_2'$ are the resulting injective partial mappings of
          nodes.

        \item $v'$ is a value in $G'$ satisfying the conditions stated
          in~Section~\ref{app:joinValues}.

      \end{itemize}

  \end{itemize}

  \textbf{Method:}\begin{enumerate}[nosep]
    \item If $v_1 = v_2$, return $(s, G_1, G_2, G, m_1, m_2, v_1)$.  In this
      case, the given pair of values matches trivially, which happens whenever
      a shared value is reached during abstraction.

    \item If $m_1(v_1) = m_2(v_2) = v \ne \bot$, return $(s, G_1, G_2, G, m_1,
      m_2, v)$.  In this case, the pair of~values is already joined.

    \item If both $v_1$ and $v_2$ are non-address values, i.e., $P_1(v_1) =
      \UNDEF$ and $P_2(v_2) = \UNDEF$, then:
      \begin{itemize}[nosep]

        \item If $m_1(v_1) \ne \UNDEF$ or $m_2(v_2) \ne \UNDEF$, return $\bot$.

        \item Create a new value node $v \in V$ such that $\level(v) =
          max(\level_1(v_1),\level_2(v_2))$.
          
        \item Extend the mapping of nodes such that $m_1(v_1) = m_2(v_2) = v$.

        \item If $\level_1(v_1) - \level_2(v_2) < l_{\diff}$,
          let $s := \mathit{updateJoinStatus}(s, \jsubset)$.

        \item If $\level_1(v_1) - \level_2(v_2) > l_{\diff}$,
          let $s := \mathit{updateJoinStatus}(s, \jsupset)$.

        \item Return $(s, G_1, G_2, G, m_1, m_2, v_1)$.

      \end{itemize}

    \item If $P_1(v_1) = \UNDEF$ or $P_2(v_2) = \UNDEF$, return $\bot$.

    \item Let $res := \mathit{joinTargetObjects}(s, G_1, G_2, G, m_1, m_2, v_1,
      v_2, l_{\diff})$.\\
      If $res = \bot$, return $\bot$.
      If $res \ne\;\hookleftarrow$, then return $res$.

    \item Let $o_1 := o(P_1(v_1))$ and $o_2 := o(P_2(v_2))$.

    \item If $\kind_1(o_1) = \dls$, let $res = \mathit{insertLeftDlsAndJoin}(s, 
      G_1, G_2, G, m_1, m_2, v_1, v_2, l_{\diff})$.  If~$res = \bot$, return 
      $\bot$. If $res \ne\;\hookleftarrow$, then return $res$.

    \item If $\kind_2(o_2) = \dls$, let $res = \mathit{insertRightDlsAndJoin}(s, 
      G_1, G_2, G, m_1, m_2, v_1, v_2, l_{\diff})$.  If~$res \in \{ \bot, 
      \hookleftarrow \}$, return $\bot$. Otherwise, return $res$.

  \end{enumerate}
\end{algorithm}

If the input values are identical ($v_1 = v_2$), the resulting value $v$ is the
same identical value, say $v_1$, which prevents shared nodes from being
processed as nested data structures during abstraction. If both values are
mapped to the same value node in the destination SMG, i.e., $m_1(v_1) =
m_2(v_2)$, the node $m_1(v_1)$ (or, equivalently, $m_2(v_2)$) to which they are
mapped is returned since such a pair of values has already been successfully
joined before. A pair of non-address values visited for the first time, i.e., a
pair of values for which $m_1(v_1) = \UNDEF = m_2(v_2)$, is joined by creating a
fresh value node $v$ in $G'$ with the appropriate nesting level\footnote{In case
the difference of $\level_1(v_1)$ and $\level_2(v_2)$ differs from $l_{\diff}$,
the join status is appropriately updated. This reflects the fact that, e.g., a
single unknown (or interval) abstract value that is used as the value of
multiple fields of some region (for instance, an array) that is more nested than
the region may concretise to \emph{different} concrete values for each of the
fields while it must concretise to the \emph{same} concrete value if the
abstract value is on the same level as the given region.}, and the mapping of
nodes is extended such that $m_1(v_1) = m_2(v_2) = v$. Non-address values are
never joined with addresses---if such a~situation occurs, the whole join
operation fails.\footnote{Note that the handling of non-address values is kept
simple since the basic notion of SMGs does not distinguish any special kinds of
non-address values (numbers, intervals, etc.), but there is still room for
improvement, especially in conjunction with the extensions described in Section
\ref{sec:extensions}.} For addresses seen for the first time, the algorithm
tries to join their targets with each other with three~possible~outcomes:
\begin{enumerate}

  \item The join succeeds, and $\mathit{joinValues}$ then succeeds too.

  \medskip
  \item The join fails in a recoverable way (denoted by the result being
  $\hookleftarrow$). Intuitively, this happens when trying to join addresses
  that are found to be obviously different, i.e., addresses that are found to be
  different without going deeper into the sub-SMGs rooted at them---e.g., due to
  they point to some object with different offsets, different target specifiers,
  incompatible levels, size, validity, linking fields (for DLSs), or when the
  objects they point to are already mapped to some other objects. In case of
  the recoverable failure, if at least one of the target objects is a DLS, the
  algorithm tries to virtually \emph{insert a DLS} in one of the SMGs, which
  allows it to create a $0+$ DLS in the destination SMG and continue by joining
  the appropriate successor values. If this fails too, the whole join operation
  fails.\footnote{As mentioned already in Section~\ref{sec:join}, the recovery
  could be tried even when the impossibility of joining two addresses is
  discovered much later during joining the sub-SMGs rooted at the given
  addresses. Then, however, back-tracking would be necessary, which we try to
  avoid for efficiency reasons.}

  \medskip
  \item The join fails in an irrecoverable way (denoted by the result being
  $\bot$) in which case $\mathit{joinValues}$ fails too.

\end{enumerate}

\enlargethispage{5mm}

\vspace*{-6mm}\subsection{Join of Target Objects} \label{app:joinTargetObjects}
\vspace*{-3mm}

\begin{algorithm}[t]

  \caption{\small$\mathit{joinTargetObjects}$($s$, $G_1$, $G_2$, $G$, $m_1$,
  $m_2$, $a_1$, $a_2$, $l_{\diff}$)}
  \label{alg:joinTargetObjects}
  \smallskip

  \textbf{Input:}\begin{itemize}[nosep]

    \item Initial join status $s \in \jstat$.

    \item SMGs $G_1 = (O_1,V_1,\Lambda_1,H_1,P_1)$, $G_2 =
      (O_2,V_2,\Lambda_2,H_2,P_2)$, and $G = (O, V, \Lambda, H, P)$.

    \item Injective partial mappings of nodes $m_1, m_2$ as defined
      in~Section~\ref{app:join}.

    \item Addresses $a_1 \in V_1$ and $a_2 \in V_2$.

    \item Nesting level difference $l_{\diff} \in \zed$.

  \end{itemize}

  \textbf{Output:}\begin{itemize}[nosep]

    \item $\bot$ in case of an irrecoverable failure.

    \item $\hookleftarrow$ in case of a recoverable failure.

    \item Otherwise, a tuple $(s', G_1', G_2', G', m_1', m_2', a')$
      where:
      \begin{itemize}[nosep]

        \item $s' \in \jstat$ is the resulting join status.

        \item $G_1', G_2', G'$ are SMGs as defined in Section
          \ref{app:joinTargetObjects}.

        \item $m_1'$, $m_2'$ are the resulting injective partial mappings of
          nodes.

        \item $a'$ is an address in $G'$ satisfying the conditions stated
          in~Section~\ref{app:joinTargetObjects}.

      \end{itemize}

  \end{itemize}

  \textbf{Method:}\begin{enumerate}[nosep]

    \item If $\of(P_1(a_1)) \ne \of(P_2(a_2))$
      or $\level_1(a_1) - \level_2(a_2) \ne l_{\diff}$, return $\hookleftarrow$.

    \item Let $o_1 := o(P_1(a_1))$ ~and~ $o_2 := o(P_2(a_2))$.

    \item If $\kind_1(o_1) = \kind_2(o_2)$ ~and~ $\tg(P_1(a_1)) \ne
      \tg(P_2(a_2))$, return $\hookleftarrow$.

    %

    \item \label{nullOrAlreadyMerged} If $o_1 = \nil = o_2$ or $m_1(o_1) =
      m_2(o_2) \ne \UNDEF$,\\ let $(G, m_1, m_2, a) := mapTargetAddress(G_1, G_2,
      G, m_1, m_2, a_1, a_2)$ and return $(s, G_1, G_2, G, m_1, m_2, a)$. In this
      case, the targets are already joined, and we need to create a new address
      for the corresponding object $o \in O$.
      
    \item Let $s := matchObjects(s, G_1, G_2, m_1, m_2, o_1, o_2)$.  If $s =
    \bot$, return $\hookleftarrow$.

    \item Create a new object $o \in O$.

    \item Initialize the labelling of $o$ to match the labelling of $o_1$ if
      $\kind_1(o_1) = \dls$, or to match the labelling of $o_2$ if $\kind_2(o_2) =
      \dls$, otherwise take the labelling from any of them since both $o_1$ and
      $o_2$ are equally labelled regions.

    \item If $\kind_1(o_1) = \dls$ ~or~ $\kind_2(o_2) = \dls$, ~let~ $\len(o) :=
      min(\len_1(o_1), \len_2(o_2))$.

    \item Let $\level(o) := max(\level_1(o_1), \level_2(o_2))$.

    \item\label{removalForDelayedJoin} If $m_1(o_1) \ne \UNDEF$, replace each
    edge leading to $m_1(o_1)$ by an equally labelled edge leading to~$o$,
    remove $m_1(o_1)$ together with all nodes and edges of $G$ that are
    reachable via $m_1(o_1)$ only, and remove the items of $m_1$ whose target
    nodes were removed.  Likewise for $m_2$ and $o_2$. Note that this mechanism
    is called a \emph{delayed join of sub-SMGs} and explained
    in~Section~\ref{app:DelayedJoin}.

    \item Extend the mapping of nodes such that $m_1(o_1) = m_2(o_2) = o$.

    \item Let $(G, m_1, m_2, a) := mapTargetAddress(G_1, G_2, G, m_1,
      m_2, a_1, a_2)$.

    \item Let $res := \mathit{joinSubSMGs}(s, G_1, G_2, G, m_1, m_2, o_1, o_2, o,
      l_{\diff})$. If $res = \bot$, return $\bot$.  Otherwise return $(s, G_1,
      G_2, G, m_1, m_2, a)$.

  \end{enumerate}
\end{algorithm}


The $\mathit{joinTargetObjects}$ function (cf.~Alg.~\ref{alg:joinTargetObjects})
joins a pair of sub-SMGs rooted at the given pair of addresses and returns a
single address node that represents both the input addresses in the destination
SMG. The function inputs a triple of SMGs $G_1$, $G_2$, $G$ (two source SMGs and
one destination SMG) and a pair of addresses $a_1$ and $a_2$ from $G_1$ and
$G_2$, respectively. If the function fails in joining the given addresses, it
returns $\bot$ in case of an irrecoverable failure and $\hookleftarrow$ in case
of a recoverable failure (intuitively, for efficiency reasons, this happens when
the offsets, target specifiers, nesting levels, the kinds of the target objects,
their sizes, or validity are found incompatible without going deeper in the
sub-SMGs rooted at the given addresses). If the function succeeds, it returns
a~triple of SMGs $G_1'$, $G_2'$, $G'$ and an address $a$ from $G'$ such that:
\begin{itemize} 

  \item $\MI(G_1) \subseteq \MI(G_1')$ and $\MI(G_2) \subseteq \MI(G_2')$ where
  $G_1$ and $G_2$ can differ from $G'_1$ and $G'_2$, respectively, due to an
  application of join reinterpretation on some of the pairs of objects being
  joined only.

  \item The sub-SMGs $G_1''$ and $G_2''$ of $G_1'$ and $G_2'$ rooted at $a_1$
  and $a_2$, respectively, are joined as the sub-SMG $G''$ of $G'$ rooted at $a$,
  i.e., it is required that $\MI(G_1'') \subseteq \MI(G'') \supseteq \MI(G_2'')$.

\enlargethispage{6mm}

  \item The sub-SMG $G' \setminus G''$ is exactly the sub-SMG of $G$ that
  consists of objects and values that are not removed in
  Step~\ref{removalForDelayedJoin} of the function due to using the principle of
  delayed join of sub-SMGs described in Section~\ref{app:DelayedJoin}.

\end{itemize}

The algorithm first checks compatibility of the offsets that the points-to edges
leading from $a_1$ and $a_2$ are labelled with and checks whether the difference
in the nesting depth is appropriate.
Then, provided that the target objects $o_1 = o(P_1(a_1))$ and $o_2 =
o(P_2(a_2))$ are of the same kind, the algorithm checks whether they are reached
via the same target specifiers (when one is a DLS and one a region, the
specifiers may differ since the target specifier is not important for the
region).
If these tests do not pass, the algorithm fails in a recoverable way.

Next, if both $o_1$ and $o_2$ are null or if they have already been joined with
each other (and they are now reached through a different pair of addresses
only), the function $mapTargetAddress$ (cf.~Alg.~\ref{alg:mapTargetAddress}) is
used to join the addresses $a_1$ and $a_2$ into a fresh address $a \in A$ (so
that $m_1(a_1) = m_2(a_2) = a$) and to create a points-to edge from $a$ to the
join of $o_1$ and $o_2$, i.e., the object $m_1(o_1) = m_2(o_2)$.
If the objects $o_1$ and $o_2$ have not been joined so far, the algorithm checks
through the $matchObjects$ function (Alg.~\ref{alg:matchObjects} discussed
below) whether $o_1$ and $o_2$ can be safely joined (based on their labels,
labels of their outgoing edges, and the state of their mapping so far).
If so, a fresh object $o$ in $G$ is created that is intended to semantically
cover both $o_1$ and $o_2$.

We allow DLSs to be joined with regions as well as with DLSs of a different
minimal length, which requires the minimal length of the object $o$ to be
adjusted so that it covers both cases, i.e., $\len(o) := min(\len_1(o_1),
\len_2(o_2))$. 
The nesting level must also be properly chosen: the larger value is chosen since
a path through more DLSs is more abstract and hence more general (covering the
more concrete path through less DLSs).

Line~\ref{removalForDelayedJoin} now solves the case when exactly one of the
mappings $m_1(o_1)$ or $m_2(o_2)$ is already defined---the situation when both
of them are already defined is handled on line~\ref{nullOrAlreadyMerged} and in
function $matchObjects$.
This case can occur as a consequence of the DLS insertion algorithm described
in~Section~\ref{app:insertDLS}, and it is further discussed
in~Section~\ref{app:DelayedJoin}.
For now, it is enough to note that line~\ref{removalForDelayedJoin} eliminates
the impact of the DLS insertion algorithm on $o_1$ or $o_2$ (depending on which
of them was mapped within the DLS insertion) as well as on their sub-SMGs, so
that they can subsequently be merged as though both of them were encountered for
the first time.

Next, the mapping of nodes is extended such that $m_1(o_1) = m_2(o_2) = o$,
followed by using the function $mapTargetAddress$ to map the addresses $a_1$ and
$a_2$ to a new address $a \in A$ and to create a points-to edge from $a$ to $o$
(as we have already seen above).
Finally, the $\mathit{joinSubSMGs}$ function is called recursively for the
triple $o_1, o_2, o$.

\bigskip

\enlargethispage{5mm}

While we believe that Alg.~\ref{alg:mapTargetAddress} is self-explaining, we
provide some more intuition to Alg.~\ref{alg:matchObjects} that performs a local
check whether objects $o_1$, $o_2$ from SMGs $G_1$ and $G_2$, can be joined
under the current mappings $m_1$ and $m_2$, respectively, possibly for the price
of updating the current joint status $s$.
The algorithm assumes that at least one of the objects is non-null (the case of
both them being null is handled in Alg.~\ref{alg:joinTargetObjects}).
Under this assumption, if one of the objects is null, the algorithm fails since
null can be joined with null only.

The algorithm also fails if both $o_1$ and $o_2$ are mapped but not to each
other.
Next, even if $o_2$ is not mapped, but $o_1$ is mapped to some other object
$o'_2 \in O_2 \setminus \{ o_2 \}$, a failure happens (and symmetrically with
the roles of $o_1$ and $o_2$ swapped).
The reason for the failures based on the mappings is that at most two objects
can be mapped together.
The only case when the join can succeed with one of the objects mapped is when
the other is not mapped---this scenario is a consequence of the further
discussed DLS insertion and will be handled by cancelling the mapping of the
object that has already been mapped and by mapping $o_1$ and $o_2$ to each
other.

Subsequently, the algorithm checks whether the sizes of $o_1$ and $o_2$ are
equal and whether they have the same validity status.
In case the objects are DLSs, their defining offsets are checked for equality
too.
Then, the algorithm checks for all fields that appear in both objects and whose
values are already mapped whether their values are mapped to each other.
If some of the tests does not pass, the algorithm fails.


Finally, the algorithm checks whether the lengths of the objects or their kinds
imply a need to update the join status (joining a more concrete object with a
more abstract one).
If so, the join status is updated accordingly.

\begin{algorithm}[t]
  \caption{\small$mapTargetAddress$($G_1$, $G_2$, $G$, $m_1$, $m_2$, $a_1$, $a_2$)}
  \label{alg:mapTargetAddress}
  \smallskip

  \textbf{Input:}\begin{itemize}[nosep]

    \item SMGs $G_1 = (O_1,V_1,\Lambda_1,H_1,P_1)$, $G_2 =
      (O_2,V_2,\Lambda_2,H_2,P_2)$, and $G = (O, V, \Lambda, H, P)$ with
      the corresponding sets of addresses $A_1, A_2, A$.

    \item Injective partial mappings of nodes $m_1, m_2$ as defined
      in~Section~\ref{app:join}.

    \item Addresses $a_1 \in A_1$, $a_2 \in A_2$ referring with the same offset
    $\of =  \of(P_1(a_1)) = \of(P_2(a_2))$ to objects $o_1 = o(P_1(a_1))$, $o_2
    = o(P_2(a_2))$, respectively, which are already joined into an object $o =
    m_1(o_1) = m_2(o_2)$ and which are accessible via target specifiers such
    that $\kind_1(o_1) = \kind_2(o_2) \Rightarrow \tg(P_1(a_1)) =
    \tg(P_2(a_2))$.
    
  \end{itemize}

  \textbf{Output:}\begin{itemize}[nosep]

    \item A tuple $(G', m_1', m_2', a)$ where:
      \begin{itemize}[nosep]

        \item $G'$ is an SMG obtained from $G$ by extending its set of addresses
        by an address $a$~representing the join of $a_1$ and $a_2$ (unless $G$
        already contains this address) together with a~points-to edge from $a$
        to $o$ representing the join of the points-to edges between $a_1$, $a_2$
        and $o_1$, $o_2$, respectively.
        
        \item $m_1'$ and $m_2'$ are the resulting injective partial mappings of
        nodes that are either identical to $m_1$ and $m_2$ (if $a$ already
        exists in $G$) or obtained from $m_1$ and $m_2$ by extending them such
        that $m_1'(a_1) = m_2'(a_2) = a$.

    \end{itemize}

  \end{itemize}

  \textbf{Method:}\begin{enumerate}[nosep]

    \item Let ~$o_1 := o(P_1(a_1))$, ~$\of := \of(P_1(a_1))$.

    \item If $o_1 = \nil$,~ let ~$o := \nil$. ~Otherwise, let ~$o := m_1(o_1)$.


    \item If $\kind_1(o_1) = \dls$, let $\tg := \tg(P_1(a_1))$.
      Otherwise, let $\tg := \tg(P_2(a_2))$.

    \item If there is an address $a \in A$ such that $P(a) = (\of,
    \tg, o)$, return $(G, m_1, m_2, a)$.

    \item Extend $A$ by a fresh address $a$, then extend $P$ by a new points-to
    edge $a \pointsto{\of, \tg} o$.

    \item Extend the mapping of nodes such that $m_1(a_1) = m_2(a_2) = a$.

    \item Return $(G, m_1, m_2, a)$.

  \end{enumerate}
\end{algorithm}

\begin{algorithm}[t]
  \caption{\small$matchObjects$($s$, $G_1$, $G_2$, $m_1$, $m_2$, $o_1$, $o_2$)}
  \label{alg:matchObjects}
  \smallskip

  \textbf{Input:}\begin{itemize}[nosep]

    \item Initial join status $s \in \jstat$.

    \item SMGs $G_1 = (O_1,V_1,\Lambda_1,H_1,P_1)$ and $G_2 =
      (O_2,V_2,\Lambda_2,H_2,P_2)$.

    \item Injective partial mappings of nodes $m_1, m_2$ as defined
      in~Section~\ref{app:join}.

    \item Objects $o_1 \in O_1$ and $o_2 \in O_2$ such that $o_1 \neq \nil \vee
    o_2 \neq \nil$.

  \end{itemize}

  \textbf{Output:}\begin{itemize}[nosep]

    \item $\bot$ in case $o_1$ and $o_2$ cannot be joined.

    \item Otherwise, $s \in \jstat$ reflecting the impact of the labels of $o_1$
    and $o_2$ on the relation of the semantics of $G_1$ and $G_2$.

  \end{itemize}

  \textbf{Method:}\begin{enumerate}[nosep]

    \item If $o_1 = \nil$ ~or~ $o_2 = \nil$, ~return $\bot$.

    \item If $m_1(o_1) \ne \UNDEF \ne m_2(o_2)$ and $m_1(o_1) \ne m_2(o_2)$,
      return $\bot$.

    \item If $m_1(o_1) \ne \UNDEF$ and $\exists o_2' \in O_2: \; m_1(o_1) =
      m_2(o_2')$, return $\bot$.

    \item If $m_2(o_2) \ne \UNDEF$ and $\exists o_1' \in O_1: \; m_1(o_1') =
      m_2(o_2)$, return $\bot$.

    \item If $\size_1(o_1) \ne \size_2(o_2)$ or $\valid_1(o_1) \ne \valid_2(o_2)$,
    return $\bot$.

    \item If $\kind_1(o_1) = \kind_2(o_2) = \dls$, then:
      \begin{itemize}[nosep]

        \item If $\nfoBS_1(o_1) \ne \nfoBS_2(o_2)$,~ $\pfoBS_1(o_1) \ne 
          \pfoBS_2(o_2)$, ~or~ $\hfoBS_1(o_1) \ne \hfoBS_2(o_2)$, return~$\bot$.

      \end{itemize}

    \item Collect the set $F$ of all pairs $(\of,\ty)$ occurring in has-value 
      edges leading from $o_1$ or $o_2$.
      
    \item For each field $(\of,\ty) \in F$ do:
      \begin{itemize}[nosep]

        \item Let $v_1 = H_1(o_1, \of, \ty)$ and $v_2 = H_2(o_2, \of, \ty)$.

        \item If $v_1 \ne \UNDEF \ne v_2$ and $m_1(v_1) \ne \UNDEF \ne m_2(v_2)$
          and $m_1(v_1) \ne m_2(v_2)$, return $\bot$.
          
      \end{itemize}

    \item If $\len_1'(o_1) < \len_2'(o_2)\;\;$
      or $\;\;\kind_1(o_1) = \dls \;\wedge\; \kind_2(o_2) = \reg$,\\
      let $s := \mathit{updateJoinStatus}(s, \jsupset)$.
      
    \item If $\len_1'(o_1) > \len_2'(o_2)\;\;$
      or $\;\;\kind_1(o_1) = \reg \;\wedge\; \kind_2(o_2) = \dls$,\\
      let $s := \mathit{updateJoinStatus}(s, \jsubset)$.
      
    \item Return $s$.

  \end{enumerate}

\end{algorithm}

\enlargethispage{5mm}

\vspace*{-5mm} \subsection{DLS Insertion} \label{app:insertDLS} \vspace*{-3mm}

Assume that a pair of addresses $a_1$ and $a_2$ from SMGs $G_1$ and $G_2$ is to
be joined in order to allow $G_1$ and $G_2$ to be joined into an SMG $G$.
Further, assume that the objects that the addresses refer to cannot be joined,
but at least one of them is a DLS---in particular, assume that it is the object
with the address $a_1$, denote it $d_1$, and denote the other object $o_2$ (the
other possibility being symmetric). As mentioned already at the beginning of
Section~\ref{app:join}, in such a case, we proceed as though $a_2$ pointed to a
$0+$ DLS $d_2$ preceding $o_2$ and labelled equally as $d_1$ up to its length.
We then join $d_1$ and $d_2$ into a~single 0+ DLS $d$ in $G$ and continue by
joining the addresses $a_{next}$ and $a_2$ where $a_{next}$ is the value stored
in the next/prev pointer of $d_1$ (depending on whether we came to $d_1$ via the
$\first$ or $\last$ target specifier, respectively). We call this mechanism a
\emph{DLS insertion} because it can be seen as if the join of objects was
preceded by a virtual insertion of~a~DLS from one of the SMGs into the other
SMG. This extension is possible since the semantics of a~0+~DLS includes the
empty list, which can be safely assumed to appear anywhere, compensating a
missing object in one of the SMGs.

\enlargethispage{5mm}

Algorithm~\ref{alg:insertLeftDlsAndJoin} implements the DLS insertion. The
algorithm first checks whether the DLS $d_1$ from $G_1$ that we would like to
virtually insert into $G_2$ has not been processed by the join algorithm already
in the past. If this is the case and there is some object $o$ in $G_2$ that has
been joined with $d_1$ into $d$, the join fails since the $d_2$ segment
(possibly represented by a region as its concrete instance) is not missing, but
it is not connected to the rest of the SMG in a~way compatible with $d_1$ (at
least not for the current way $G_1$ and $G_2$ are being joined). If no such
object $o$ exists, the DLS $d$ to which $d_1$ is mapped in $G$ is used as the
result of joining $d_1$ with the virtually added segment $d_2$, and unless even
the address $a_1$ has already been processed, the join continues by the
addresses $a_{next}$ and $a_2$. Intuitively, this case arises when inserting
a~single missing segment that should be reachable through several paths in the
SMG. Note that such a situation is, in fact, quite usual since a DLS can be
reached both forward and backward. The algorithm, however, has to insert a
single virtual segment $d_2$ for all such paths.

If $d_1$ has not yet been processed, the algorithm checks whether there is some
hope that the virtual insertion of $d_2$ could help (or whether it is better to
try to proceed with the join in some other way: e.g., try to insert a DLS from
$G_2$ into $G_1$ in case both of the addresses $a_1$ and $a_2$ point to DLSs or
fall-back to introducing a 0/1 abstract object as mentioned
in~Section~\ref{sec:extensions}). However, unlike in the function
$\mathit{joinTargetObjects}$, if we do not want to go deeper in the SMGs (which
we do not want for efficiency reasons), there is not so many properties to check
since we do not have two objects whose labelling we could compare, but the
single DLS $d_1$ whose counterpart we want to virtually insert into $G_2$ only.
So, we at least check that there is no conflict of the successor addresses
$a_{next}$ and $a_2$ according to the current mapping of nodes, i.e., we require
$m_1(a_{next}) = \UNDEF \;\vee\; m_2(a_2) = \UNDEF \;\vee\; m_1(a_{next}) =
m_2(a_2)$. 

If the above checks pass, the DLS insertion proceeds as follows: Let $F$ be the
set of the linking fields of $d_1$ that are oriented forward wrt the direction
of the traversal.  In particular, let $F = \{\nfoBS_1(d_1)\}$ if $\tg(P_1(a_1))
= \first$, and $F = \{\pfoBS_1(d_1)\}$ if $\tg(P_1(a_1)) = \last$. First, the
DLS $d$ representing the join of $d_1$ and the virtually inserted $d_2$ is
created in the destination SMG with the same labelling as that of $d_1$ up to
$\len(d) = 0$. Together with basically copying the DLS $d_1$ from $G_1$ to $G$,
we also copy the $F$-restricted sub-SMG rooted at it from $G_1$ into $G$,
excluding the nodes for which $m_1$ is already defined (these were already
reached through some other paths in the past, and the newly copied part of the
$F$-restricted sub-SMG rooted at $d_1$ is just linked to them). Subsequently,
the algorithm extends the mapping $m_1$ for the nodes newly inserted to $G$,
creates the appropriate address node $a \in A$ as well as the points-to edge
leading from $a$ to $d$, and extends the mapping of addresses such that
$m_1(a_1) = a$. The algorithm then continues by joining the pair of successor
values $a_{next}$ and $a_2$.

\begin{algorithm}[t]
  \caption{\small$\mathit{insertLeftDlsAndJoin}$($s$, $G_1$, $G_2$, $G$, $m_1$,
  $m_2$, $a_1$, $a_2$, $l_{\diff}$)}
  \label{alg:insertLeftDlsAndJoin}
  \smallskip

  \textbf{Input:}\begin{itemize}[nosep]

    \item Initial join status $s \in \jstat$.

    \item SMGs $G_1 = (O_1,V_1,\Lambda_1,H_1,P_1)$, $G_2 =
      (O_2,V_2,\Lambda_2,H_2,P_2)$, and $G = (O, V, \Lambda, H, P)$.

    \item Injective partial mappings of nodes $m_1, m_2$ as defined
      in~Section~\ref{app:join}.

    \item Values $a_1 \in V_1$ such that $\kind_1(o(P_1(a_1))) = \dls$ and 
      $a_2 \in V_2$.

    \item Nesting level difference $l_{\diff} \in \zed$.

  \end{itemize}

  \textbf{Output:}\begin{itemize}[nosep]

    \item $\bot$ in case of an unrecoverable failure.

    \item $\hookleftarrow$ in case of a recoverable failure.

    \item Otherwise, a tuple $(s', G_1', G_2', G', m_1', m_2', a')$
      where:
      \begin{itemize}[nosep]

        \item $s' \in \jstat$ is the resulting join status.

        \item $G_1', G_2', G'$ are SMGs as defined in Section
          \ref{app:joinTargetObjects}.

        \item $m_1'$, $m_2'$ are the resulting injective partial mappings of
          nodes.

        \item $a'$ is an address in $G'$ satisfying the conditions stated
          in~Section~\ref{app:joinTargetObjects}.

      \end{itemize}

  \end{itemize}

  \textbf{Method:}\begin{enumerate}[nosep]

    \item Let $(d_1, \of, \tg) := P_1(a_1)$.

    \item If $\tg = \first$, let $nf := \nfoBS_1(d_1)$; if $\tg = \last$, 
      let $nf := \pfoBS_1(d_1)$; otherwise return $\hookleftarrow$.

    \item Let $a_{next} := H_1(d_1, nf, \ptr)$.

    \item If $m_1(d_1) \ne \UNDEF$, then:
      \begin{itemize}[nosep]
        \item Let $d := m_1(d_1)$.

        \item If $\exists o \in O:\; m_2(o) = d$, ~return
          $\hookleftarrow$.

        \item If $m_1(a_1) = \UNDEF$, create a new value node $a \in V$ and a
          new edge $a \pointsto{\of, \tg} d$ in $P$, and extend the mapping
          of~nodes such that $m_1(a_1) = a$.  Otherwise let $a := m_1(a_1)$ and
          return $(s, G_1, G_2, G, m_1, m_2, a)$.

        \item Let $res := \mathit{joinValues}(s, G_1, G_2, G, m_1, m_2, a_{next}, 
          a_2, l_{\diff})$.  If $res = \bot$, return $\bot$. Otherwise, let 
          $(s, G_1, G_2, G, m_1, m_2, a) := res$.

      \end{itemize}

    \item If $m_1(a_{next}) \ne \UNDEF$ and $m_2(a_2) \ne \UNDEF$ and
      $m_1(a_{next}) \ne m_2(a_2)$, ~return $\hookleftarrow$.

    \item Let $s' := (\len_1(d_1) = 0) ~? ~\jsupset ~: ~\join$.  ~Let $s :=
      \mathit{updateJoinStatus}(s, s')$.

    \item Extend $G$ by a fresh copy of the $\{nf\}$-restricted sub-SMG of $G_1$
      rooted at $d_1$, but excluding the nodes that are already mapped in $m_1$
      such that the copy of $d_1$ is a DLS $d$.  Then extend the mapping $m_1$
      such that the newly created nodes in $O \cup V$ are mapped from the
      corresponding nodes of $O_1 \cup V_1$.

    \item Initialize the labelling of $d$ to match the labelling of $d_1$ up to
      the minimum length, which is fixed to zero, i.e., $\len(d) = 0$.

    \item Let $a \in V$ be the address such that $P(a) = (\of, \tg, d)$ if such an
      address exists in $G$.  Otherwise, create a new value node $a \in V$ and a
      new edge $a \pointsto{\of, \tg} d$ in $P$, and extend the mapping of~nodes
      such that $m_1(a_1) = a$.

    \item Let $res := \mathit{joinValues}(s, G_1, G_2, G, m_1, m_2, a_{next}, a_2,
      l_{\diff})$.  If $res = \bot$, return $\bot$.  Otherwise let $(s, G_1,
      G_2, G, m_1, m_2, a') := res$.

    \item Introduce a new has-value edge $d \hasvalue{nf, \ptr} a'$ in $H$.

    \item Return $(s, G_1, G_2, G, m_1, m_2, a)$.

  \end{enumerate}

\end{algorithm}

\enlargethispage{5mm}

\vspace*{-4mm} \subsection{Delayed Join of Sub-SMGs} \label{app:DelayedJoin}
\vspace*{-3mm}

The mechanism of DLS insertion increases the chances for two SMGs to be
successfully joined, but it brings one complication to be taken care of. Assume
that SMGs $G_1$ and $G_2$ are being joined into an SMG $G$. When applying the
DLS insertion mechanism on a~DLS $d_1$ from $G_1$, the not yet traversed sub-SMG
$G_1'$ of $G_1$ reachable from $d_1$ is copied into $G$ too (likewise for the
symmetric case). However, some nodes of $G_1'$ that are inserted into $G_2$ may
in fact exist in $G_2$ and be reachable through some other address than the
address at which the DLS insertion is started. Note that this \emph{may} but
\emph{needs not} happen, and at the time when the DLS insertion is run, it is
unknown which of the two cases applies. In theory, a backward traversal through
the SMGs could be used here, but we chose not to use it since we were afraid of
its potential bad impact on the performance. That is why, we always insert a DLS
together with the sub-SMG rooted at it, and as soon as we realize that some DLS
$d_1$ from $G_1$ whose counterpart was inserted into $G_2$ does have a real
counterpart in $G_2$, the result of the join of $d_1$ with the inserted DLS is
deleted (together with all the values and objects reachable from that DLS only)
and a proper join---which we denote as the so-called \emph{delayed join}---is
used instead (cf. Point~\ref{removalForDelayedJoin} of the function
$\mathit{joinTargetObjects}$). Note that running the delayed join is indeed
necessary since the insertion of a DLS is optimistic in that its sub-SMG is
either missing too, or if it is not missing, it is the same as the sub-SMG of
the inserted DLS. This needs, however, not to be the case.

\enlargethispage{6mm}

\vspace*{-5mm} \subsection{Join of SPCs} \vspace*{-4mm}

The $\mathit{joinSPCs}$ function (cf.~Alg.~\ref{alg:joinSPCs}) is the top-level
of the join algorithm used when reducing the number of SPCs generated for
particular basic blocks of the program being analysed. It inputs a pair of
garbage-free SPCs \mbox{$C_1=(G_1,\nu_1)$}, $C_2=(G_2,\nu_2)$ where
$range(\nu_1) = range(\nu_2) = \Var$ is the common set of program variables, and
for each $x \in \Var$, the objects $\nu_1(x)$ and $\nu_2(x)$ are labelled
equally (this condition necessarily holds for SPCs generated for the same
program). The algorithm either fails and returns~$\bot$, or it returns the
resulting join status and an SPC $C = (G, \nu)$ where $range(\nu) = \Var$, and
the triple $G_1$, $G_2$, $G$ satisfies the assertions about joined SMGs stated
in~Section~\ref{sec:join}.

\enlargethispage{5mm}

The function starts by initializing the mappings of nodes $m_1$ and $m_2$ to the
empty set and the join status $s$ to $\iso$. Then, for each program variable $x
\in \Var$, a fresh region $r$ is created in $G$, labelled equally as $r_1$ (or
$r_2$), and the mappings are extended such that $\nu(x) = m_1(r_1) = m_2(r_2) =
r$ where $r_1 = \nu_1(x)$ and $r_2 = \nu_2(x)$. Next, for each program variable
$x \in \Var$, the $\mathit{joinSubSMGs}$ function is called with the
corresponding triple of objects $\nu_1(x), \nu_2(x), \nu(x)$. The value of
$m_1$, $m_2$, and $s$ gets propagated between each pair of subsequent calls.
Subsequently, the $\mathit{joinSPCs}$ function checks whether there was not
created any cycle consisting of 0+ DLSs only in $G$, and if so, the algorithm
fails since the DLS consistency requirement would be broken this way (cf.
Section~\ref{sec:defOfSMGs}).

\begin{algorithm}[t]

  \caption{\small$\mathit{joinSPCs}(C_1, C_2)$}

  \label{alg:joinSPCs}

  \smallskip

  \textbf{Input:}\begin{itemize}[nosep]

    \item Garbage-free SPCs $C_1=(G_1,\nu_1)$, $C_2=(G_2,\nu_2)$
      with SMGs\\$G_1 = (O_1,V_1,\Lambda_1,H_1,P_1)$, $G_2 =
      (O_2,V_2,\Lambda_2,H_2,P_2)$\\
      where $range(\nu_1) = range(\nu_2) = \Var$, and for each $v \in \Var$,\\
      the labelling of $\nu_1(v)$ is equal to the labelling of $\nu_2(v)$.

  \end{itemize}

  \textbf{Output:}\begin{itemize}[nosep]

    \item $\bot$ in case $C_1$ and $C_2$ cannot be joined.

    \item Otherwise, a tuple $(s, C)$ where:
      \begin{itemize}[nosep]

        \item $s \in \jstat$ is the resulting join status.

        \item $C = (G,\nu)$ where $range(\nu) = \Var$ and the SMG $G$
          satisfies the condition\\$\MI(G_1) \subseteq \MI(G) \supseteq \MI(G_2)$.

      \end{itemize}

  \end{itemize}

  \textbf{Method:}\begin{enumerate}[nosep]

    \item Let $G$ be an empty SMG $G$, ~$\nu := m_1 := m_2 := \emptyset$, 
      ~$s := \; \iso$.

    \item For each program variable $v \in \Var$:
      \begin{itemize}[nosep]

        \item Let $r_1 := \nu_1(v)$ and $r_2 := \nu_2(v)$.
          
        \item Create a fresh region $r \in O$, initialize its labelling to match
          the labelling of $r_1$.
          
        \item Extend the mappings such that $m_1(r_1) = m_2(r_2) = \nu(v) = r$.

      \end{itemize}

    \item For each program variable $v \in \Var$:
      \begin{itemize}[nosep]
        \item Let $r_1 := \nu_1(v)$, $r_2 := \nu_2(v)$, and $r := \nu(v)$.

        \item Let $res := \mathit{joinSubSMGs}(s, G_1, G_2, G, m_1, m_2, r_1, 
          r_2, r, 0)$.

        \item If $res = \bot$, return $\bot$. Otherwise, let $(s, G_1, G_2, G,
          m_1, m_2) := res$.
      \end{itemize}

    \item If there is any cycle consisting solely of 0+ DLSs in $G$, return
      $\bot$.

    \item Return $(s, C)$ where $C = (G, \nu)$.
  \end{enumerate}

\end{algorithm}

\vspace*{-5mm} \subsection{Join of Sub-SMGs within Abstraction}
\label{app:mergeSubSMGs} \vspace*{-3mm} 

The $joinSubSMGsForAbstraction$ function (cf.~Alg.~\ref{alg:mergeSubSMGs})
implements the core functionality of the elementary merge operation used as a
part of our abstraction mechanism (Section~\ref{sec:abstraction}). It inputs an
SMG $G = (O, V, \Lambda, H, P)$, a pair of objects $o_1, o_2 \in O$, and a
triple of binding offsets $\hfo$, $\nfo$, $\pfo \in \nat$. If it succeeds, it
returns an SMG $G' = (O', V', \Lambda', H', P')$ and a fresh DLS $d \in O'$
which represents the merge of $o_1$ and $o_2$ in $G'$ and which is the entry
point of the sub-SMG representing a join of the sub-SMGs rooted at $o_1$ and
$o_2$.  What remains to be done in the elementary merge operation is the
reconnection of the pointers surrounding $o_1$ and $o_2$ to $d$ (apart from
those related to their nested sub-SMGs), cf. Section~\ref{sec:abstraction}.  As
an auxiliary result (used in the algorithm of searching for longest mergeable
sequences), the algorithm returns the join status $s \in \jstat$ comparing the
semantics of the sub-SMGs rooted at $o_1$ and $o_2$ as well as the sets $O_1,
O_2 \subseteq O'$ and $V_1, V_2 \subseteq V'$ that contain those objects and
values whose join produced the sub-SMG nested below $d$. Note that the join
status returned by $\mathit{mergeSubSMGs}$ is not affected by the kinds of $o_1,
o_2$ and the values in their next/prev fields since the loss of information due
to merging $o_1$ and $o_2$ into a~single list segment is deliberate.

The function proceeds as follows. Using the offsets $\nfo$, $\pfo$, the values
of the next/prev fields of $o_1$ and $o_2$ are remembered and temporarily
replaced by $0$ (in order for the subsequently started join of sub-SMGs not to
go through these fields). Then a fresh DLS $d$ is created in $O$ and labelled by
the given offsets $\hfo$, $\nfo$, $\pfo$ and the minimum length equal to
$\len_1(o_1) + \len_2(o_2)$, other labels are taken from $o_1$ (or $o_2$ since
the other labels are equal). The mapping of objects is initialized such that
$m_1(o_1) = m_2(o_2) = d$ and the nesting level difference is initialized based
on $\kind(o_1)$ and $\kind(o_2)$ using the rules stated
in~Section~\ref{app:joinSubSMGs}. The generic algorithm $\mathit{joinSubSMGs}$
is then called on the triple $o_1, o_2, d$. If it fails or the resulting SMG
contains any cycle consisting of 0+ DLSs only, the algorithm exits
unsuccessfully. If it succeeds, the values of the next/prev fields in $o_1,
o_2$, which were temporarily replaced by $0$, are restored to their original
values. If $\kind(o_1) = \kind(o_2) = \reg$, the level of each node that appears
in the image of $m_1$ (or $m_2$) is increased by one (since these nodes are now
recognized as nested), and all points-to edges leading to $d$ are relabelled by
the $\all$ target specifier.

\enlargethispage{5mm}

The resulting sets of nodes are computed as follows: $O_1 := range(m_1) \cap O$,
~$O_2 := range(m_2) \cap O$, ~$V_1 := range(m_1) \cap V$, and ~$V_2 = range(m_2)
\cap V$. The resulting join status is the status returned by
$\mathit{joinSubSMGs}$.

\begin{algorithm}[t]

  \caption{\small$\mathit{joinSubSMGsForAbstraction}$($G$, $o_1$, $o_2$, $\hfo$,
  $\nfo$, $\pfo$)}

  \label{alg:mergeSubSMGs}

  \smallskip

  \textbf{Input:}\begin{itemize}[nosep]

    \item SMGs $G = (O, V, \Lambda, H, P)$.

    \item Objects $o_1, o_2 \in O$ that are the roots of the $\{ \nfo, \pfo
    \}$-restricted sub-SMGs $G_1$ and $G_2$ of $G$ that are to be joined and
    that are such that $level(o_1) = level(o_2)$ and $size(o_1) = size(o_2)$.

    \item Candidate DLS offsets $\hfo, \nfo, \pfo \in \nat$.

  \end{itemize}

  \textbf{Output:}\begin{itemize}[nosep]

    \item $\bot$ in case $G_1$ and $G_2$ cannot be joined.

    \item Otherwise, a tuple $(s, G', d, O_1, V_1, O_2, V_2)$
    where:\begin{itemize}[nosep]

      \item $s \in \jstat$ is the resulting join status (determines the cost of
        joining $G_1$ and $G_2$).

      \item $G'$ is an SMG obtained from the input SMG $G$ by extending it with
      a new DLS $d$ below which the join of $G_1$ and $G_2$ is nested.

      \item $O_i \subseteq O$ and $V_i \subseteq V$ for $i = 1,2$ are sets of
      non-shared objects and values of $G_1$ and $G_2$,
      respectively.

    \end{itemize}

  \end{itemize}

  \textbf{Method:}\begin{enumerate}[nosep]

    \item Let $a_p := H(o_1, \pfo, \ptr)$, $a_n := H(o_2, \nfo, \ptr)$, $a_1
    := H(o_1, \nfo, \ptr)$,\\ and $a_2 := H(o_2, \pfo, \ptr)$.

    \item Replace each has-value edge of $H$ leading from $o_1$ or $o_2$ and
    labelled by $(\nfo, \ptr)$ or $(\pfo, \ptr)$ by a has-value edge leading to
    $0$ and having the same label.

    \item Extend $O$ with a fresh valid DLS $d$ and label it with the head,
    next, and prev offsets $\hfo$, $\nfo$, and $\pfo$, the minimum length
    $\len(o_1) + \len(o_2)$, level $\level(o_1)$, and the size $\size(o_1)$.

    \item If $\kind(o_1)=\kind(o_2)$, let $l_{\diff} := 0$. Otherwise, let
    $l_{\diff} := (\kind(o_1) = \dls) ? 1 : -1$.

    \item Let $res := \mathit{joinSubSMGs}(\iso, G_1, G_2, G, \{ (o_1,d) \}, \{
    (o_2,d) \}, o_1, o_2, d, l_{\diff})$.\\ If $res = \bot$, return $\bot$.
    Otherwise let $(s, \_, \_, G, m_1, m_2) := res$.

    \item If $G$ contains any cycle consisting of 0+ DLSs only, return $\bot$.

    \item Drop the temporarily created has-value edges of $H$ leading from $o_1$
    and $o_2$ to $0$ and labelled by $(\nfo, \ptr)$ or $(\pfo, \ptr)$ and restore
    the original has-value edges $o_1 \hasvalue{\pfo, \ptr} a_p$, $o_1
    \hasvalue{\nfo, \ptr} a_1$, $o_2 \hasvalue{\pfo, \ptr} a_2$, and $o_2
    \hasvalue{\nfo, \ptr} a_n$.

    \item If $\kind(o_1)=\kind(o_2)=\reg$, increase by one the level of each
    object and value that appears in the image of $m_1$ or $m_2$, and
    relabel all points-to edges leading to $d$ by the $\all$ target specifier.

    \item Return $(s, G, d, O \cap range(m_1), V \cap range(m_1), O \cap
      range(m_2), V \cap range(m_2))$.

\end{enumerate}

\end{algorithm}

\vspace*{-5mm} \section{Longest Mergeable Sequences} \label{app:longestMergSeq}
\vspace*{-3mm}

In this appendix, we formalize the notion of longest mergeable sequences
informally introduced in Section~\ref{sec:abstraction}. Assume an SPC
$C=(G,\nu)$ where $G = (O,V,\Lambda,H,P)$ is an SMG with the sets of regions
$R$, DLSs $D$, and addresses $A$. The \emph{longest mergeable sequence} of
objects given by a candidate DLS entry $(o_c, \hfoBS_c, \nfoBS_c, \pfoBS_c)$
where $o_c \in O$ is the longest sequence of distinct valid heap objects of $G$
whose first object is $o_c$; all objects in the sequence are of level $0$; all
DLSs that appear in the sequence have $\hfoBS_c$, $\nfoBS_c$, and $\pfoBS_c$ as
their head, next, and prev offsets; and the following holds for any two
neighbouring objects $o_1$ and $o_2$ in the sequence:\begin{enumerate}

  \item The objects $o_1$ and $o_2$ are doubly-linked, i.e., there are addresses
  $a_1, a_2 \in A$ such that $o_1 \hasvalue{\nfoBS_c,\ptr} a_1
  \pointsto{\hfoBS_c,\tg_2} o_2$ for $\tg_2 \in \{ \first, \reg \}$ and
  \mbox{$o_2 \hasvalue{\pfoBS_c,\ptr} a_2 \pointsto{\hfoBS_c,\tg_1} o_1$} for
  $\tg_1 \in \{ \last, \reg \}$.

  \item The $\{ \nfoBS_c, \pfoBS_c \}$-restricted sub-SMGs $G_1$, $G_2$ of $G$
  rooted at $o_1$ and $o_2$ can be joined using the extended join algorithm that
  yields the sub-SMG $G_{1,2}$ to be nested below the join of $o_1$ and $o_2$ as
  well as the sets $O_1$, $V_1$ and $O_2$, $V_2$ of non-shared objects and
  values of $G_1$ and $G_2$, respectively, whose join gives rise to $G_{1,2}$.
  
  \item The non-shared objects and values of $G_1$ and $G_2$ (other than $o_1$
  and $o_2$ themselves) are reachable via $o_1$ or $o_2$, respectively, only.
  This is, $\forall a \in A \setminus V_1 ~\forall o' \in O_1 \setminus \{ o_1
  \}: o(P(a)) \neq o', ~\forall o' \in O \setminus O_1 ~\forall v \in V_1:~ v
  \notin H(o')$, and likewise for $o_2$, $V_2$, and $O_2$. Moreover, the sets
  $O_1$ and $O_2$ contain heap objects only.

  \item The objects $o_1$ and $o_2$ are a part of an uninterrupted sequence.
  Therefore:\begin{enumerate}

    \item Regions that are not the first nor last in the sequence can be pointed
    to their head offset from their predecessor, successor, or from their
    non-shared restricted sub-SMG only. Formally, if $o_1 \in R \setminus \{ o_c
    \}$, then \mbox{$\neg~ \exists o \in O \setminus (O_1 \cup \{ o_2, o' \})$}
    $\exists a \in A ~\exists \of \in \nat:~ o \hasvalue{\of,ptr} a
    \pointsto{\hfoBS_c,\reg} o_1$ where the object $o'$ is the predecessor of
    $o_1$, i.e., $o' = o(P(H(o_1,\pfoBS_c,\ptr)))$.\footnote{Note that no
    special formal treatment is needed for $o_2$ since it will take the role of
    $o_1$ when checking the next neighbouring pair in the sequence. The above
    also implicitly ensures that pointers to the head offset of the last object
    are not restricted.}

    \item If $o_1$ ($o_2$) is a DLS, the only object that can point to its end
    (beginning)~is~$o_2$~($o_1$), resp. Formally, \mbox{$\text{if } o_1 \in
    D\text{, then } \neg~ \exists o \in O \setminus \{ o_2 \}~ \exists a \in A$}
    $\exists \of \in \nat: o \hasvalue{\of,\ptr} a \pointsto{\hfoBS_c,\last}
    o_1$. $\text{If } o_2 \in D \text{, then } \neg~ \exists o \in O \setminus
    \{ o_1 \}~ \exists a \in A$ $\exists \of \in \nat: o \hasvalue{\of,\ptr} a
    \pointsto{\hfoBS_c,\first} o_2$.

    \item Finally, only non-shared objects of $G_1$ and $G_2$ can point to
    non-head offsets of $o_1$ and $o_2$, respectively. Formally, \mbox{$\neg~
    \exists o \in O \setminus O_1 ~\exists a \in A$} $\exists \of, \ofprime \in
    \nat ~\exists \tg \in \ptt:~ o \hasvalue{\of,\ptr} a
    \pointsto{\ofprime\backspace,\tg} o_1 ~\wedge~ \ofprime \ne \hfoBS_c$, and
    likewise for $o_2$ and $O_2$.

  \end{enumerate}

\end{enumerate}

\vspace*{-5mm} \section{Symbolic Execution of Conditional Statements}
\label{app:condSymExec} \vspace*{-3mm}

Checking equality of values is trivial in SMGs since it reduces to identity
checking. To check non-equality, we propose a sound, efficient, but incomplete
approach as mentioned already in Section~\ref{sec:inconsistency-check}. This
approach is formalized in the function $proveNeq$ shown as
Alg.~\ref{alg:proveNeq} that contains a number of comments to make it
self-explaining. The algorithm uses the $lookThrough$ function
(Alg.~\ref{alg:lookThrough}) for traversing chains of $0+$ DLSs while looking
for objects whose existence is guaranteed and whose unique addresses can serve
as a basis for a non-equality proof. Note that the algorithms do not allow for
comparing values of fields with incompatible types. Hence, we require all
type-casts to be explicitly represented as separate instructions of the
intermediate code so that the fields being \mbox{compared are always of
compatible types.}

\enlargethispage{5mm}

If neither equality nor inequality of a pair of values $v_1$ and $v_2$, which
are compared in a~conditional statement, can be established, the symbolic
execution must follow both branches of the conditional statement. For each of
the branches, we attempt to reflect the condition allowing the execution to
enter that branch in the SMG $G$ to be processed in the branch, effectively
reducing the semantics of $G$. However, for efficiency reasons, we do again not
reflect all consequences of the branch conditions that could in theory be
reflected, but only the easy to handle ones, which is sound, and it suffices in
all the case studies that we have considered. In particular, we restrict SMGs
according to the branch conditions as follows (if none of the below described
cases applies, the SMGs are not modified):

\begin{itemize}

  \item If $v_1$ and/or $v_2$ are non-address values, one of them is replaced by
  the other in the $v_1=v_2$ branch (a~non-address value can be replaced by an
  address but not vice versa).

  \item If there is a chain of $0+$ DLSs connected into a doubly-linked list in
  the given SMG such that the $\first$ address of the first DLS is $v_1$ and the
  last DLS contains $v_2$ in its next field (or vice versa), the chain is
  removed by calling the DLS removal algorithm repeatedly in the $v_1=v_2$
  branch. In~the $v_1 \ne v_2$ branch, the computation is split to as many cases
  as the number of $0+$~DLSs in the chain is, and in each of the cases, the
  minimum length of one of the DLSs is incremented (reflecting that at least one
  of them must be non-empty).

  \item If $v_1$ points to a DLS $d$ with the $\first$ target specifier and
    $v_2$ points to $d$ with the $\last$ target specifier
    (or vice versa), it is clear that $\len(d) < 2$ since
  otherwise $proveNeq$ would succeed in proving the inequality between $v_1$ and
  $v_2$. In this case, the following two modifications of the encountered SMGs
  can be applied in the different branches of the encountered conditional
  statement:\begin{itemize}
  
    \item In the $v_1=v_2$ branch, if $\len(d) = 1$, the DLS $d$ is replaced by
    an equally labelled region (excluding the DLS-specific labels) since $d$
    must consist of exactly one concrete node in this case.

    \item In the $v_1 \ne v_2$ branch, if $\len(d) = 1$ or the value of the next
    address is equal to the value of the prev address, i.e., $H(d,\pfo(d),\ptr) =
    H(d,\nfo(d),\ptr)$, then the minimum length of $d$ is increased to $2$ since
    $d$ must consist of at least two concrete nodes in this case.

  \end{itemize}

\end{itemize}

Besides equality checking, we also allow for comparisons of addresses using the
\emph{less than} or \emph{greater than} operators in case both of the addresses
point to the same (concrete) object---we simply compare the offsets. This
functionality is needed for successful verification of the NSPR-based case
studies mentioned in~Section~\ref{sec:experiments}.

\begin{algorithm}[t]

  \caption{\small$proveNeq(G, v_1, v_2)$}

  \label{alg:proveNeq}

  \smallskip

  \textbf{Input:}\begin{itemize}[nosep]

    \item An SMG $G = (O,V,\Lambda,H,P)$.

    \item A pair of values $v_1, v_2 \in V$ such that $\level(v_1) = \level(v_2) =
      0$.

  \end{itemize}

  \textbf{Output:}\begin{itemize}[nosep]

    \item $\true$ if the inequality between the given pair of values was proven,
      $\false$ otherwise.

  \end{itemize}

  \textbf{Method:}\begin{enumerate}[nosep]
    \item Let $(v_1, O_1) := lookThrough(G, v_1)$.

    \item Let $(v_2, O_2) := lookThrough(G, v_2)$.

    \item If $v_1 = v_2$ or $O_1 \cap O_2 \ne \emptyset$, return $\false$.
      \comment{possible sharing of values}

    \item If $v_1 \notin A$ or $v_2 \notin A$, return $\false$.
      \comment{simplified handling of data values}

    \item Let $o_1 := o(P(v_1))$ and $o_2 := o(P(v_2))$.

    \item If $o_1 = o_2$:
      \begin{itemize}[nosep]
        \item If $\tg(P(v_1)) = \tg(P(v_2))$, return $\true$.
          \tabto{5.9cm}\textcolor{gray}{// same object, different offsets}

        \item If $\tg(P(v_1)) = \first$ and $\tg(P(v_2)) = \last$, return
          $\len'(o_1) \ge 2$.

        \item If $\tg(P(v_1)) = \last$ and $\tg(P(v_2)) = \first$, return
          $\len'(o_1) \ge 2$.

        \item Otherwise return $\false$.

      \end{itemize}

    \item If $\of(P(v_1)) < 0$ or $\of(P(v_2)) < 0$, return $\false$.
      \comment{out of bounds}

    \item If $v_1 \ne 0$ and $\size(o_1) \le \of(P(v_1))$, return $\false$.
      \comment{out of bounds}

    \item If $v_2 \ne 0$ and $\size(o_2) \le \of(P(v_2))$, return $\false$.
      \comment{out of bounds}

    \item If $v_1 = 0$ or $v_2 = 0$, return $\true$.
      \comment{$0$ and a valid address of an object}

    \item Return $\valid(o_1) \wedge \valid(o_2)$.
      \comment{addresses of allocated objects}

  \end{enumerate}

\end{algorithm}

\begin{algorithm}[H]

  \caption{\small$lookThrough(G, v)$}

  \label{alg:lookThrough}

  \smallskip

  \textbf{Input:}\begin{itemize}[nosep]

    \item An SMG $G = (O,V,\Lambda,H,P)$.

    \item A value $v \in V$ such that $\level(v) = 0$.

  \end{itemize}

  \textbf{Output:}\begin{itemize}[nosep]

    \item A pair $(v', Visited)$ where:
      \begin{itemize}[nosep]

        \item $v' \in V$ is the value reached after all 0+ DLSs are traversed.

        \item $Visited$ is the set of all 0+ DLSs reachable from $v$ in the
          forward direction without traversing any other object.

      \end{itemize}

  \end{itemize}

  \textbf{Method:}\begin{enumerate}[nosep]

    \item Let $Visited := \emptyset$.

    \item Let $o := o(P(v))$.

    \item If $o \notin \{ \UNDEF, \nil \}$ and $\len'(o) = 0$, then:

      \begin{itemize}[nosep]
        \item Let $Visited := Visited \cup \{ o \}$.

        \item If $\tg(P(v)) = \first$, let $v := H(o, \nfo(o), \ptr)$ and continue
          with step 2.

        \item If $\tg(P(v)) = \last$, let $v := H(o, \pfo(o), \ptr)$ and continue
          with step 2.

      \end{itemize}

    \item Return $(v, Visited)$.

  \end{enumerate}

\end{algorithm}

\enlargethispage{6mm}

\vspace*{-9mm} \section{Usage of Predator} \label{sec:tutorial} \vspace*{-3mm}

The simplest way of running Predator is to use either the \texttt{slgcc} or
\texttt{slllvm} scripts as shown below with \verb|SOURCE.c| being the C program
that Predator should analyse:\vspace*{1mm}

\verb|    /PATH_TO_predator_DIRECTORY/sl/slgcc  SOURCE.c|

\verb|    /PATH_TO_predator_DIRECTORY/sl/slllvm SOURCE.c|

\smallskip

The run of Predator can be influenced by a number of options summarised in
Table~\ref{tab:options}. The options can be passed to Predator via the
environment variable \verb|SL_OPTS|). Further settings can then be provided
via the \texttt{config.h} file from the Predator distribution as discussed
below.

Predator can also be invoked directly through the chosen compiler. In
particular, one can proceed as follows with the GCC compiler:

\begin{center}\verb|gcc [CFLAGS] -fplugin=LIBSL [SL_OPTS] SOURCE.c|
\end{center}

Here, \verb|LIBSL| represents a path to the Predator plug-in, which always
ends with the \texttt{.so} suffix for GCC. For instance, when starting
Predator from its main directory, \verb|LIBSL| should be replaced by
\verb|./sl_build/libsl.so|. When launching Predator this way, one can use
compiler options \verb|CFLAGS|, such as \mbox{-m32/-m64}, as well as Predator
options \verb|SL_OPTS| described in Table~\ref{tab:options}.

For the Clang/LLVM compiler, one has to first create the bitcode file and
then start the analysis:

\bigskip

\verb| clang [CFLAGS] -Xclang -fsanitize-address-use-after-scope\|

\verb|       -g -S -emit-llvm SOURCE.c -o SOURCE.bc|

\verb| opt SOURCE.bc -lowerswitch -load LIBSL -sl [SL_OPTS]|

\bigskip

A further possibility not requiring additional options but available on Linux
only is the following:\vspace*{1.5mm}

\verb|clang -g SOURCE.c -Xclang -load -Xclang LIBSL|

\vspace*{1.5mm} For Clang/LLVM, the Predator plug-in has the \texttt{.so} suffix
on Linux and the \texttt{.dylib} suffix on Darwin.

\vspace*{-6mm} \subsection{Deeper Configuration of Predator} \vspace*{-3mm}

Apart from using the above described options, one can further configure the
behaviour of Predator using the \texttt{config.h} file from its distribution.
After that, Predator must, of course, be recompiled. Moreover, usage some of the
above described options can be replaced by changing the \texttt{config.h} file.

\enlargethispage{5mm}

Via the \texttt{config.h} file, one can control various aspects of the
\emph{abstractions} used in Predator. For instance, one can say whether the
abstraction should be performed at each basic block or at loop points and/or
returns from function calls only, whether abstraction to singly-/doubly-linked
list segments is allowed, what the various abstraction thresholds are, set an
additional cost for introducing list segments by abstraction, or decide whether
the abstraction of SMGs should be applied to longest sequences of SMG nodes that
are amenable to abstraction only or whether it can be applied to shorter
sequences too. One can also decide whether \emph{integer intervals} may be used
to represent values and/or offsets and whether they can widened. 

\begin{table}[p]
  \catcode`\-=12
  \centering

\caption{Predator plug-in options. For the GCC-plugin, use with the prefix
\texttt{-fplugin-arg-libsl}.}

\begin{tabular*}{\textwidth}{p{0.43\textwidth}|p{0.5\textwidth}}
  \bf Options                    & \bf Description\\
\hline
  {\tt-help}                     & Help\\
  {\tt-verbose=<uint>}           & Turn on verbose mode \\
  {\tt-pid-file=<file>}          & Write PID of self to <file> \\
  {\tt-preserve-ec}              & Do not affect the exit code \\
  {\tt-dry-run}                  & Do not run the analysis \\
  {\tt-dump-pp[=<file>]}         & Dump linearised CL code \\
  {\tt-dump-types}               & Dump also type info \\
  {\tt-gen-dot[=<file>]}         & Generate CFGs \\
  {\tt-type-dot=<file>}          & Generate type graphs \\
  {\tt-args=<peer-args>}         & Arguments given to the analyser (see below)\\
  &\\
  \bf Peer arguments             & \\
  {\verb| track_uninit|}         & Report usage of uninitialised values \\
  {\verb| oom|}                  & Simulate possible shortage of memory 
                                   (\texttt{malloc} can fail) \\
  {\verb| no_error_recovery|}    & No error recovery, stop the analysis as
                                     soon as an error is detected\\
  {\verb| memleak_is_error|}     & Treat memory leaks as an error \\
  {\verb| exit_leaks|}           & Report memory leaks while executing
                                     a no-return function\\
  {\verb| verifier_error_is_error|} & Treat reaching of
                                     \verb|__VERIFIER_error()| as an error \\
  {\verb| error_label:<string>|} & Treat reaching of the given label as an error \\
  {\verb| int_arithmetic_limit:<uint>|} & The highest integer number
                                     Predator can count to\\
  {\verb| allow_cyclic_trace_graph|} & Create a node with two parents on
                                     entailment \\
  {\verb| forbid_heap_replace|}  & Do not replace a previously tracked node if
                                     entailed by a new one \\
  {\verb| allow_three_way_join[:<uint>]|}  & Using the general join of
                                             possibly incomparable SMGs
                                             (so-called three-way join) 
     \begin{itemize}[nosep,after=\vspace{-\baselineskip}]
                 \item[0] never
                 \item[1] only when joining nested sub-heaps
                 \item[2] also when joining SPCs if considered useful
                 \item[\bf 3] always 
     \end{itemize}\\

  {\verb| join_on_loop_edges_only[:<int>]|}  &  
     \vspace{-\topsep}\begin{itemize}[nosep,after=\vspace{-\baselineskip}]
     \item[-1] never join, never check for entailment,
               always check for isomorphism
     \item[0] join SPCs on each basic block entry
     \item[1] join only when traversing a loop-closing edge,
              entailment otherwise
     \item[2] join only when traversing a loop-closing edge,
              isomorphism otherwise
     \item[\bf 3] same as 2 but skips the isomorphism check if possible
     \end{itemize}\\

  {\verb| state_live_ordering[:<uint>]|} & On the fly ordering of 
                                           SPCs to be processed 
     \begin{itemize}[nosep,after=\vspace{-\baselineskip}]
     \item[0] do not try to optimise the order of heaps 
     \item[1] reorder heaps when joining 
     \item[\bf 2] reorder heaps when creating their union (list of SMGs) too 
    \end{itemize}\\

  {\verb| no_plot|}              & Do not generate graphs (ignore all calls of
                 \verb|__sl_plot*()| and  \verb|__VERIFIER_plot()|) \\
  {\verb| dump_fixed_point|}     & Dump SPCs of the obtained fixed-point \\

  {\verb| detect_containers|}    & Detect low-level implementations of high-level
  list containers and operations over them (such as various initialisers,
  iterators, etc.) \cite{containers16}. \\

\end{tabular*}
  
  \label{tab:options}
\end{table}

One can enable/disable the \emph{call cache} implementing a table of summaries
and set its various parameters (e.g., whether matches in the table are sought
using isomorphism or entailment, how the cache should be pruned, etc.). One can
choose the \emph{scheduler} choosing basic blocks to be explored (corresponding
to a depth-first search, breadth-first search, or a load-driven search choosing
blocks with fewest SPCs waiting to be explored).

One can specify various \emph{limits} of the analysis such as the maximum call
depth, the maximum integer constant to be tracked, or the maximum integer to be
used as the minimum length of a list segment. One can also control whether parts
of the generated SPCs should be shared using a \emph{copy-on-write} mechanism.
One can control how much Predator should try to \emph{recover} after a bug is
found. Further, one can also enable/disable a static pre-analysis trying to
detect variables that are \emph{dead} at certain program locations whose results
may subsequently be used to prune variables tracked by the main analysis. 

The \texttt{config.h} file can be used to activate various \emph{debugging
outputs} too. Further information about what can be configured and how can be
found directly in the \texttt{config.h} file, which contains many explanatory
comments too.

\vspace*{-5mm} \subsection{Installing and Using PredatorHP} \vspace*{-3mm}

\enlargethispage{5mm}

In order to be able to use the Predator Hunting Party (PredatorHP), one has to
download its binary version or source code from
\url{https://www.fit.vutbr.cz/research/groups/verifit/tools/predatorhp}. When building from sources, one has
to make sure that \texttt{git}, \texttt{python}, and all dependencies for
Predator itself are installed. In the directory with PredatorHP, one can use the
script \verb|build-all.sh|. To analyse a single program using PredatorHP, one
can use the following script (whose options are described in
Table~\ref{tab:hpoptions}):

\verb|    predatorHP.py --propertyfile=<prpfile> [--witness=<file>]|

\noindent\verb|                    [--compiler-options=CFLAGS] SOURCE.c|

\begin{table}[h]
  \catcode`\-=12
  \centering
  \caption{Options of the Predator Hunting Party.}
\begin{tabular*}{\textwidth}{p{0.43\textwidth}|p{0.5\textwidth}}
  \bf Options                  & \bf Description\\
\hline
  {\verb|-h, --help|}              & Help\\
  {\verb|-v, --version|}           & Show the program's version number \\
  {\verb|--propertyfile=<prpfile>|} & A <prpfile> specifying properties to
                                      be verified according to SV-COMP rules\\
  {\verb|--compiler-options=CFLAGS|} & Specify options given to compiler
                                   (e.g.~\verb|--compiler-options="-m32 -g"|)\\
  {\verb|--witness=<file>|}        & Write the witness trace in XML to <file> \\

\end{tabular*}
  \label{tab:hpoptions}
  \vspace*{0mm}
\end{table}
\end{document}